\documentclass{jfm}

\usepackage{graphicx}
\usepackage{newtxtext}
\usepackage{newtxmath}
\usepackage{natbib}
\usepackage{hyperref}
\hypersetup{
    colorlinks = true,
    urlcolor   = blue,
    citecolor  = black,
}

\newcommand{\RomanNumeralCaps}[1]
\linenumbers

\usepackage{xcolor}
\definecolor{kimlightblue}{RGB}{95,178,194}	
\usepackage{subcaption} 
\usepackage{tikz}

\graphicspath{
    {./FiguresHL/}
}

\newcommand{\mean}[1]{\langle #1 \rangle}
\newcommand{\swi}{second wind-induced}
\DeclareUnicodeCharacter{03BC}{$\mu$}

\shorttitle{Exploring the Weber dependency of jet fragmentation}
\shortauthor{R. Vallon, M. Abid and F. Anselmet}

\title{Exploring the Weber dependency of jet fragmentation: a Direct Numerical
Simulation investigation}

\author{Romain Vallon \corresp{\email{romain.vallon@centrale-marseille.fr}},
Malek Abid \and Fabien Anselmet\aff{1}}

\affiliation{\aff{1} Aix Marseille University, CNRS, Centrale Méditerranée, IRPHE,
Marseille, France} 

\begin{document}

\maketitle

\begin{abstract} 

  Jet fragmentation is investigated through a Direct Numerical Simulation
  campaign using Basilisk \citep{popinet_basilisk_2013}. The simulations span
  over one order of magnitude of gaseous Weber numbers (13 to 165), i.e. over the 
  \swi{} and atomization regimes, and the jets develop over distances up to 28
  nozzle diameters. The study focuses on the size and velocity distributions of
  droplets, as well as their joint distribution. Two models derived from
  different theoretical backgrounds, the statistical description of the
  turbulence intermittency \citep{novikov_distribution_1997} and the empirical
  description of the ligament-mediated fragmentation
  \citep{villermaux_ligament-mediated_2004}, are compared for describing the
  droplet size distribution close to the nozzle. The characteristics of the
  size-velocity joint distribution are explained using the vortex ring theory
  \citep{saffman_vortex_1992} which highlights two sources of fragmentation.
  Finally, the joint histogram of the particulate Reynolds and Ohnesorge numbers
  is analysed and a normalisation is suggested. It reveals that the
  delimitations of the droplet phase space, once properly normalised, are
  self-similar and independent of the gaseous Weber number, both numerically and
  experimentally.

\end{abstract}


\section{Introduction \label{sec:dns.intro}}

Jet fragmentation occurs in numerous natural mechanisms and industrial
applications. It can appear in the form of an ocean or lava spray when waves
crash on the shore or during volcanic eruption, yet it is more common to find
this physical mechanism in  medication sprays, fuel injection systems of
combustion engines or agricultural sprinkling. Jet fragmentation can be a
challenging configuration to be studied numerically. Fragmentation flows of high
Reynolds and Weber numbers present a large diversity of scales and fluid objects
whose dynamics are partly governed by the surface tension and the turbulent
characteristics of the flow, which gives them a high complexity. Their Direct
Numerical Simulation (DNS) requires to solve the two phase Navier Stokes
equations with surface tension. A fine resolution of the interfaces is of utmost
importance and can be achieved with an optimized use of computing resources
thanks to adaptive grids.  Those multiphase flows result from the injection of a
dense phase at a velocity $U_{inj}$ into a lighter phase through a nozzle of
diameter $d_n$ and produce a polydisperse spray. The phases are denoted by
subscript $i$ which takes the value 1 for the injected dense phase and 2 for the
lighter phase. Both phases are respectively renamed liquid and gas in the
following.  With the Reynolds ($\textit{Re}$) and Weber ($\textit{We}$) numbers,
the Ohnesorge ($\textit{Oh}$) number completes the list of governing
dimensionless numbers.  The first one represents the ratio of inertia over
viscosity and the second one the ratio of inertia over surface tension. The
latter relates to the droplet deformation and represents the ratio of viscosity
over the product of the surface tension and inertia. Their respective
expressions follow:

\begin{equation}  \textit{Re}_i = \frac{\rho_i U_{inj}
d_n}{\mu_i},~~~~\textit{We}_i=\frac{\rho_i U_{inj}^2
d_n}{\sigma},~~~~\textit{Oh}_1=\frac{\mu_1}{\sqrt{\rho_1 d_n \sigma}}
\label{eq:dns.dimensionless_numbers}
\end{equation}

\noindent where $\rho_i$ and $\mu_i$ denote the density and the dynamic
viscosity of the phase $i$ and $\sigma$ the surface tension between the two
phases, taken as constant.

\begin{figure}
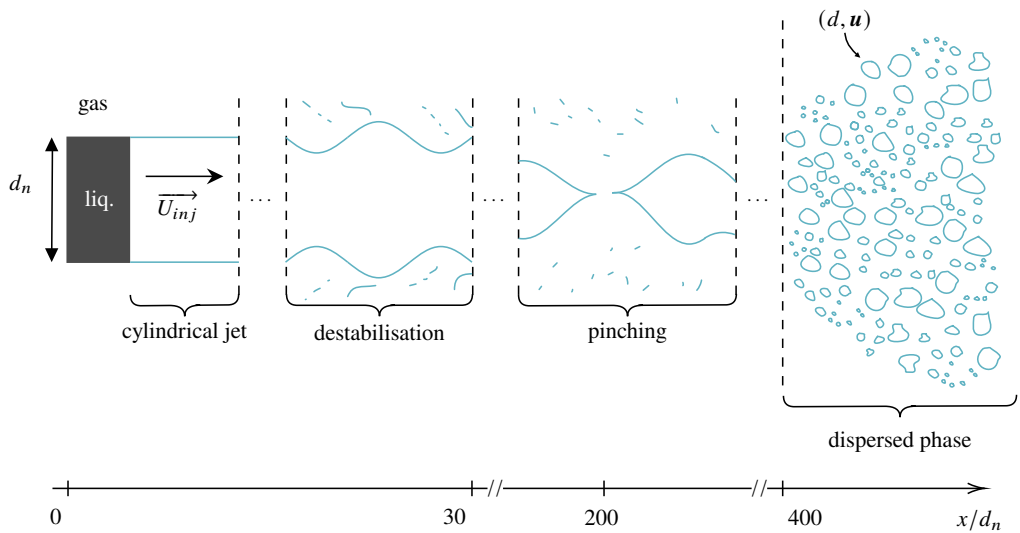

  \centering
  \resizebox{\textwidth}{!}{%

    \tikzset{every picture/.style={line width=0.75pt}} 



  }
  \caption{
    Scheme of the fragmentation in the \swi{} regime of a
    liquid jet into a dispersed phase composed of droplets with size $d$ and
    velocity vector $\boldsymbol{u}$. The distances indicated here correspond to
    the ones observed in the experiments of \citet{felis_experimental_2020},
    for which the jet lies in the \swi{} regime of fragmentation
    with $We_2 = 24$.
  }
  \label{fig:scheme_frag}
\end{figure}

\citet{lefebvre_atomization_2017} categorized five fragmentation regimes for
non-assisted fragmentation of round jets, whose delimitations mainly depend on
the Weber number.  The focus is given here on two of them: the \swi{} regime for
which $\textit{We}_2\in[13,40.3]$ and the so-called atomisation regime for which
$\textit{We}_2>40.3$.  Complementary, the jet configurations are distinguished
between large jets, $d_n>1 \textnormal{mm}$, and small jets, $d_n<1
\textnormal{mm}$.  In addition, the fragmentation of a jet is often split into
several breakup types: the primary and the secondary breakups.  The former
corresponds to the generation of elements only coming from the dense core, while
the latter considers large elements dumped from the core which undergo further
fragmentation, illustrated in figure \ref{fig:scheme_frag}. Thus, the physical
border of the two breakup types is the location where the dense core pinches off
and generates large scale elements, which are unstable in flows of moderate or
large liquid Reynolds number $\textit{Re}_1$ and gaseous Weber number
$\textit{We}_2$.

Numerical studies of jet fragmentation mainly focus on the primary breakup
region, close to the nozzle, due to limitations on computational resources and
numerical challenges
\citep{gorokhovski_modeling_2008,fuster_numerical_2009,tryggvason_direct_2011,popinet_numerical_2018}.
\citet{zandian_planar_2017} realised DNS to study the evolution of a planar jet
and specifically focused on the development of three-dimensional instabilities.
\citet{ling_spray_2017} studied a quasi planar gas-liquid mixing layer at
moderate density ratio ($\rho_1/\rho_2=20$, $Re_1=160000$, $We_2=20$) thanks to
finely resolved DNS. They were able to explain precisely the development of
instabilities on the sheet interface.  They captured the development of Taylor
Culick instabilities \citep{taylor_dynamics_1959,culick_comments_1960} as well
as the fragmentation of a ligament into droplets and finally compared the
droplet size distribution obtained for different grid refinements with the
logarithmic normal and $\Gamma$ laws.

On the side of round liquid jets, the latest studies rely on DNS using the code
Basilisk \citep{popinet_basilisk_2013} or the SPH method.
\citet{chaussonnet_three-dimensional_2018} used the latter to explore the
droplet population produced by a twin-fluid atomizer at high pressure up to
$x/d_n\approx10$ ($\rho_1/\rho_2=93$, $Re_1=1.27\times10^7$, $We_2=1375$).
\citet{ling_direct_2017} used Basilisk to observe the influence of viscosity on
the fragmentation of a round biodiesel jet ($\rho_1/\rho_2=78.2$, $Re_1=1450$,
$We_2=12.9$) developing up to $x/d_n\approx20$ while testing different grid
refinements.  \citet{zhang_modeling_2020} observed the fragmentation of a round
diesel jet injected through a solid G-spray injector, developing up to
$x/d_n\approx20$ as well ($\rho_1/\rho_2=233$, $Re_1=13400$, $We_2=177$).
Through their study, the authors were able to observe the fragmentation of the
liquid core into droplets as well as the spatial distribution of the vortices
along the core.  In addition, the authors modeled the droplet size distribution
relative to the azimuthal angle by a hyperbolic tangent function.  Finally, both
studies relying on Basilisk compared the logarithmic stable and the $\Gamma$
laws, the former being derived in the context of turbulence
\citep{novikov_infinitely_1994,novikov_distribution_1997} and the latter in the
context of ligament mediated fragmentation
\citep{villermaux_ligament-mediated_2004,villermaux_fragmentation_2020}, to fit
the droplet size distribution and concluded on the better performance of the fit
with the logarithmic normal law in linear mode, i.e. fitting the signal as it
is.

Later experimental studies \citep{stevenin_flow_2016,felis_experimental_2020}
used specific droplet tracking velocimetry (DTV) and laser Doppler velocimetry
(LDV) apparatus to explore the dispersed zone of agricultural-like jets
($\rho_1/\rho_2=828.5$, $Re_1=41833$, $We_2=24$). The measurements were carried
far away from the nozzle, $x\geqslant400~d_n$, in the zone where the liquid core
is fully atomized and where only the secondary breakup occurs. Based on those
joint size-velocity measurements, \citet{vallon_multimodal_2021} highlighted the
multimodal nature of the droplet size distribution along with the existence of
droplet subgroups, each of them being characterised by a specific pair of size
and velocity.

The present paper aims to complete the experimental campaigns by studying
numerically the field close to the nozzle in similar flow conditions up to
$x/d_n=28$ in order to have a more global view of the fragmentation process that
agricultural like jets undergo as well as to compare the logarithmic and
$\Gamma$ laws for describing the droplet size distribution. To do so, section
\ref{sec:dns.flowmodel_prmframe} presents the flow modeling and the parameter
framing. Section \ref{sec:dns.flowcharac_dropstat} is dedicated to the analysis
of the overall flow characteristics. Section \ref{sec:dns.two_speed_frag}
focuses on the analysis of the droplet population statistics and the mechanisms
from which they are generated, while section \ref{sec:ccl} puts in perspective
the conclusions and opens up on the study of the droplet topography.

\section{Flow modeling and parameter framing \label{sec:dns.flowmodel_prmframe}}

This section presents the governing equations, the numerical methods, the choice
of the physical configurations, the numerical configuration and the computation
cost. It finally introduces the selection of the most unstable mode of the jet,
in order to stimulate the jet fragmentation. 

\subsection{Governing equations \label{sec:dns.governing_eq}}

Direct Numerical Simulations (DNS) aim to resolve all time and length scales by
solving the Navier-Stokes equations. However, this resolution is often limited
by the available computational resources. The fragmentation mechanism under
consideration occurs at low Mach numbers neglecting gravitational forces and
involves two immiscible, incompressible fluids. The flow dynamics is then
governed by the unsteady Navier-Stokes equations and can be expressed in the
theoretical framework of a one fluid flow with variable density and viscosity
as:

\begin{equation}  \frac{\partial \rho \textbf{u}}{\partial t} +
  (\textbf{u} \cdot \pmb{\nabla})(\rho \textbf{u}) = -\nabla p + \pmb{\nabla} \cdot \big(\mu
  (\pmb{\nabla}  \textbf{u} + \pmb{\nabla}^T \textbf{u}) \big) + \textbf{T}_\sigma
\label{eq:dns.Navier_Stokes}
\end{equation}

\begin{equation}
  \frac{\partial \rho}{\partial t} + \pmb{\nabla} \cdot(\rho\boldsymbol{u}) = 0,
  \label{eq:dns.eq_continuity}
\end{equation}

\begin{equation}
  \pmb{\nabla} \cdot \textbf{u} = 0
\end{equation}

 \noindent where $\textbf{u}$ is the velocity vector, $p$ the pressure and
 $\textbf{T}_\sigma$ the surface tension force, only defined on the liquid-gas
 interfaces. The two phases are taken into account in the one fluid framework
 through the phase indicator, named fraction field and denoted $\alpha$ in the
 following. The fraction equals 1 if a cell only contains liquid and 0 if it
 only contains gas. The one fluid viscosity and density are computed over the
 phase quantities following $\mu= \alpha\mu_1 + (1-\alpha)\mu_2$ and
 $\rho=\alpha\rho_1 + (1-\alpha)\rho_2$. Injecting the expression of $\rho$ and
 $\mu$ in Eq.  \ref{eq:dns.eq_continuity} and noting that
 $\partial_t\rho_1=\partial_t\rho_2=0$ lead to reformulate the continuity
 equation in terms of $\alpha$:

\begin{equation}
  \partial_t \alpha + \pmb{\nabla}\cdot(\alpha \textbf{u}) = 0
  \label{eq:dns.eq_continuity_alpha}
\end{equation}

\noindent which can also be seen as the advection equation of the volume
fraction. Instead of resolving the Navier Stokes equations for each phase, this
approach enables the resolution of only a single set of equations. This, though,
implies the implicit assumption that the velocity field $\textbf{u}$ evolves
continuously in space.

 \subsection{Numerical methods \label{sec:dns.num_methods}}

The DNS under consideration are computed with the solver developed by the
Basilisk community. Basilisk is an open source project which aims to develop
efficient solvers and methods which can be adapted to a wide range of
configurations \citep{popinet_basilsik-atomisation_2016}. This project is mainly
led by St\'{e}phane Popinet and benefits from the contribution of all the
Basilisk community. The present study largely relies on the atomisation code
available on the wiki of the project \citep{popinet_basilsik-atomisation_2016}.

The Navier-Stokes equations are solved for a biphasic flow with a constant
surface tension using numerical schemes similar to those of
\citet{popinet_gerris_2003} and \citet{lagree_granular_2011}.  The resolution of
the equations relies on time steps limited by the Courant-Friedrichs-Lewy (CFL)
condition, the advection scheme of Bell-Collela-Glaz
\citep{bell_second-order_1989} and an implicit solver for the viscosity. The
gas-liquid interface is tracked with a Volume-Of-Fluid (VOF) scheme which is
geometric, conservative and non-diffusive
\citep{lopez-herrera_electrokinetic_2015}.  Regarding the surface tension, the
interfacial force is calculated as $\textbf{T}_\sigma = \sigma \kappa
\textbf{n}\delta_s$, where $\kappa$ is the interface curvature, $\delta_s$ is
the interface Dirac function and $\textbf{n}$ is the normal vector to the
interface pointing outward. Considering the Continuum-Surface-Force (CSF) method
and the Peskin immersed boundary method, the interfacial force can be
approximated by $\textbf{T}_\sigma = \sigma \kappa \nabla \alpha$ where $\kappa$
is computed by the use of a height function
\citep{abu-al-saud_conservative_2018}.  A projection method is used to compute
the centered pressure gradient and the acceleration field.  The VOF scheme is
combined with an octree adaptive grid \citep{agbaglah_parallel_2011} while the
grid adaptation algorithm relies on a wavelet estimated discretization error,
described by \citet{popinet_quadtree-adaptive_2015} and used for atmospheric
boundary layer simulations by \citet{van_hooft_towards_2018}. Such grids present
the advantage of finely resolving the gas-liquid interface while having a
coarser resolution away from the interfaces, and thus decreasing the time needed
for computing the DNS. Finally, the droplet detection is achieved by a tag
function which associates a different tag to each neighbourhood of connected
cells respecting a threshold condition on the fraction field, set to
$\alpha>1\times10^{-3}$ in our DNS.

\subsection{Physical configuration and parameters \label{sec:dns.phys_config}}

The domain is a cubic box of dimension $L_x$. A liquid round jet is injected
into a quiescent gas at a mean velocity $U_{inj}$, directed along the $x$-axis,
through a disc of diameter $d_n$ and length $l_x=d_n/3.2$. The latter disc is called
nozzle in the following. The injection condition is set on the disc face located
at $x=0$ while a free stream condition is imposed at the location $x=L_x$. In
addition, a Neumann condition on the normal velocity is imposed on the lateral
faces. A sinusoidal perturbation is superimposed on the injection velocity in
order to accelerate the development of the Kelvin Helmholtz instability on the
interface. The perturbation has an amplitude $A$ and a frequency $f$ such
that the injection velocity follows $u_{inj}=U_{inj}\big(1+A \sin(2\pi f
t)\big)$. Finally, the advection timescale is defined by $T_{a}=d_n/U_{inj}$.

One aim of this study is to draw comparisons with the experiments of
\citet{felis_experimental_2020}. First and foremost, the turbulent property of
the experimental inlet velocity profile is let aside and the numerical injection
profile is set as laminar. Real world parameter values cannot be picked because
the current computational resources do not allow to compute such configurations.
For instance, the numerical constraints prohibit large values for the density
ratio, $\rho_1/\rho_2<200$, the Reynolds number, $\max(Re)=O(10^4)$ and the
surface tension, $\sigma=O(10^{-5})\textnormal{N/m}$. Those constraints are
denoted $\mathcal{C}_0$, $\mathcal{C}_1$ and $\mathcal{C}_2$.  Even if the real
world values are unreachable, a specific attention can be set on reproducing
configurations with dimensionless numbers close to the experimental ones. The
latter study carried out DTV and LDV measurements on a water jet lying in the
\swi{} fragmentation regime \citep{lefebvre_atomization_2017}.
This regime is characterised by sharp limits on the gas Weber number:
$13<We_2<40.3$. The atomisation regime is also defined on the basis of the Weber
number, $We_2>40.3$. The first priority is thus to make the DNS Weber numbers
evolve over this range of values, which defines a third constraint,
$\mathcal{C}_3$. In order to reproduce similar deformation regimes undergone by
the droplets, considering the Ohnesorge number is relevant. Experimentally,
$Oh_1=3.4 \times 10^{-3}$, reproducing the same order of values makes a fourth
constraint $\mathcal{C}_4$.  Having a density ratio of $O(10^3)$, as for
water injection in air, is impossible. Conserving the experimental viscosity
ratio $\nu_2/\nu_1=15$ could be interesting but it would slow down the
fragmentation process, which goes against the optimisation of computer
resources. One could then have a look at the conservation of the ratio
$\gamma=\mu_1/\mu_2=(\rho_1\nu_1)/(\rho_2\nu_2)$, where it is worth noting that
the quantity $\rho_i\nu_i d_n$ is homogeneous to a mass flow rate \footnote{Indeed,
  $[\rho_i\nu_id_n]=\frac{\textnormal{kg}}{\textnormal{m}^{3}}\times\frac{\textnormal{m}^2}{\textnormal{s}}\times\textnormal{m}=\textnormal{kg}\times\textnormal{s}^{-1}$.
Also, $\rho_i\nu_id_n = \rho_iU_{inj}d_n^2/Re_i$.}.  Furthermore, $\gamma$
rewrites as $\rho_1\rho_2^{-1}/(\nu_2\nu_1^{-1})=We_1 We_2^{-1} / (Re_1
Re_2^{-1}) = We_1 Re_1^{-1} / (We_2 Re_2^{-1}) = Ca_1/Ca_2$, where $Ca_i$ is the
Capillary number of the phase $i$.  Experimentally, $\gamma$ equals 55 and can
be seen such that the mass flow rate of the liquid phase is 55 times higher than the
mass flow rate of the gas phase, or equivalently $Ca_1=55Ca_2$.  Respecting this ratio
makes a fifth constraint $\mathcal{C}_5$. The list of constraints necessary to
produce configurations close to the experiments is thus:

\begin{equation} 
  \left\{ 
    \begin{array}{ll}
      \mathcal{C}_0: & \rho_1/\rho_2 < 200 \\[2pt]
      \mathcal{C}_1: & \max(Re)=O(10^4) \\[2pt] 
      \mathcal{C}_2: & \sigma=O(10^{-5})~\textnormal{N}/\textnormal{m} \\[2pt]
    \end{array}
  \right.  
  ~~\&~~~
  \left\{ 
    \begin{array}{ll}
      \mathcal{C}_3: & We_2\in[13,40.3]~\textnormal{or}~We_2> 40.3 \\[2pt]
      \mathcal{C}_4: & Oh_1=O(10^{-3}) \\[2pt] 
      \mathcal{C}_5: & \rho_1\nu_1 / \rho_2\nu_2 = 55
    \end{array}
  \right.  
\end{equation}

\noindent which let the parameters $\rho_1$, $\rho_2$, $\nu_1$, $\sigma$,
$U_{inj}$ and $d_n$ free to choose. In order to keep a constant geometry between
different DNS, the nozzle diameter is set as constant and only the injection
velocity varies to cover the range of Weber and Reynolds of interest.
Table \ref{tab:dns.phys_prm} gives the values chosen for the parameters along
with the corresponding Ohnesorge number. Table \ref{tab:dns.uinj_we_re_f_sr} lists
the chosen injection velocities and the corresponding gaseous Weber and liquid
Reynolds numbers. Note that the Ohnesorge number is constant over all the
configurations. Thus, for all the DNS, the critical breakup Weber number for a
given droplet size is constant \citep{hinze_fundamentals_1955}.  Additionally,
the breakup regimes in the secondary atomisation are defined on the same range
of Weber numbers \citep{faeth_structure_1995} for the 10 DNS.  \textit{In fine},
the breakup regimes of the droplets are set and identical for any pair
$(Re_1,We_1)$ and the DNS explore different breakup regimes of the jet by
ranging from low to moderate $Re$ and $We$ numbers.

\begin{table} 
  \begin{center} 
    \def~{\hphantom{0}} 
    \begin{tabular}{ccccccc}
      $\rho_1~(\textnormal{kg}/\textnormal{m}^{3})$  &
      $\rho_2~(\textnormal{kg}/\textnormal{m}^{3})$ &
      $\nu_1~(\textnormal{m}^2/\textnormal{s})$ & $\nu_2
      ~(\textnormal{m}^2/\textnormal{s})$ &
      $\sigma~(\textnormal{N}/\textnormal{m})$ & $d_n~(\textnormal{m})$ & $Oh_1$
      \\[3pt]
      1 & $1/55$ & $10^{-6}$ & $10^{-6}$ & $10^{-5}$ &
      $4.48\times10^{-3}$ & $4.725\times 10^{-3}$\\ 
\end{tabular} 
\caption{
  Fixed parameters and the corresponding Ohnesorge number.
}
\label{tab:dns.phys_prm} 
\end{center} 
\end{table}


\begin{table} 
  \begin{center} 
    \def~{\hphantom{0}}
    \begin{tabular}{ccccccc|ccccccc}
    DNS & $U_{inj}$ & $We_2$ & $Re_1$ & $f$ & $St$ &&&
    DNS & $U_{inj}$ & $We_2$ & $Re_1$ & $f$ & $St$\\
    & $(\textnormal{m}/\textnormal{s})$ & & & $(\textnormal{kHz})$ & &&& &
    $(\textnormal{m}/\textnormal{s})$ & & & $(\textnormal{kHz})$ & \\[4pt]
    1 & 1.357 & 15 & 6079 & 0.340 & 1.12&&& 6 & 2.216 & ~40~~ & ~9928 & 0.901 & 1.82\\
    2 & 1.567 & 20 & 7020 & 0.454 & 1.30&&& 7 & 3.0~~ & ~73.3 & 13440 & 1.666 & 2.49\\
    3 & 1.787 & 26 & 8004 & 0.587 & 1.47&&& 8 & 3.5~~ & ~99.8 & 15680 & 2.272 & 2.91\\
    4 & 1.919 & 30 & 8598 & 0.676 & 1.58&&& 9 & 4.0~~ & 130.3 & 17920 & 2.972 & 3.33\\
    5 & 2.073 & 35 & 9287 & 0.795 & 1.72&&& 10 & 4.5~~ & 165~~ & 20160 & 3.730 &3.71\\
  \end{tabular}
  \caption{
    Injection velocities and corresponding gas Weber and liquid Reynolds
    numbers along with the frequency $f$ of the most unstable mode and the
    corresponding forcing Strouhal number $St$.
  } 
\label{tab:dns.uinj_we_re_f_sr} 
\end{center} 
\end{table}

\subsection{Most unstable mode for triggering the jet fragmentation
\label{sec:dns.unstable_mode}}

In order to trigger the jet fragmentation the earliest and save computational
resources, it is worth destabilizing the jet interface.  Following the work of
\citet{yang_asymmetric_1992} on the growth of waves in round jets, it is
possible to characterise the most unstable axisymmetric mode. In this work, the
author studied the stability of an infinitesimal perturbation on the surface of
a round jet of radius $a$ and derived the expression of the nondimensional
temporal growth rate for the $m$-th transversal mode, $(\alpha_r^*)^2_m$. This
derivation is recalled in appendix \ref{app.unstable_mode}.

Figure \ref{fig:dns.unstabmode_growthrate} gives the evolution of this
nondimensional growth rate for an axisymmetric perturbation, selected with $m=0$,
$U_{inj}=3.0\textnormal{m}/\textnormal{s}$ and a zero gas velocity injection.
The wavelength $ka$ of the most unstable mode is such that $(\alpha_r^*)^2_m$ is
maximum. The mode pulsation $\omega$ is then given by the imaginary part of the
growth rate. During the computation, $a$ was set to $d_n$ instead of $d_n/2$.
The difference between the pulsation of the mode computed for $a=d_n$ and
$a=d_n/2$  is of $O(1\textnormal{rad}/\textnormal{s})$ while the pulsations are
of $O(10^3\textnormal{rad}/\textnormal{s})$. The relative difference is thus of
$O(10^{-3})$, which seems acceptable to the authors. Table
\ref{tab:dns.uinj_we_re_f_sr} lists the most unstable mode frequencies $f$ for
each configuration defined in section \ref{sec:dns.phys_config} as well as the
Strouhal number based on the forcing frequency, $St=f\times d_n/U_{inj}$.


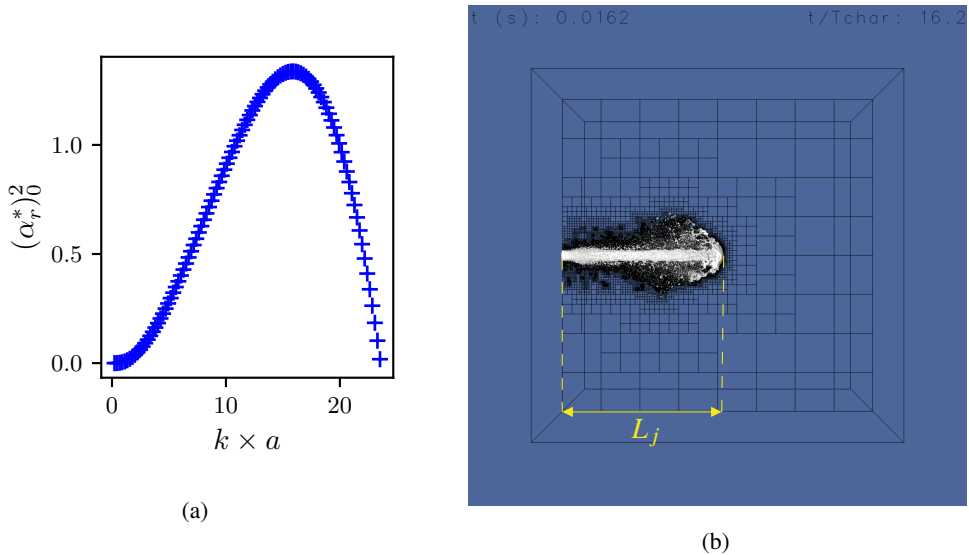
\begin{figure}
  \centering
  \begin{subfigure}[c]{0.48\textwidth}
            \centering
            \includegraphics{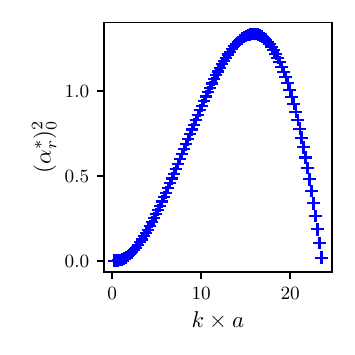}
            \caption{}
            \label{fig:dns.unstabmode_growthrate}
  \end{subfigure}%
  \hfill
  \begin{subfigure}[c]{0.50\textwidth}
            \centering
            \begin{tikzpicture}[x=0.75pt,y=0.75pt,yscale=-1,xscale=1]

            \draw (225,135) node  {\includegraphics[width=187.5pt,height=187.5pt]{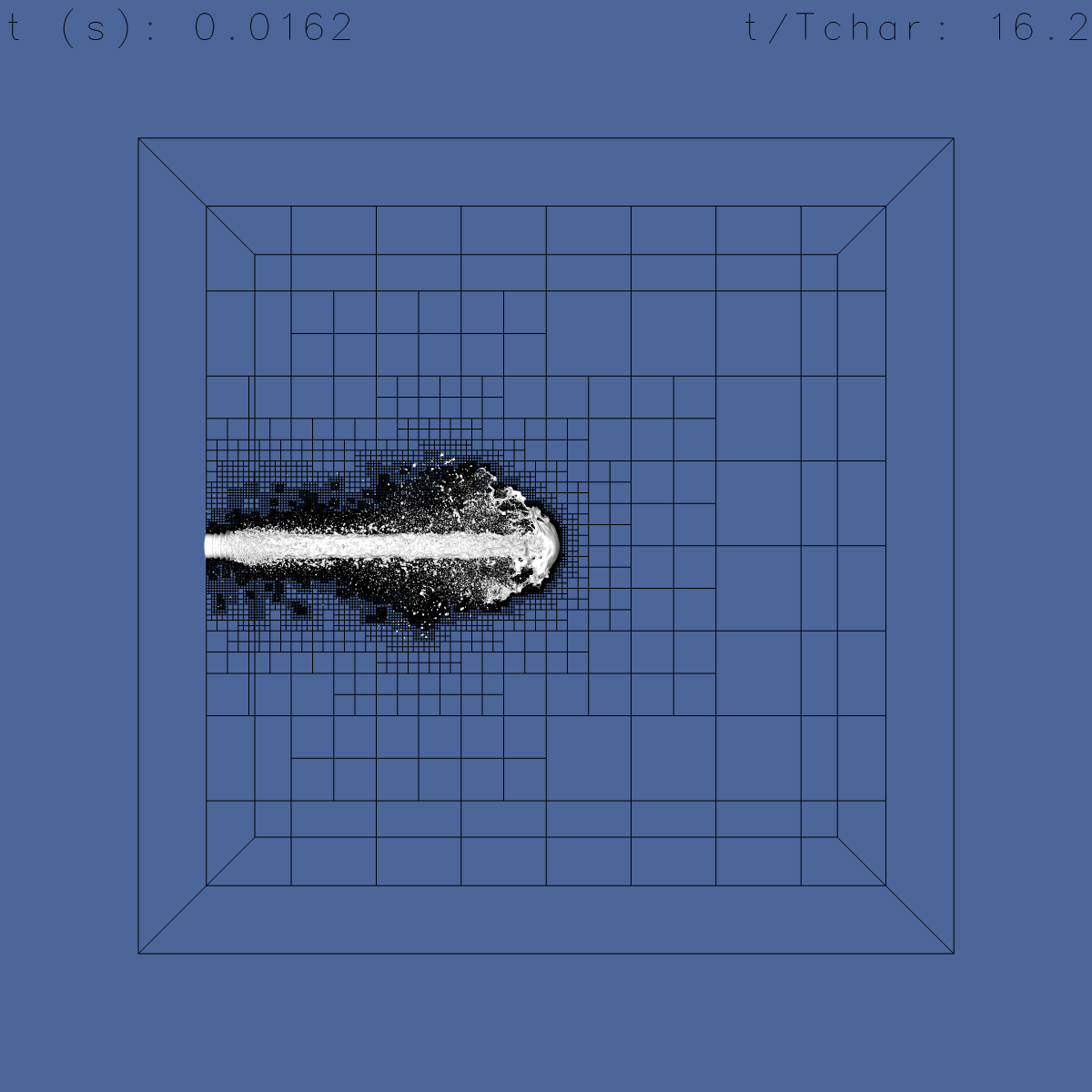}};
            \draw [color={rgb, 255:red, 248; green, 231; blue, 28 }  ,draw opacity=1 ]   (150,213) -- (224,213) ;
            \draw [shift={(227,213)}, rotate = 180] [fill={rgb, 255:red, 248; green, 231; blue, 28 }  ,fill opacity=1 ][line width=0.08]  [draw opacity=0] (5.36,-2.57) -- (0,0) -- (5.36,2.57) -- cycle    ;
            \draw [shift={(147,213)}, rotate = 0] [fill={rgb, 255:red, 248; green, 231; blue, 28 }  ,fill opacity=1 ][line width=0.08]  [draw opacity=0] (5.36,-2.57) -- (0,0) -- (5.36,2.57) -- cycle    ;
            \draw [color={rgb, 255:red, 248; green, 231; blue, 28 }  ,draw opacity=1 ] [dash pattern={on 4.5pt off 4.5pt}]  (147,136) -- (147,213) ;
            \draw [color={rgb, 255:red, 248; green, 231; blue, 28 }  ,draw opacity=1 ] [dash pattern={on 4.5pt off 4.5pt}]  (228,135) -- (227,213) ;

            \draw (180,215) node [anchor=north west][inner sep=0.75pt]  [font=\large,color={rgb, 255:red, 248; green, 231; blue, 28 }  ,opacity=1 ]  {$L_{j}$};

            \end{tikzpicture}

            \caption{}
            \label{fig:dns.10_jet_grid}
  \end{subfigure}%
  \caption{
    (a) Evolution of the nondimensional temporal growth rate $(\alpha_r^*)_0^2$
    for $U_{inj}=3.0~\textnormal{m}/\textnormal{s}$. (b) Atomisation and
    adaptive grid for $We_2=165$ (DNS 10) at $t/T_a=16.2$.
  }
  \label{fig:dns.unstabmode_dns10jet}
\end{figure}

\subsection{Numerical configuration and computational cost \label{sec:dns.num_config_cost}}

The refinement level is set to 12 and the minimum cell size in an adaptive grid
is given by $\Delta_{min}=L_x/2^{12}$. Hence the minimum cell size is
$\Delta_{min}=30.5~\mu m$ and $d_n/\Delta_{min}=146.8$. The time step is set by
the CFL condition, $|u_{max}|\Delta t / \Delta_{min}<C$, where the Courant
number is initially set to 0.8. Running the 10 DNS summed a total of 511~896
scalar hours of computation. Each of the DNS 3 to 10 ran for
$60~480~\textnormal{h}$ while DNS 1 and 2 respectively ran for $12~600~$h and
$15~456~$h. The computational performances can be tracked by checking the total
number of cells used for each DNS, $\mathcal{C}_{tot}$, the mean numerical
velocity, $\overline{\mathcal{V}_{num}}$, the maximal physical time reached by
the simulations, $t_{max}/T_a$, the maximum jet elongation, $L_{j,max}/d_n$ and
the total number of detected droplets, $N_{tot}$. The latter three are
summarized in table \ref{tab:dns.num_performances} while the other performances
parameters are given in appendix \ref{app.num_perf}. All the DNS are split into
3 runs and were computed in parallel on the Occigen HPC (CINES, France),
typically on 840 cores.  An example of atomisation produced at $We_2=165$ (DNS
10) is given in figure \ref{fig:dns.10_jet_grid}.

\begin{table} 
  \begin{center} 
    \def~{\hphantom{0}} 
    \begin{tabular}{cccccc|cccccc}
      DNS & $We_2$ & $t_{max}/T_a$ & $L_{j,max}/d_n$ & $N_{tot}$ &&&
      DNS & $We_2$ & $t_{max}/T_a$ & $L_{j,max}/d_n$ & $N_{tot}$ \\[3pt]
      1 & 15 & 34 & 28~~ & ~~~70 &&& 6 & ~40~~ & 34~~~ & 28~~ & ~3545 \\
      2 & 20 & 34 & 28~~ & ~~459 &&& 7 & ~73.3 & 24.2~ & 21.5 & ~9725 \\
      3 & 26 & 34 & 28~~ & ~1949 &&& 8 & ~99.8 & 19.75 & 17.0 & 18478 \\
      4 & 30 & 34 & 28~~ & ~2182 &&& 9 & 130.3 & 18.5~ & 14.4 & 32922 \\
      5 & 35 & 34 & 28~~ & ~2448 &&& 10& 165~~ & 16.25 & 14.2 & 45046 \\
    \end{tabular} 
    \caption{
      Maximum normalised time $t_{max}/T_a$ along with the corresponding
      normalised maximum jet elongation $L_{j,max}/d_n$ and total number of
      droplets $N_{tot}$. 
    }
  \label{tab:dns.num_performances} 
  \end{center} 
\end{table}


\section{Overall flow characteristics and droplet statistics \label{sec:dns.flowcharac_dropstat}} 

This section characterises the Turbulent Kinetic Energy (TKE) in the domain, has
a glance on the jet interface and introduces the statistics and PDFs of the
droplet population. In the following, the evolution of several variables
relatively to $t/T_a$ is analysed. 

\begin{figure}
  \centering
  \includegraphics[width=0.9\textwidth]{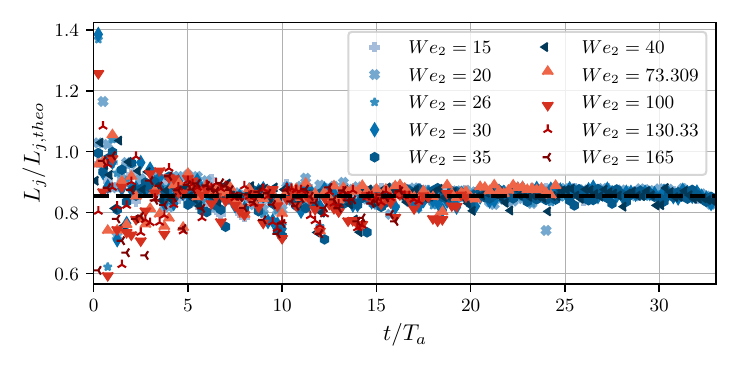}
  \caption{
    Color on-line.Temporal evolution of the jet length $L_j$ compared to the
    theoretical jet length $L_{j,theo}$ for the 10 DNS. The black dashed line
    represents the mean value $L_{jet,max}/L_{jet,theo}=0.853$ averaged over
    $t/T_a\in[0,33]$.  The blue colours denote the DNS in the \swi{} regime and
    the red colours the DNS in the atomisation regime.
  }
  \label{fig:dns.front_evolution}
\end{figure}

If the liquid core motion was the one of a solid cylinder, then the jet length
would theoretically be $L_{j,theo}=d_n\times t/T_a$, i.e. $L_{j,theo}/d_n =
t/T_a$. However, a lag of the jet tip relatively to this theoretical position is
observed. In order to link $t/T_a$ and the actual jet length $L_{j}$, figure
\ref{fig:dns.front_evolution} gives the temporal evolution of
$L_{j}/L_{j,theo}$. Here, $L_j$ is defined as the 99\% quartile of the axial
positions of the interface cells, $\alpha\in]0,1[$, and not the maximum
position. Doing so enables to exclude liquid cells which would
exist on the upstream face of the jet tip as well as to smooth out the effect of
the grid refinement.  Thus, the length of the jet equals in average 85\% of
the theoretical length, $L_j/d_n \approx 0.85\times t/T_a$, and the velocity of
the jet front equals $0.85~U_{inj}$. Note as well that $t/T_a=33$
corresponds to the instant when the jet exits the computational domain. 

\subsection{Turbulent kinetic energy \label{sec:dns.tke}}

One aim of this study is to draw conclusions on the statistics of the droplet
population. To ensure converged statistics, the flow needs to reach a
statistically steady state. Looking at the turbulent kinetic energy $k_i$
enables to conclude on this, primarily the one of the gas phase. As shown in
Table \ref{tab:dns.num_performances}, the jet extension observed for
$We_2\in[73, 165]$ (DNS 7 to 10) is smaller than the length $L_x$ of the domain.
Thus, a statistically steady state at the scale of the domain cannot be
achieved. Even so, it is possible to slice the domain in different sections
along the $x$-axis and conclude on the flow steadiness in each section.

\begin{figure}
  \centering
  \includegraphics{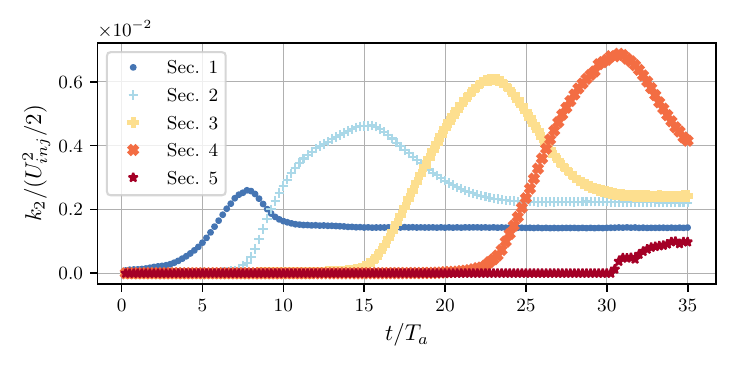}
  \caption[
    Turbulent kinetic energy in the gas phase for $We_2=40$.
  ]{
    Turbulent kinetic energy in the gas phase for $We_2=40$ (DNS 6). 
  }
  \label{fig:dns.tke_gas}
\end{figure}

The domain of length $L_x$ is sliced in 5 sections along the $x$-axis. The fifth
section represents the outlet side of the domain and its length is set to $d_n$.
The rest of the domain is evenly sliced with a slice thickness equal to
$(L_x-d_n)/4$. The sections are denoted from 1 to 5, going from the nozzle to
the outlet face. The turbulent kinetic energy is computed for both the gas and
the liquid following $k_i=\frac{1}{2}\int_{V_s}(u_{x,i}^{'2} + u_{y,i}^{'2} +
u_{z,i}^{'2})dV$ with $V_s$ the volume of the section under consideration.
Figure \ref{fig:dns.tke_gas} shows the time evolution  of the TKE for $We_2=40$
(DNS 6).  The evolution is similar in each slice: $k$ increases when the jet
head enters the section, reaches a maximum, decreases when the head enters the
following section and finally reaches a plateau.  The time sampling is set with
a step $\Delta(t/T_a)=1/4$ and smooths the fluctuations out of the plateau
region. The increase of the $k_2$ maximum and asymptote values between the
slices is due to the ongoing fragmentation and the newly created droplets, which
increases the agitation of the gas phase. Thus, once the jet head fully exits a
section, the flow reaches a statistically steady state. One could expect that
the TKE around the jet head, measured from a Lagrangian point of view, would
reach an asymptote as well and, thus, a statistically steady state.

\subsection{Close up on the jet interface \label{sec:dns.interfaces}}

This section explores qualitatively the interface of the jet in two regions of
interest: close to the nozzle where the unstable mode develops and around the
tip of the jet where the front extends and fragments.

\subsubsection{Development of the unstable mode \label{sec:dns.dev_unstable_mode}}

\begin{figure}
  \centering

  \begin{subfigure}[c]{0.40\textwidth}
            \centering
            \includegraphics[width=\textwidth]{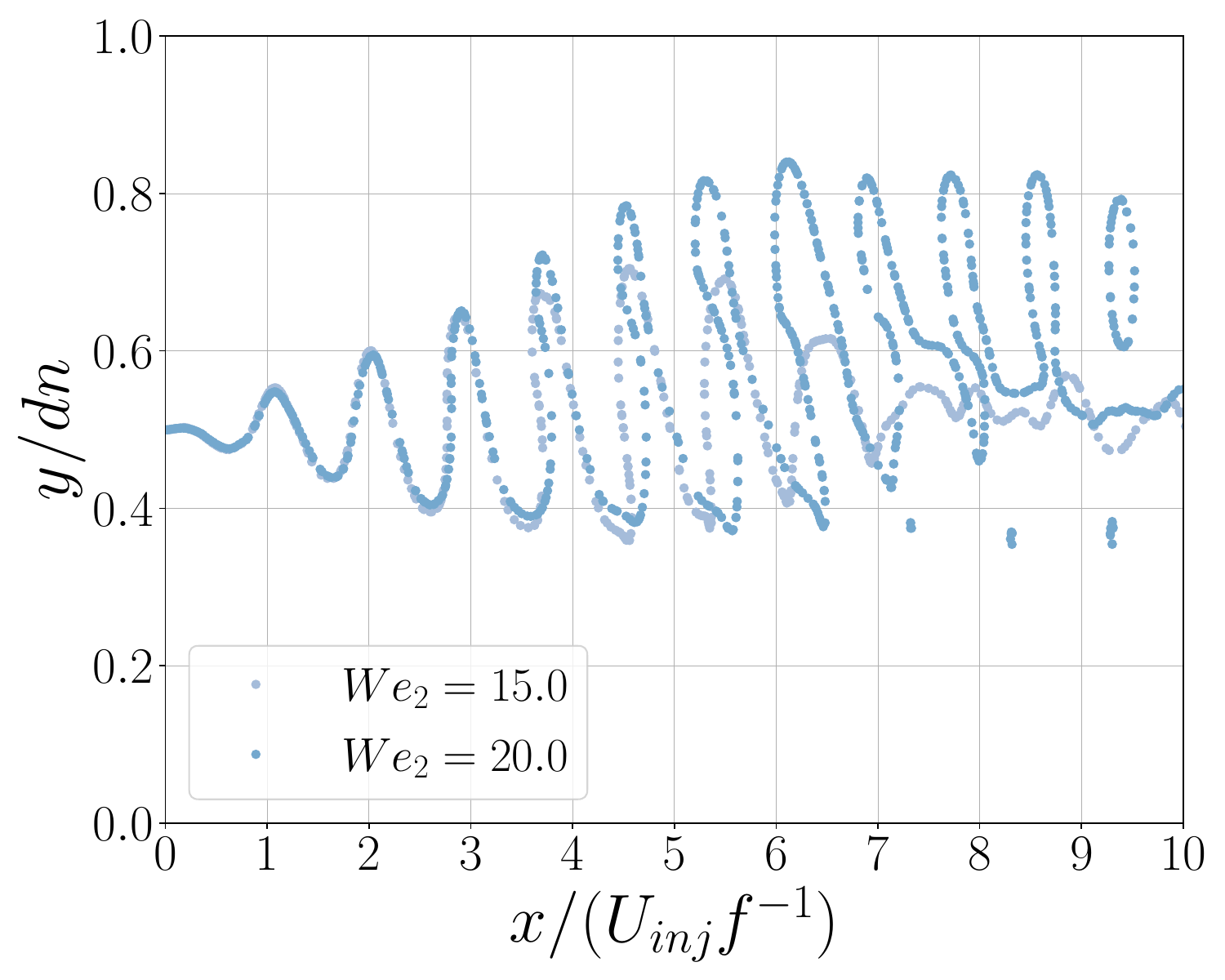}
            \caption{$We_2\in\{15,20\}$ (DNS 1 and 2)}
                \label{fig:dns.wavesuperposition_SWI1}
  \end{subfigure}%
  \hfill
  \begin{subfigure}[c]{0.40\textwidth}
            \centering
            \includegraphics[width=\textwidth]{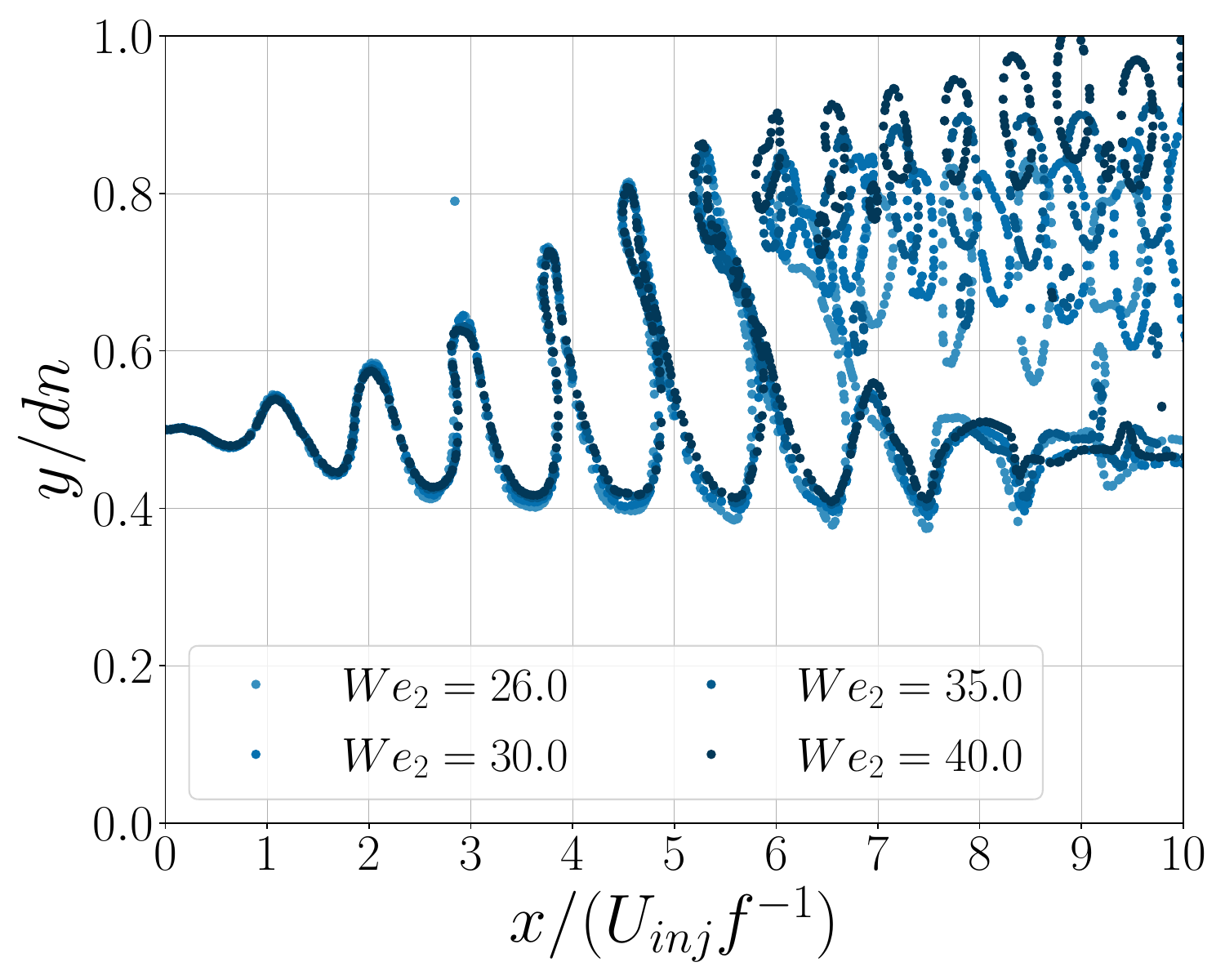}
            \caption{$We_2\in[26,40]$ (DNS 3 to 6)}
                \label{fig:dns.wavesuperposition_SWI2}
  \end{subfigure}%

  \begin{subfigure}[c]{0.40\textwidth}
            \centering
            \includegraphics[width=\textwidth]{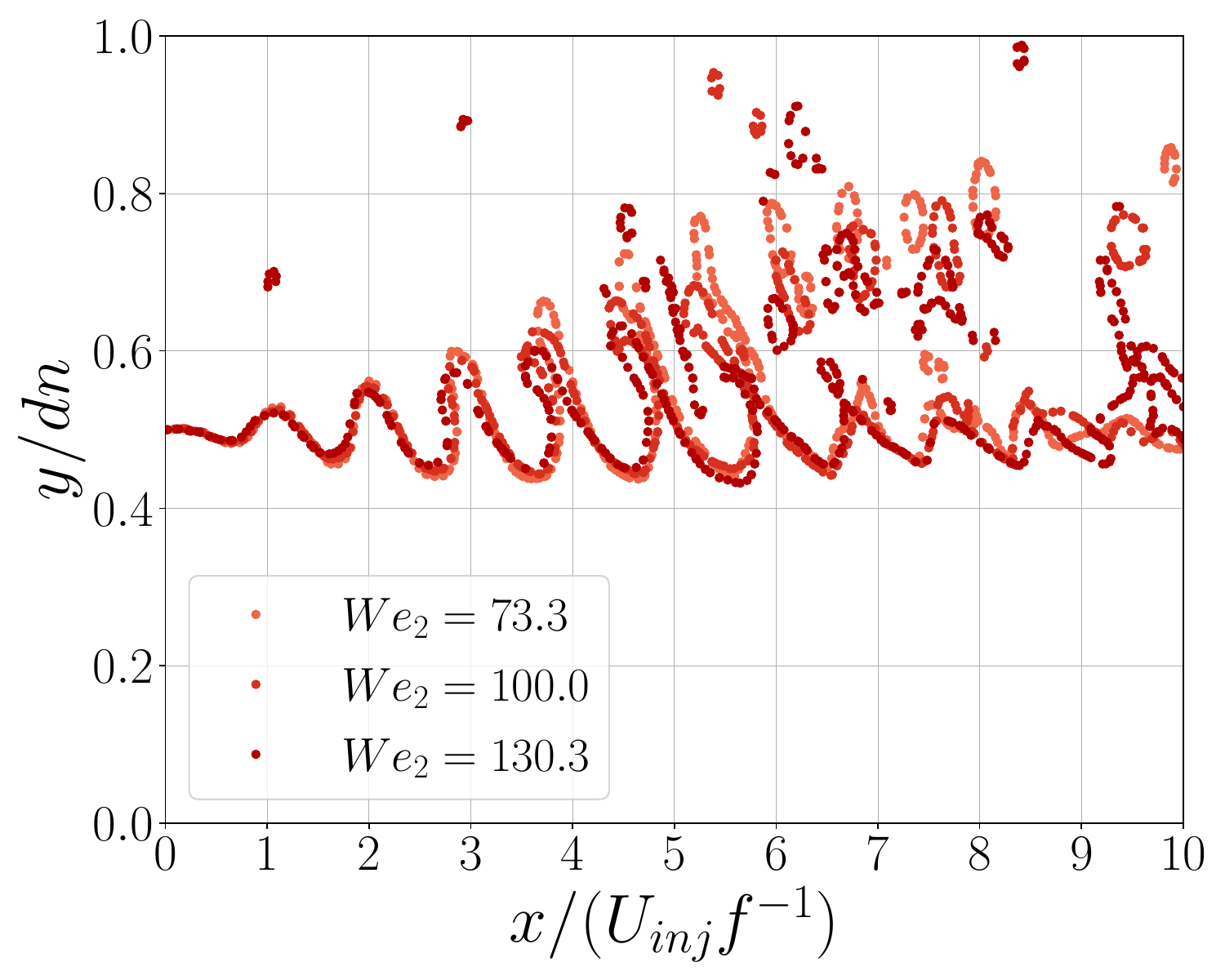}
            \caption{$We_2\in[73,130]$ (DNS 7 to 9)}
                \label{fig:dns.wavesuperposition_ATO1}
  \end{subfigure}%
  \hfill
  \begin{subfigure}[c]{0.40\textwidth}
            \centering
            \includegraphics[width=\textwidth]{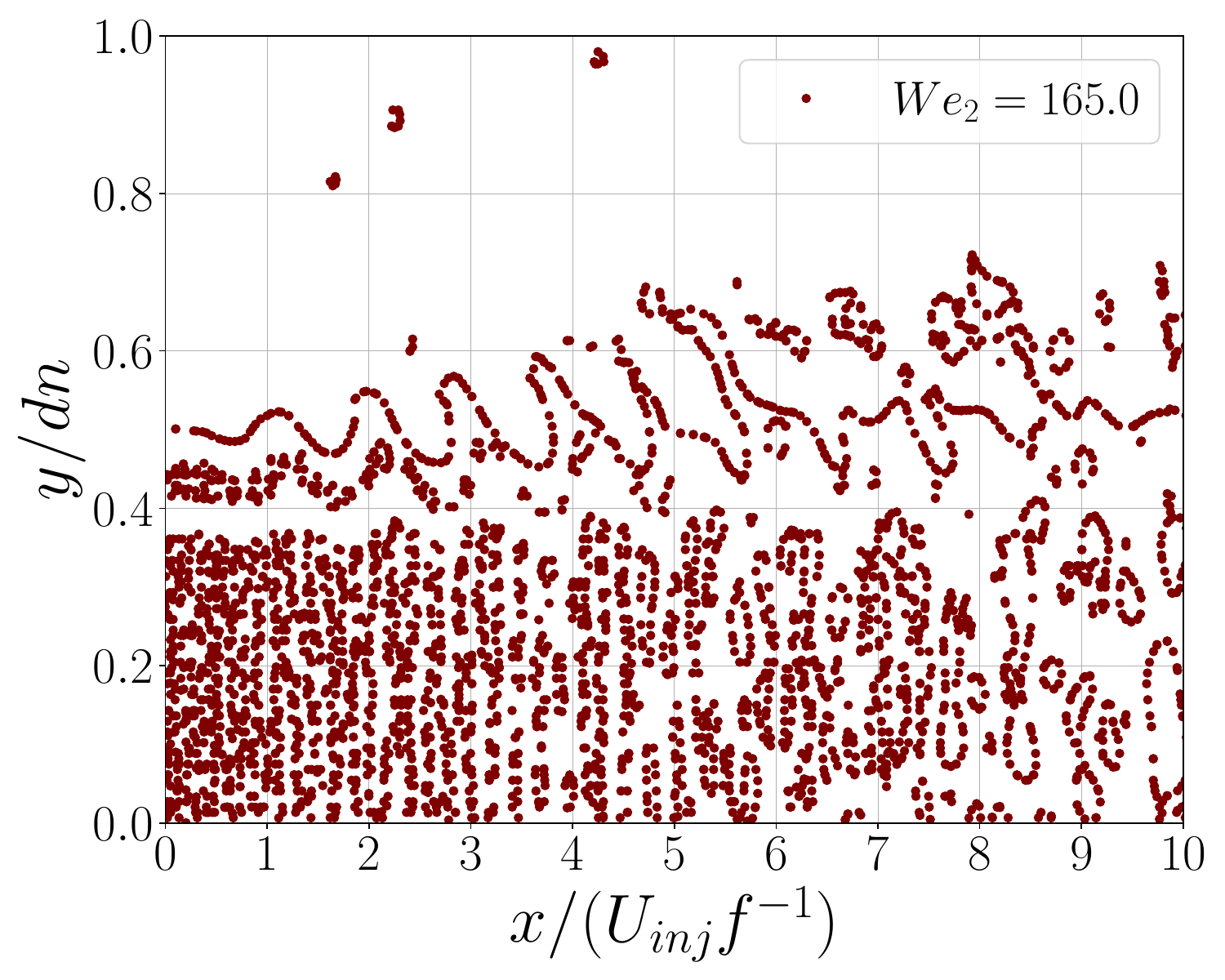}
            \caption{$We_2=165$ (DNS 10)}
                \label{fig:dns.wavesuperposition_ATO2}
  \end{subfigure}%

  \caption{
    Color on-line. Superposition of the interface sliced at $z=0$ in the region
    of the nozzle in the \swi{} regime (a,b) and the atomisation regime (c,d).
    The blue and red colors indicate the \swi{} and atomisation regimes,
    respectively. The physical times are chosen such that the sinusoidal
    perturbations are in phase. As a reminder, $x/(U_{inj}f^{-1})=(x/d_n)St$.
  }
  \label{fig:dns.wavesuperposition}
\end{figure}

In order to check qualitatively the forcing implemented in the simulations and
its outcome, one could have a look on the jet interface in the region of the
nozzle, where the unstable mode excited by the forcing should develop. To
compare the interface evolution between the different DNS, the $x$ coordinate
needs to be normalised by the characteristic length scale of the forcing, i.e.
$U_{inj}f^{-1}$. Note here that $U_{inj}f^{-1} = St / d_n$, so
$x/(U_{inj}f^{-1})=(x/d_n)St$, where $St$ is the Strouhal number based on the
forcing, given in Table \ref{tab:dns.uinj_we_re_f_sr}. Furthermore,
the physical times chosen for the comparison have to be in phase relatively to
the sinusoidal perturbation, i.e. the physical times should be chosen such that
the perturbation waves superimpose on each other. Figure
\ref{fig:dns.wavesuperposition} shows the jet interface sliced at $z=0$ and for
$y/d_n>0$, normalised as explained.

For the 10 DNS, the perturbation waves collapse well after normalising by
$U_{inj}f^{-1}$ and picking in-phase physical times. Consider first the \swi{}
regime. The jet interfaces for $We_2\in\{15,20\}$ (DNS 1 and 2) are represented
separately from those for $We_2\in[26,40]$ (DNS 3 to 6), figure
\ref{fig:dns.wavesuperposition_SWI1} and \ref{fig:dns.wavesuperposition_SWI2},
to highlight the different behavior of the forcing between them, even if DNS 1
to 6 lie in the \swi{} regime.  For the DNS 1 and 2, the development of the mode
leads to waves which only break in large elements in the latter, while they are
attenuated in the former.  Contrarily, the perturbation in the DNS 3 to 6 leads
to the development of shorter waves which break into a wider droplet population.
Here, the wave develops in the radial direction. While the wave extends
radially, up to $y/d_n\approx 0.8$, its outskirt forms a rim and the space
between the liquid core and the outer rim forms a sheet. The sheet becomes
thinner the more the wave extends, before fragmenting for $x/(U_{inj}f^{-1}) \in
[5.5,7]$.  Once the sheet has fragmented, the rim destabilizes and fragments as
well. A similar wave development can be observed for $We_2\in[73,130]$ (DNS 7 to
9), except that the  wave extension is smaller than previously, up to
$y/d_n\approx0.6$, that the wave sheet fragments earlier, for
$x/(U_{inj}f^{-1})\in[5,6]$, and that the rim fragments faster for DNS 7 or even
hardly exists for DNS 8 and 9. Finally, no rim is created when $We_2=130$ (DNS
9).  Specific attention is required for $We_2=165$ (DNS 10). Figure
\ref{fig:dns.wavesuperposition_ATO2} indicates the presence of interface in the
liquid core, meaning that the core is populated by some volume made of the
lighter phase, i.e. bubbles. The ``bubbles'' are generated from $t/T_a\approx10$
and are likely to originate from a numerical artefact related to the volume
fraction threshold used for droplet detection. Note that bubbles also appear,
but later on timewise, in the DNS 9. The presence of bubbles changes the fluid
dynamics inside the core but appears to modify only slightly the interface
dynamics in the time scope of the study and any perturbation would be smoothed
out by considering the overall droplet population. 

\subsubsection{Development of the jet head \label{sec:dns.dev.jet_head}}

\begin{figure}
  \centering
  \begin{subfigure}[c]{0.49\textwidth}
            \centering
            \includegraphics[width=\textwidth]{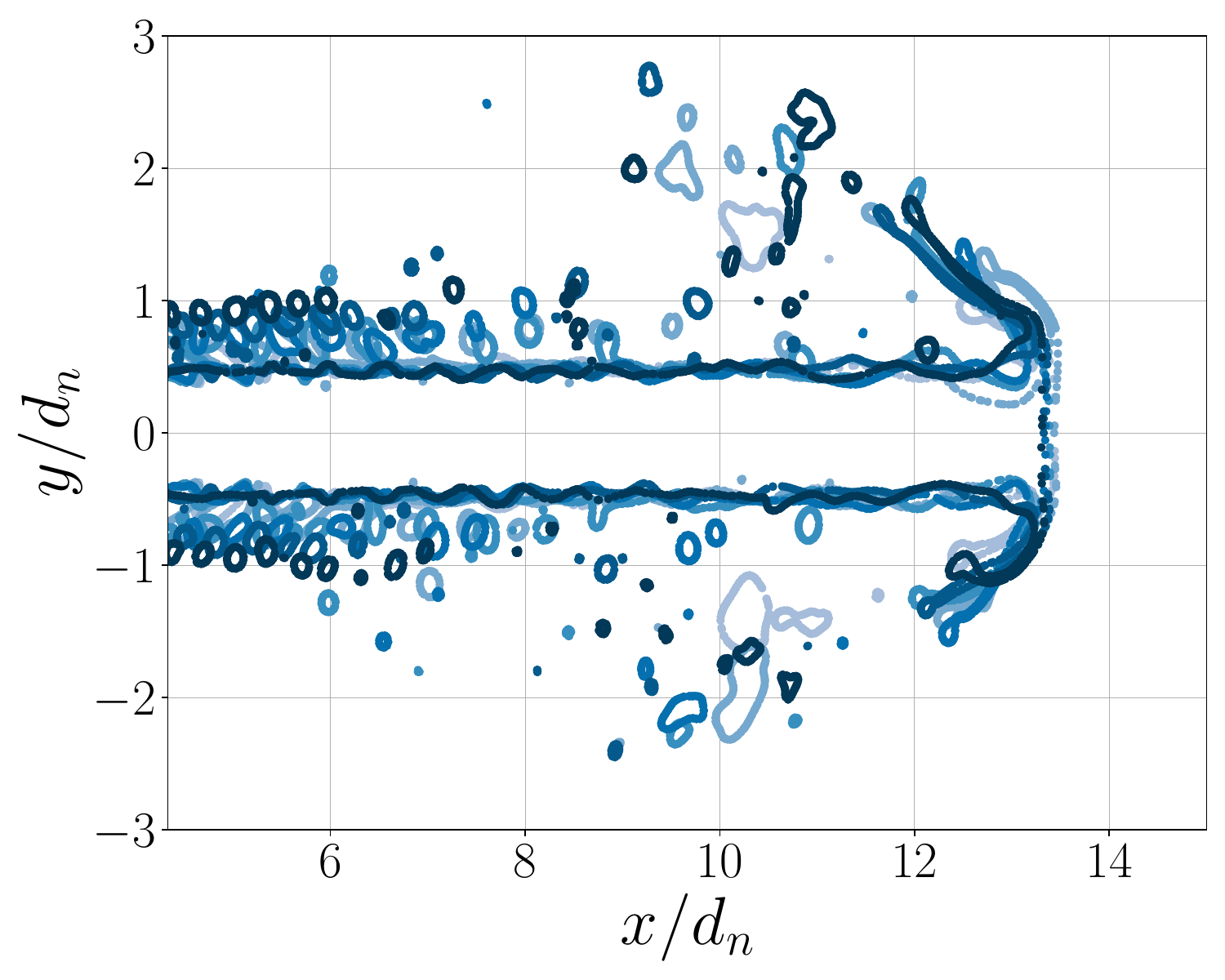}
                \caption{}
                \label{fig:dns.jethead_superposition_SWI15}
  \end{subfigure}%
  \begin{subfigure}[c]{0.49\textwidth}
            \centering
								\includegraphics[width=\textwidth]{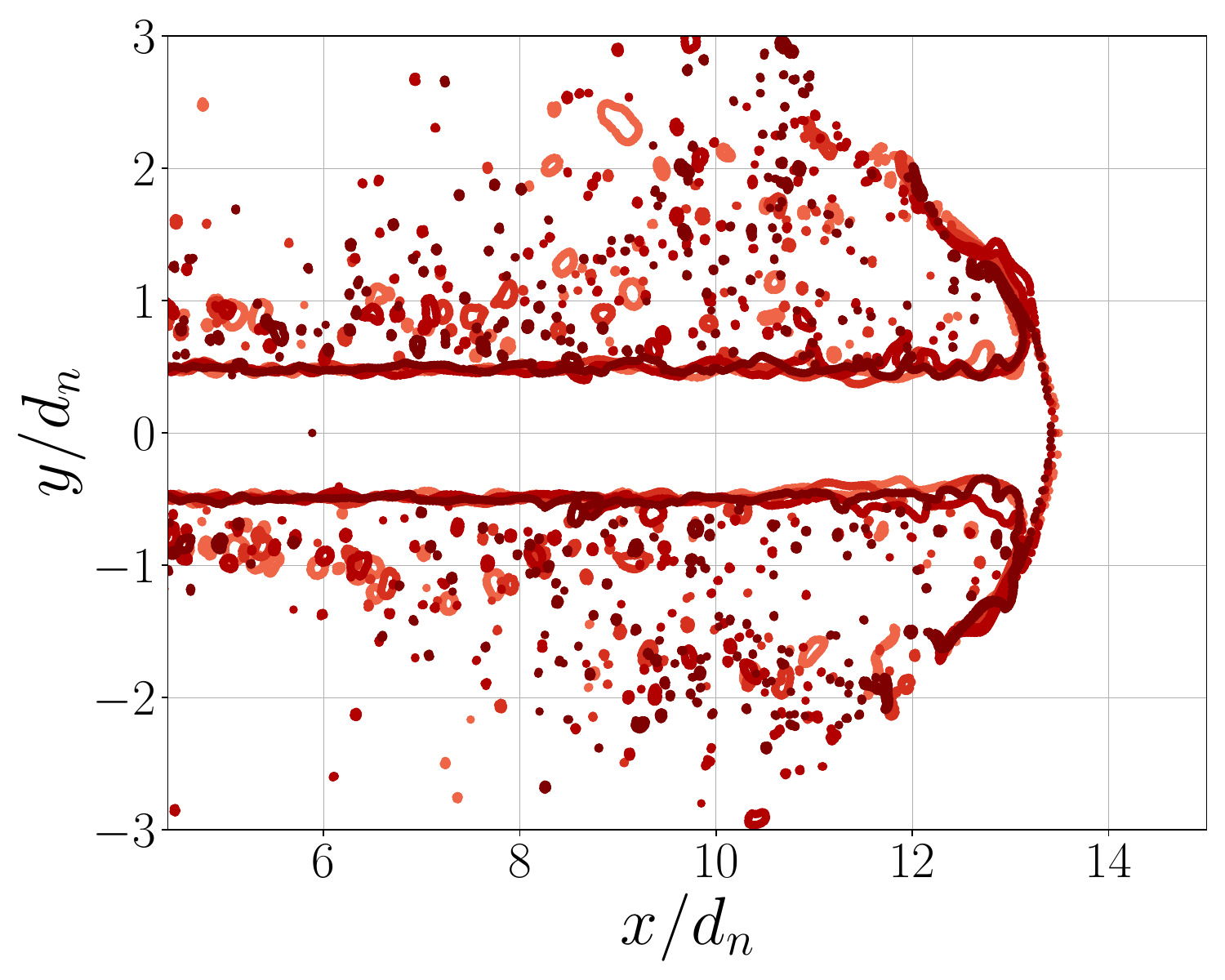}
                \caption{}
                \label{fig:dns.jethead_superposition_ATO15}
  \end{subfigure}%

  \caption{
    Color on-line. Superposition of the interface sliced at $z=0$ in the region
    of the jet head in the \swi{} regime (a) and in the atomisation regime (b)
    at $t/T_a=15$.
  }
  \label{fig:dns.headsuperposition}
\end{figure}

In addition to the development of the wave perturbation, it is possible to
have a glance on the head of the jet. Figure \ref{fig:dns.headsuperposition}
presents the jet interface sliced at $z=0$ in the region of the head of the jet
for both regimes at the same physical time, $t/T_a=15$. What appears at first is
the difference of geometry of the front between the two regimes. In the \swi{}
regime, the front is plane while it is parabolic in the atomisation regime. This
difference results from the force equilibrium between the liquid and gas phases
depending on the injection velocity. In both regimes, the head extends up to
$y/d_n\approx2$ and experiences piercing (data not shown here) which could be
due to the Taylor Culick instability
\citep{taylor_dynamics_1959,culick_comments_1960}. However, the dynamics of the
head extension is quite different. In the \swi{} regime, the head extension can
produce thick ligaments able to extend over distances of the order of $d_n$
while, in the atomisation regime, the ligaments fragment once they are detached
from the head sheet. The difference in the resulting droplet population is
qualitatively visible in figure \ref{fig:dns.headsuperposition} where the
droplets appear to be more numerous in the atomisation regime than in the \swi{}
regime.

\subsection{Statistics of the droplet population \label{sec:dns.dropstat}}

\begin{figure}
  \centering
  \begin{subfigure}[c]{0.49\textwidth}
            \centering
            \includegraphics[width=\textwidth]{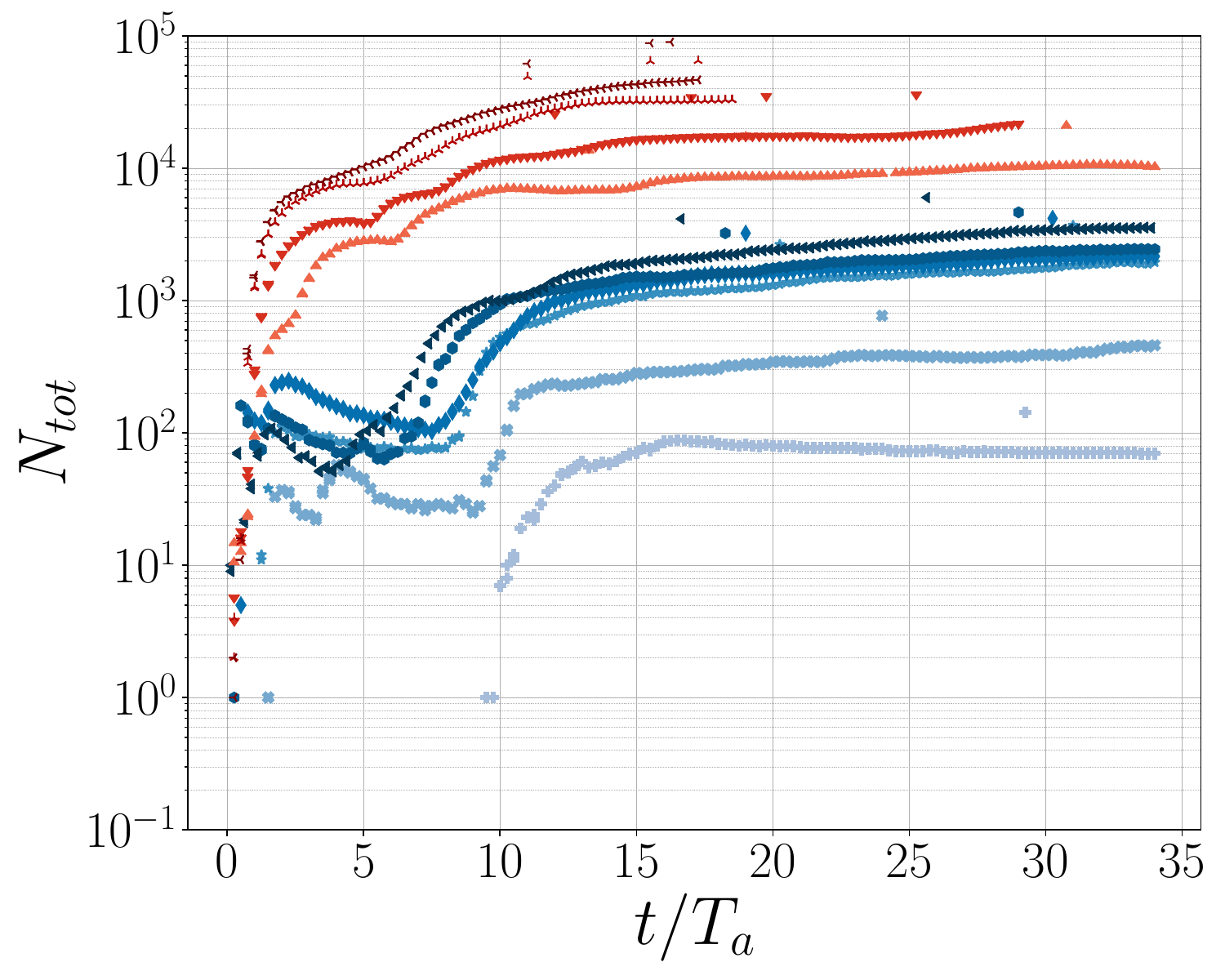}
                \caption{}
                \label{fig:dns.drop_obs_not_normalised}
  \end{subfigure}%
  \begin{subfigure}[c]{0.49\textwidth}
            \centering
            \includegraphics[width=\textwidth]{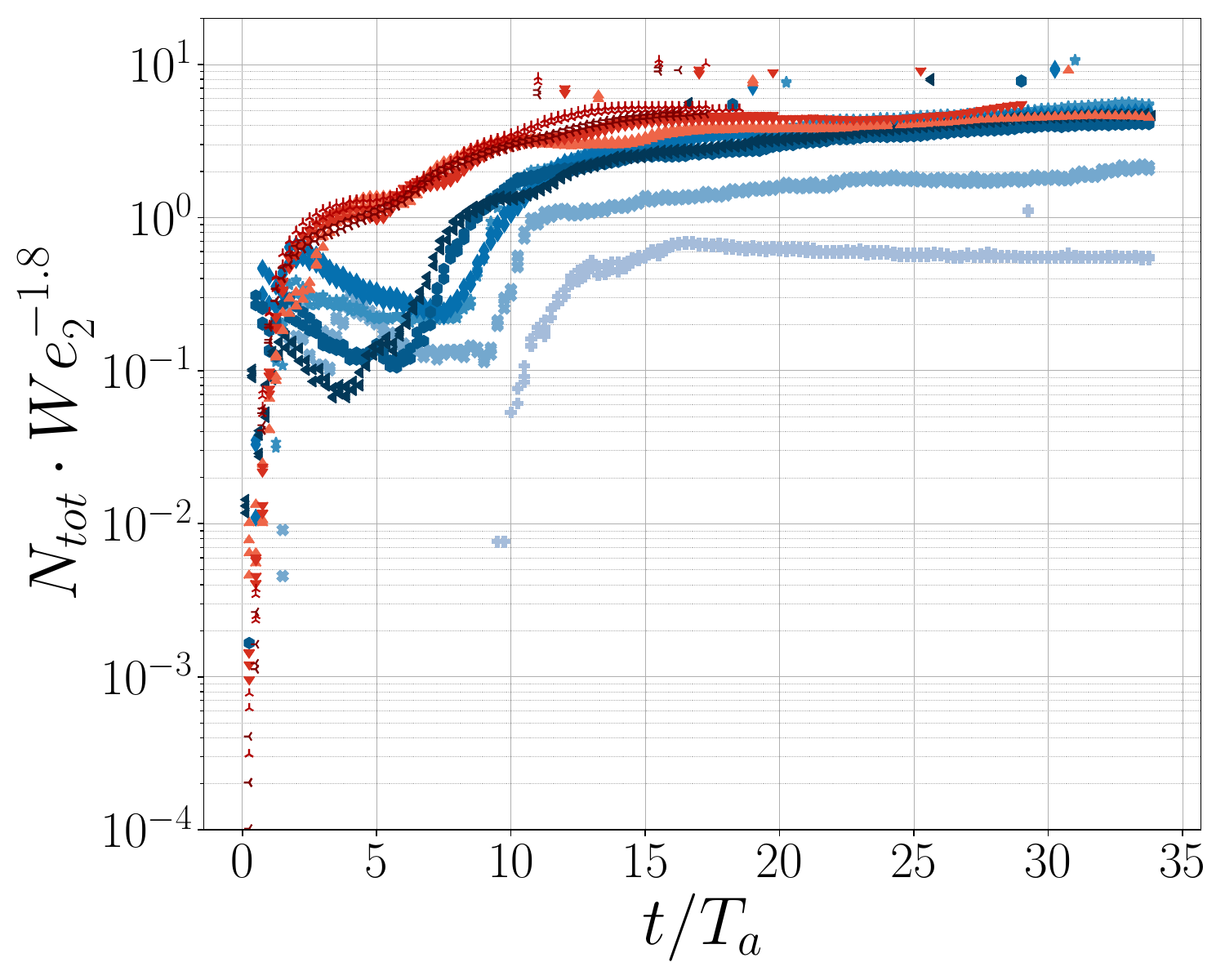}
                \caption{}
                \label{fig:dns.drop_obs_normalised}
  \end{subfigure}%
  \caption{
    Total number of detected droplets, (a) unscaled and (b) scaled by
    $We_2^{-1.8}$. The blue and red colours denote the \swi{} and atomisation
    regimes. The color code denoting the DNS is the same as in Figure
    \ref{fig:dns.front_evolution}. 
  }
  \label{fig:dns.drop_observations}
\end{figure}

Figure \ref{fig:dns.drop_observations} presents the evolution of $N_{tot}$, the
number of droplets detected by the tag function implemented in Basilisk. First
and foremost, the droplets produced for $We_2\in\{15,20\}$ (DNS 1 and 2) do not
exceed 1000 elements, which is not enough to draw conclusions on the statistics
of those two populations. Thus, the DNS 1 and 2 are discarded in the following.
All the other DNS show a total number of elements larger than 1000, which enable
to carry out a statistical analysis. The two regimes distinguish from each other
by the total number of produced droplets. The total number is of $O(10^3)$ in
the \swi{} regime whereas it is of $O(10^4)$ in the atomisation regime, reaching
up to $5\times10^{4}$ elements for $We_2=165$ (DNS 10). Even so, after rescaling
by $We_2^{1.8}$, the numbers of elements for $We_2\in[26,165]$ (DNS 3 to 10)
collapse all together and $N_{tot}$ tends toward $5~We_2^{1.8}$ for both
regimes, excepted DNS 1 and 2.  The transition to a steady production of
droplets differs between the two regimes. In the \swi{} regime, the total number
of elements quickly increases and drops before reaching a steady rate. The
observed decrease could be due to the interactions between the jet head
development and the corollas induced by the mode forcing, interactions which
bring the droplets back to the liquid core.  All the 10 DNS are close to a
steady regime with a slight departure because of the difference in the maximum
physical time reached by each DNS. The temporal evolution of the mean value for
the size, axial and radial velocity as well as a proposal of scaling are given
in appendix \ref{app.temp_evol_mean}.

\begin{figure} 
  \centering 
  \includegraphics{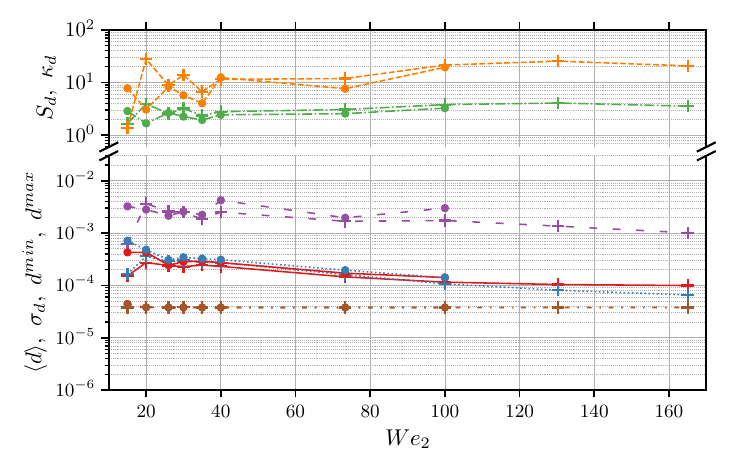} 
  \caption{
    Color on-line. Evolution of $\langle \cdot \rangle$ (red), $\sigma$ (blue),
    $S$ (green), $\kappa$ (orange), the minimum (purple) and maximum (brown)
    against $We_2$ for the size $d$. The pluses ($+$) correspond to $t/T_a=15$
    and the bullets ($\bullet$) to $t/T_a=25$. Note that $S$ and $\kappa$ are
    both dimensionless and that the dimensional variables are expressed with the
    SI base units.  
  }
  \label{fig:dns.stat_vs_weg} 
\end{figure}

The arithmetic mean operator and the standard deviation are respectively denoted
$\langle \cdot \rangle$ and $\sigma$ while the skewness and the excess kurtosis
are respectively denoted $S$ and $\kappa$. Here, we considered the excess
kurtosis equal to the kurtosis subtracted by 3 such that the normal distribution
has a zero excess kurtosis.  Figure \ref{fig:dns.stat_vs_weg} gives the
evolution along $We_2$ of the four first statistical moments along with the
minimum and the maximum values for the size at the time instants $t/T_a=15$ and
$t/T_a=25$. Note that the droplet tagging function implemented in Basilisk can
return droplets with a volume $V$ smaller than $\Delta_{min}^3$, the volume of
the smallest grid cell. This behavior is expected and due to the cells having a
volume fraction $f$ between $10^{-3}$ and 1 and being disconnected from any
liquid neighbourhood. To ensure physical consistency regarding the grid
characteristics, all the droplets with a volume smaller than or equal to the
minimum cell volume are discarded, i.e. any droplets such that
$V\leqslant\Delta_{min}^3$. Assuming spherical droplets, this condition implies
a minimum droplet diameter $d_{min}=\sqrt[3]{6/\pi}\Delta_{min} \approx
37.8~\mu\textnormal{m}$.

Let us consider the statistical moments of the droplet size. Details about the
evolution of the moments for the axial and radial velocities are given in
appendix \ref{app.stat_mom_vel}. Globally, both the mean and standard deviation
decrease with $We_2$ and are of $O(100~\mu\textnormal{m})$. Similarly, the
maximum value decreases but is one order larger, $O(1~\textnormal{mm}$).
Meanwhile, the skewness and the excess kurtosis slightly increase and are respectively
of $O(1)$ and $O(10)$. An increase in $We_2$ corresponds to an increase of the
inertial forces relatively to the surface tension forces. Then, the larger
$We_2$, the more likely the droplets undergo fragmentation and fewer sizes are
stable. Thus, the mean diameter and the maximum diameter decrease and more
droplets group around the mean diameter which, as a consequence, reduces the
standard deviation. A decrease of the minimum diameter should also occur, but the
condition on the minimal droplet volume numerically filters out any diameter
smaller than $37.8~\mu \textnormal{m}$. In parallel, the increase in the
skewness value, which is positive, points out that the droplet size distribution
is slightly more right-tailed with higher $We_2$.  This indicates that the
right-hand tail, towards sizes larger than $\langle d \rangle$, exists on a size
range larger than the one on which the left-hand tail exists. Finally, the
positive sign and the increase of the excess kurtosis with $We_2$ indicates that the
distributions are leptokurtic for all the values of $We_2$ and that the
distribution tail increases in length relatively to the mean and the standard
deviation.  Equivalently, the range of rare very large sizes, relatively to the
mean, is both broader with higher $We_2$ and larger than the Gaussian
distribution for which $\kappa=0$. As the flow geometry remains the same for the
different values of $We_2$, the increase in the excess kurtosis and the skewness is due
to the depletion of large sizes in the benefits of small sizes, grouped around
the mean diameter.  However, even if $d_{max}$ decreases, the large values of
$S$ and $\kappa$ show that the larger droplet sizes do not disappear completely
from the flow and still exist at higher values of $We_2$. 

Finally, the values of the skewness and the excess kurtosis of the size distribution
are very large and one order of magnitude larger than those of the velocity
distributions, see appendix \ref{app.stat_mom_vel}. This indicates a wider spanning range for the size distribution
than for the two velocity distributions and justifies the use of a loglog scale
to visualise the size distribution.

\subsection{Distributions of the size and the velocity
\label{sec:dns.pdf_size_velocity}}

\begin{figure} \centering \begin{subfigure}[c]{0.48\textwidth} \centering
  \includegraphics[width=\textwidth]{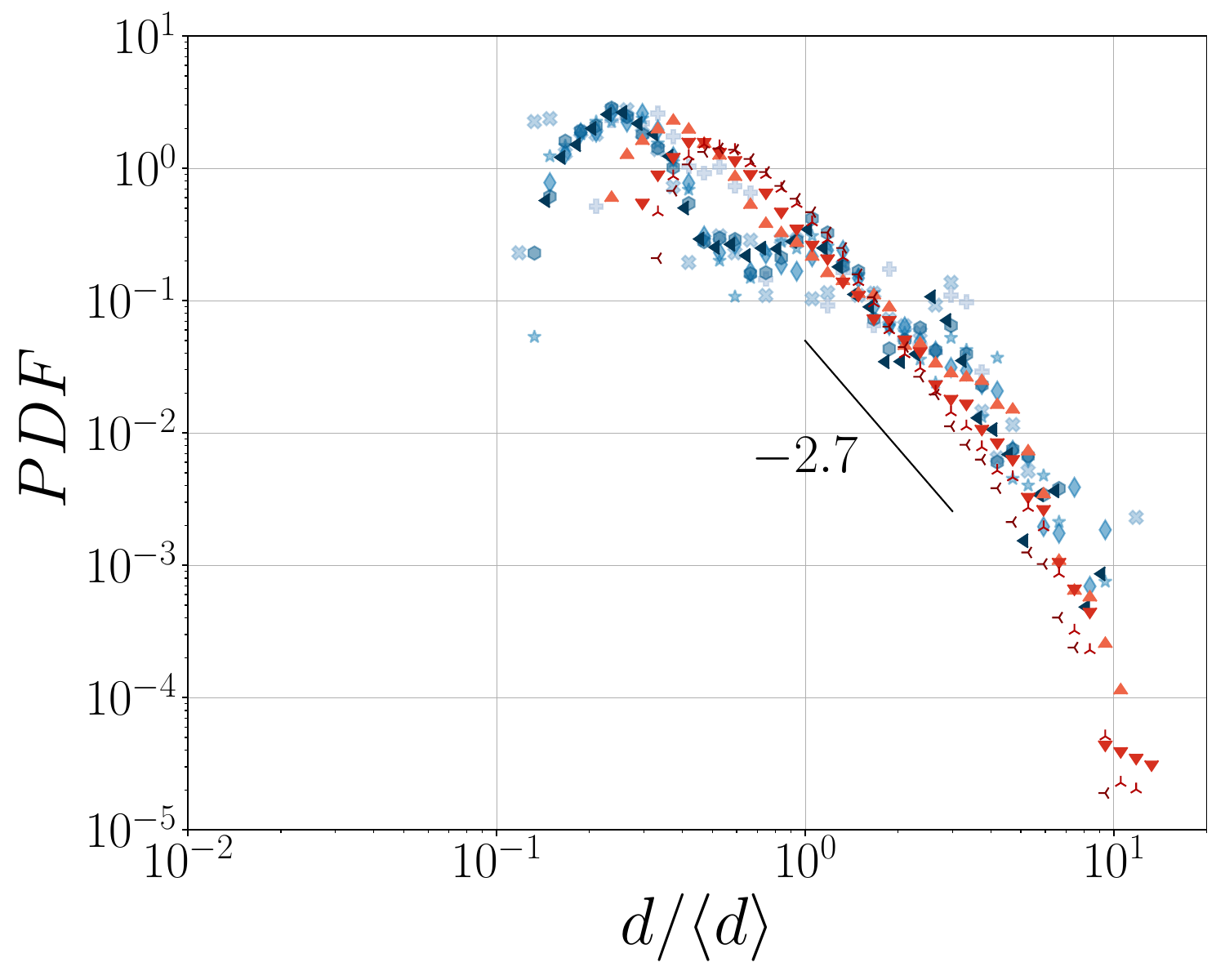} \caption{}
\label{fig:dns.droppdf_size_tm15} \end{subfigure}%
\begin{subfigure}[c]{0.48\textwidth} \centering
  \includegraphics[width=\textwidth]{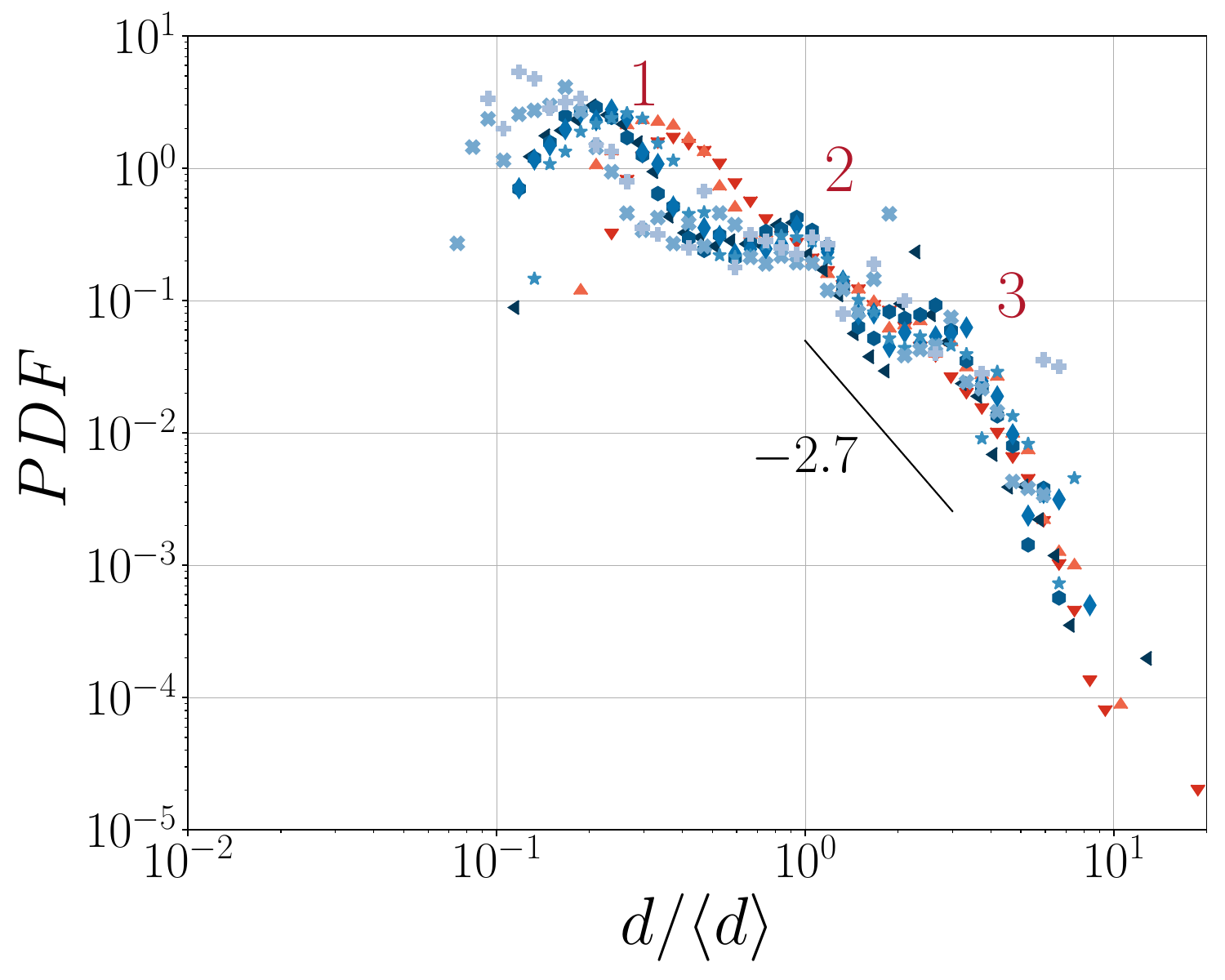} \caption{}
\label{fig:dns.droppdf_size_tm25} \end{subfigure}%

  \begin{subfigure}[c]{0.48\textwidth} \centering
  \includegraphics[width=\textwidth]{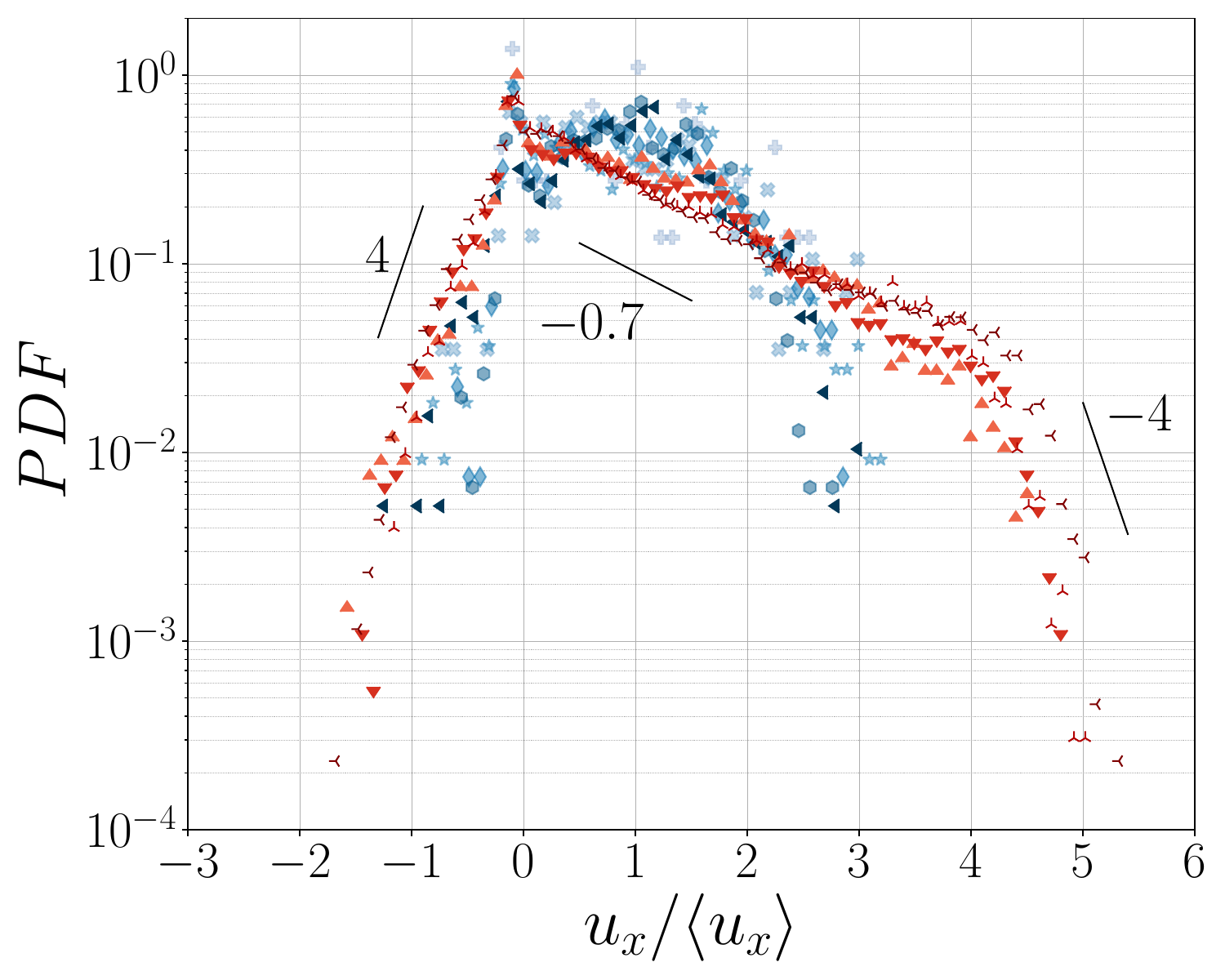} \caption{}
\label{fig:dns.droppdf_ux_tm15} \end{subfigure}%
\begin{subfigure}[c]{0.48\textwidth} \centering
  \includegraphics[width=\textwidth]{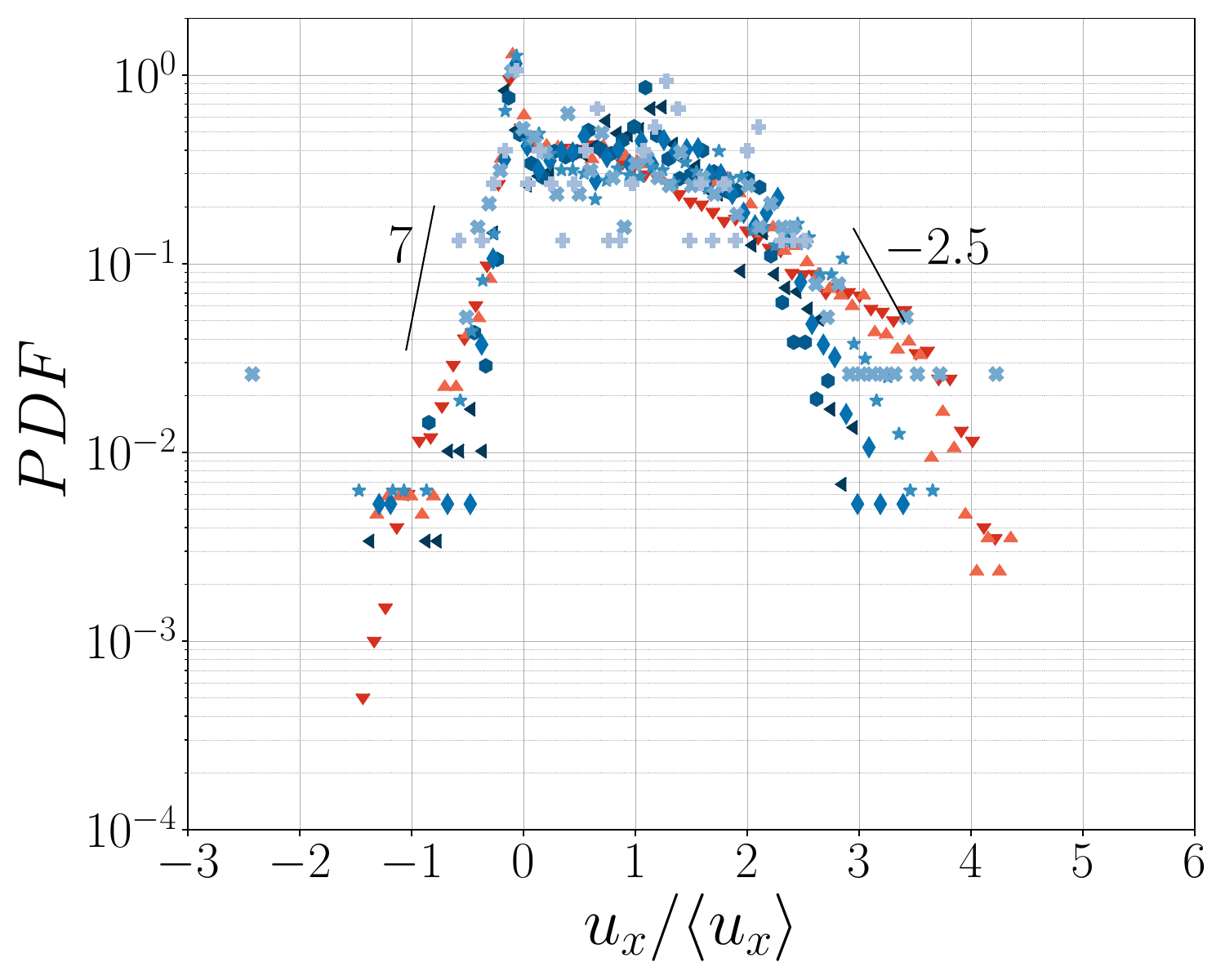} \caption{}
\label{fig:dns.droppdf_ux_tm25} \end{subfigure}%

  \begin{subfigure}[c]{0.48\textwidth} \centering
  \includegraphics[width=\textwidth]{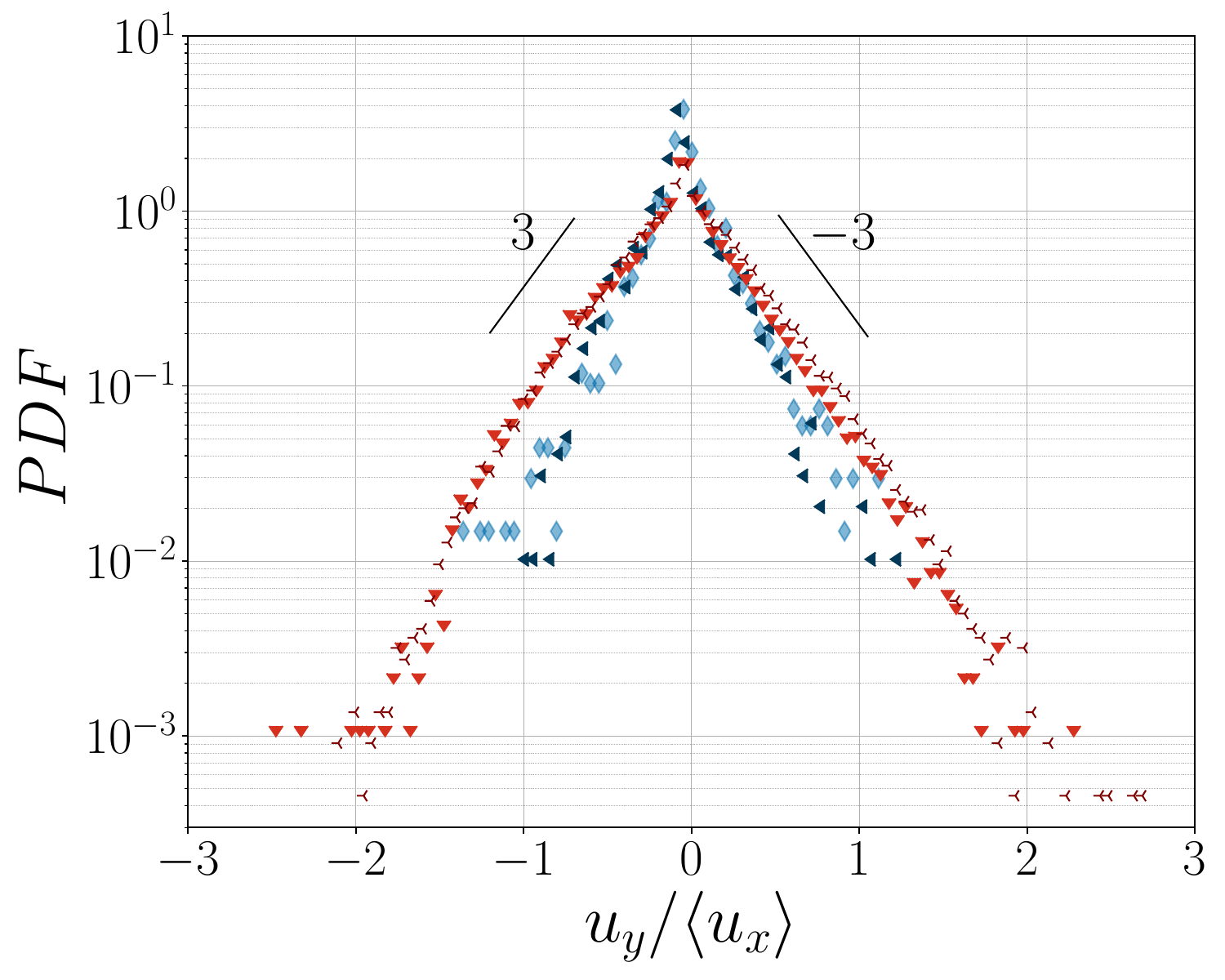} \caption{}
\label{fig:dns.droppdf_uy_tm15} \end{subfigure}%
\begin{subfigure}[c]{0.48\textwidth} \centering
  \includegraphics[width=\textwidth]{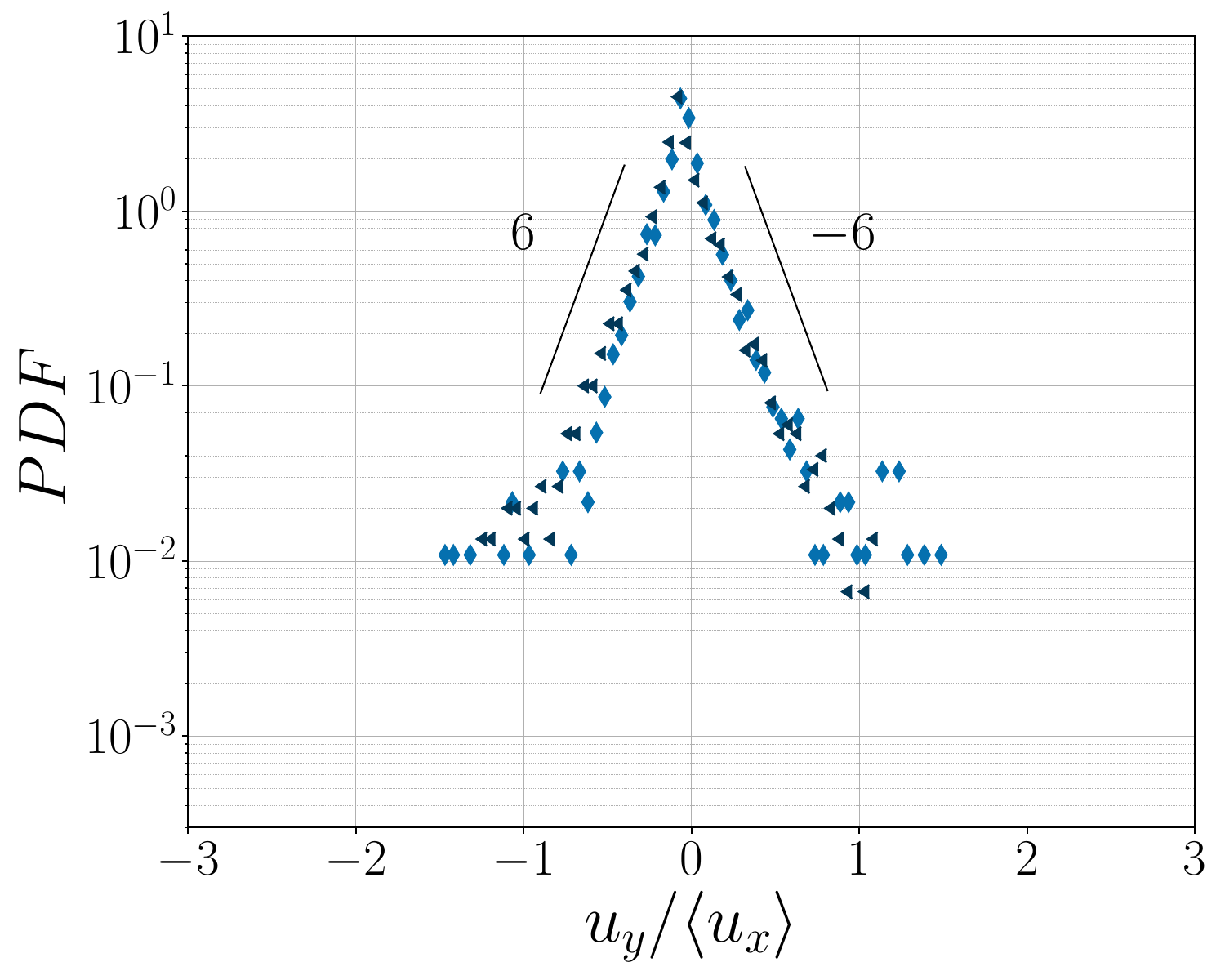} \caption{}
\label{fig:dns.droppdf_uy_tm25} \end{subfigure} 
  \caption{
    Distributions of $d/\langle d \rangle$ (a,b), $u_x/\langle u_x
    \rangle$ (c,d) and $u_y/\langle u_x \rangle$ (e,f) at $t/T_a=15$ (left) and
    $t/T_a=25$ (right). The blue and red colours denote the \swi{} and
    atomisation regimes. The colour code denoting the DNS is the same as in Figure
    \ref{fig:dns.front_evolution}.
  }
  \label{fig:dns.allvar_pdf} 
\end{figure}

Complementary to the statistical moments, it is worth looking at the
distributions of the sizes and the velocities of the droplets. For the sake of
clarity, the number PDF of any variable $\zeta$ is denoted $\mathcal{P}_\zeta$
in the following.  Even if the mean values of the size and the velocities are
not fully converged, we consider the PDF of each variable normalised by its
mean. However, $u_y$ being close to zero in average, normalising by $\langle u_y
\rangle$ is not relevant and $u_y/\langle u_x \rangle$ is considered instead.
Figure \ref{fig:dns.allvar_pdf} gives the PDFs of $d/\langle d \rangle$,
$u_x/\langle u_x \rangle$ and $u_y/\langle u_x \rangle$ at the time instants
$t/T_a=15$, where both regimes are computed, and $t/T_a=25$. First of all, it is
interesting to note that the PDFs in each regime collapse for the three variables
even if the mean values are not converged. From the three distributions, only
that for $u_y/\langle u_x \rangle $ shows a similar behavior between the two
regimes of fragmentation, excepting for the width and the slope of the tails.
For both regimes, the PDF tails scale with $\exp(\pm a\times u_y/ \langle u_x
\rangle)$ where $a$ nearly equals 6 in the \swi{} regime and
nearly equals 3 in the atomisation regime. The difference in the tail width goes
along with the difference between the exponential coefficient. Indeed, the
larger the coefficient is, the smaller the tail width is. This can be explained
once again with the increase of the relative velocity between the injection and
the gas phase, and thus the shear, when $We_2$ increases.  Note that, due to the
flow symmetry, $\mathcal{P}_{u_z/\langle u_x \rangle}$ follows a trend similar
to that of $u_y/\langle u_x \rangle $. 

Regarding the size distribution, different modes appear clearly between the two
fragmentation regimes. The size PDF derived from the atomisation regime shows
one main mode centered on $d/\langle d \rangle=0.5$ while the PDF for the \swi{}
regime shows 3 modes centered on $d/\langle d\rangle=\{0.2, 1, 2.5\}$, denoted
from 1 to 3 in figure \ref{fig:dns.droppdf_size_tm25}. Even if the main mode
appears to be shifted towards larger $d/\langle d \rangle$ when $We_2$
increases, it refers to the same range of physical sizes $d$ between
$47~\mu\textnormal{m}$ and $58~\mu\textnormal{m}$ with a mean value of $55~\mu
\textnormal{m}$, considering $We_2\in[26,165]$ (DNS 3 to 10). 

\citet{tennekes_first_1972} derived handy equations to estimate the turbulent
Reynolds number $Re_\tau$ from the ratio of the extreme scales of the flow and
the Taylor microscale Reynolds number $Re_\lambda$: $Re_\tau \propto
(d_n/\Delta_{min})^{4/3}$ and $Re_\lambda \approx \sqrt{Re_\tau}$ where
$Re_\lambda = u_{RMS}\lambda / \nu_2$ and $\Delta_{min}=30.5~\mu\textnormal{m}$,
see section \ref{sec:dns.num_config_cost}.  With the chosen configuration, the
estimation gives $Re_\tau\approx 775$ and $Re_\lambda\approx27.8$. Besides,
assuming that the turbulence intensity  is around $20\%$ of the injected
velocity, i.e. $u_{RMS} = 0.2~U_{inj}$, it is possible to estimate the Taylor
microscale $\lambda$. Table \ref{tab:dns.taylor_microscale} lists the
estimation of $\lambda$ for the 10 DNS. The Taylor microscale decreases with
the gaseous Weber number $We_2$, which is expected as $Re_\lambda$ is set by the
configuration and $u_{RMS}$ increases with the injection velocity. Physically,
the root mean square velocity increases with the injection velocity and the
Taylor microscale decreases. Most importantly, the normalised values of
$\lambda$ correspond to the third mode of $\mathcal{P}_{d/\langle d\rangle}$ at
both $t/T_a=\{15,25\}$ in the \swi{} regime, $We_2<40.3$, and
indicate that the larger droplets observed in the DNS might be related to the
most probable vortex size set by the gas turbulence.  

\begin{table}
  \begin{center} 
    \def~{\hphantom{0}} 
    \begin{tabular}{cccccc|cccccc}
      DNS & $We_2$ & $\lambda~(\mu\textnormal{m})$ & $\lambda/\langle d
      \rangle_{15}$ & $\lambda/\langle d \rangle_{25}$ &&& DNS & $We_2$ &
      $\lambda~(\mu\textnormal{m})$ & $\lambda/\langle d \rangle_{15}$ &
      $\lambda/\langle d \rangle_{25}$ \\[3pt]
      1 & 15 & 1031 & 7~~~ & 2.50 &&& 6 & ~40~~ & 632 & 2.83 & 2.37 \\
      2 & 20 & ~893 & 3.47 & 2.21 &&& 7 & ~73.3 & 467 & 3.37 & -- \\
      3 & 26 & ~784 & 3.36 & 3.28 &&& 8 & ~99.8 & 400 & 3.65 & -- \\ 
      4 & 30 & ~729 & 3.29 & 2.62 &&& 9 & 130.3 & 350 & 3.52 & -- \\
      5 & 35 & ~675 & 2.85 & 2.37 &&& 10& 165~~ & 311 & 3.28 & -- \\ 
\end{tabular}
  \caption{ 
    Estimation of the Taylor microscale $\lambda$ and its normalised values at
    $t/T_a=\{15,25\}$.  
  }
  \label{tab:dns.taylor_microscale} \end{center} \end{table}

\begin{figure} \centering \begin{subfigure}[c]{0.49\textwidth} \centering
  \includegraphics[width=\textwidth]{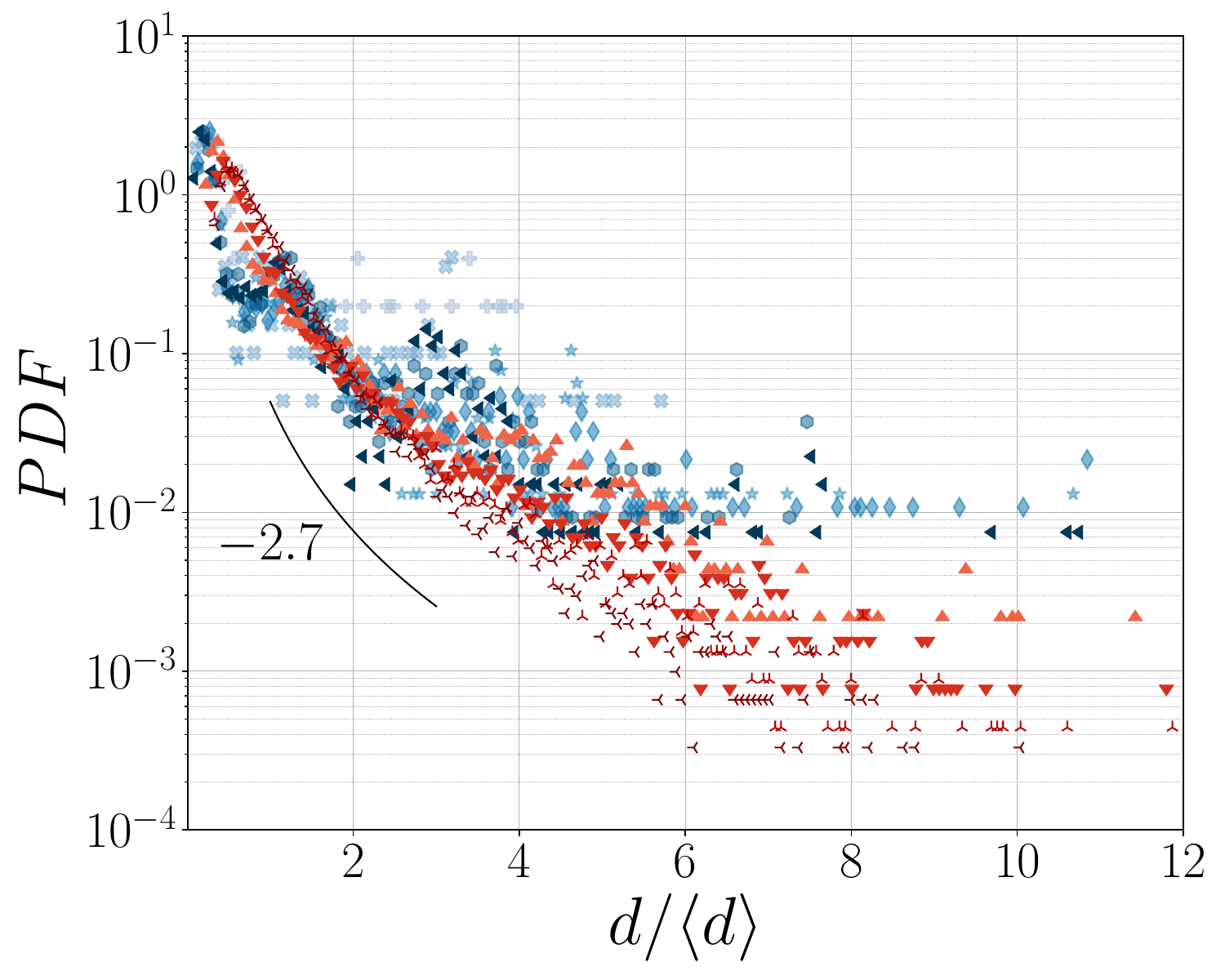}
\caption{$t/T_a=15$} \label{fig:dns.sizedistri_tm15_linlog} \end{subfigure}%
\begin{subfigure}[c]{0.49\textwidth} \centering
\includegraphics[width=\textwidth]{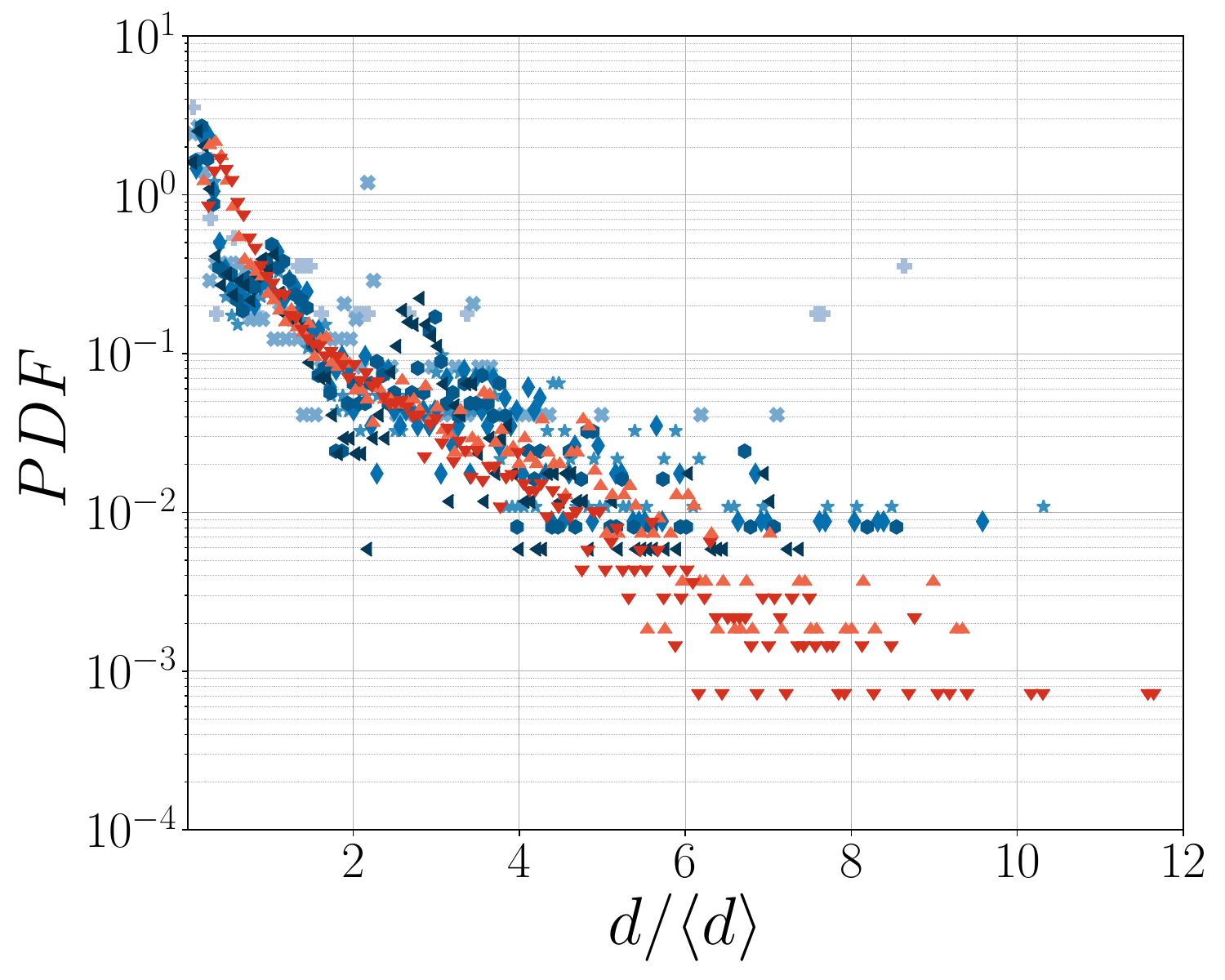}
\caption{$t/T_a=20$} \label{fig:dns.sizedistri_tm20_linlog} \end{subfigure}%

  \caption{
    Distribution of $d/\langle d\rangle$ in semi-logarithmic scale at $t/T_a=15$
    and $t/T_a=20$. The blue and red colours denote the \swi{} and atomisation
    regimes, respectively. The colour code denoting the DNS is the same as in
    Figure \ref{fig:dns.front_evolution}. The solid black line corresponds to a
    power law of coefficient $-2.7$, as in figure
    \ref{fig:dns.droppdf_size_tm15} and \ref{fig:dns.droppdf_size_tm25}.  
  }
  \label{fig:dns.sizedistri_linlog}
\end{figure}

$\mathcal{P}_{d/\langle d \rangle}$ in both regimes shows a similar tail
evolution scaling as $(d/\langle d  \rangle)^{-2.7}$, which was also observed
experimentally \citep{vallon_multimodal_2021}. This power law scaling goes
against the experimental observation of \citet{simmons_correlation_1977-1} who
remarked that the size distribution in industrial jet shows a tail scaling as an
exponential. Figure \ref{fig:dns.sizedistri_linlog} gives the size distribution
in a semi logarithmic scale. The time instant $t/T_a=20$ has been chosen over
$t/T_a=25$ in order to highlight the trend of the size distribution in the
\swi{} regime thanks to the distribution for $We_2=73$ (DNS 7). It appears
that, at both time instants, none of the size distribution follows a unique
exponential decay. Instead, the distribution in the atomisation regime follows
two exponential scalings, the first one for $d/\langle d\rangle\in[0.5,2]$ and
the second one for $d/\langle d\rangle\in[4,8]$, with a transition region
scaling as $d/\langle d\rangle^{-2.7}$ for $d/\langle d\rangle\in[2,4]$.
Following the analysis of an experimental spray by \citet{vallon_liquid_2021},
the existence of those two scalings could indicate that the distribution is
composed of two distributions originating from different fragmentation sources.
The modelling of the size PDF by theoretical distributions is achieved in
section \ref{sec:dns.sizepdf_model}. 

Regarding the distribution of the axial velocity, it is often expected in jet
fragmentation flows that the droplets show a positive axial velocity less than
or equal to the injection velocity as they are globally advected towards
increasing $x/d_n$. However, $\mathcal{P}_{u_x/\langle u_x \rangle}$ shows large
probabilities for a range of negative velocities, $u_x/\langle u_x \rangle\in
[-2,0]$, with a sharp peak at $u_x/\langle u_x \rangle = 0$. The right-hand tail
exists in both regimes on a range of velocities larger than the injection
velocity $U_{inj}$.  For instance, considering the droplet population for
$We_2=40$ (DNS 6) lying in the \swi{} regime at $t/T_a=25$ with
$\langle u_x \rangle=1~\textnormal{m/s}$, we have $\mathcal{P}(2 < u/\langle u_x
\rangle< 3)>0$, meaning that there exists droplets such that $u_x/U_{inj} \in
[0.9,1.35]$, thus being faster than the injection velocity
$U_{inj}=2.216\textnormal{m/s}$. The same conclusion can be drawn for the other
DNS.

Surprisingly, for the atomisation regime, the tail expansions in the regions of
negative velocities and velocities larger than $U_{inj}$ follow a similar trend,
scaling as $\exp(\pm a\times u_x/\langle u_x \rangle)$ with $a=4$.  However, in
the \swi{} regime, the left-hand tail and the right-hand tail present two
different scalings: the former scales as $(u_x/\langle u_x \rangle)^7$ and the
latter as $(u_x/\langle u_x \rangle)^{-2.5}$. The argument of the increasing
relative velocity between the injection and the gas phase, and consequently in
the standard deviation, can be considered to explain the difference in the tail
expansion between the two regimes. Finally, in addition to the sharp peak
for zero velocities, the axial velocity PDF is centered on $u_x/ \langle u_x
\rangle=1$ in the \swi{} regime and presents a continuous decrease scaling as
$\exp(-0.7\times u_x/\langle u_x \rangle)$ in the atomisation regime. The
specific characteristics of the velocity PDF, $u_x<0$ and $u_x\geqslant
U_{inj}$, are explored in section \ref{sec:dns.vortexring}.

\subsection{Modeling the droplet size PDF \label{sec:dns.sizepdf_model}}

When it comes to modeling the distribution of the droplet size, one theoretical
distribution is necessary to test: the $\Gamma$ law derived from the
ligament-mediated fragmentation framework \citep{villermaux_fragmentation_2020}
along with its refinement exposed by \citet{kooij_what_2018}, here after denoted
$f_\Gamma$ and $f_{\Gamma B}$.  While the former was specifically designed to
describe the droplet size PDF resulting from the breakup of a ligament, the
latter was designed to describe the size PDF resulting from the overall
fragmentation of a jet. A previous study carried out by
\citet{vallon_multimodal_2021} highlighted the limits of those two distributions
for modelling size PDF far away from the nozzle, $x/d_n\in[400,800]$ in the
context of agricultural like sprays, and the satisfying performance of the law
derived by \citet{novikov_distribution_1997} in the framework of turbulence
intermittency, denoted $f_\epsilon$ in the following. More details about each
law and the related framework are given in \citet{vallon_multimodal_2021}. The
three theoretical laws write as follows:

\begin{eqnarray} f_\Gamma: & \mathcal{P}(x=d/\langle d \rangle) & =
  \frac{n^n}{\Gamma(n)} x^{n-1}e^{-n x}, \\ f_{\Gamma B}: &
  \mathcal{P}(x=d/\langle d \rangle) & = \frac{2(mn)^{(m+n)/2}
  x^{(m+n)/1-1}}{\Gamma(m)\Gamma(n)} \mathcal{K}_{m-n}(2\sqrt{nmx}), \\
  f_{\epsilon}: & \mathcal{P}(y=-\ln(l/l_1)) & = \frac{ a^{3/2} }{ \sqrt{2\pi}
  \sigma y^{3/2}} \exp \bigg\{- \frac{a}{2\sigma^2} \big(
  ay^{-1/2}-y^{1/2}\big)^2 \bigg\}, ~y\geqslant 0.  \end{eqnarray}

\noindent In the expression of $f_\Gamma$, $n$ represents the corrugations of
a ligament before its breakup. The corrugations determine the size PDF resulting
from the breakup \citep{villermaux_ligament-mediated_2004}. The same logic takes
place in the expression of $f_{\Gamma B}$. Additionally, the ligaments can show
a large variety of sizes in the flow. This variety is taken into account by $m$
which sets the order of the ligament size distribution \citep{kooij_what_2018}.
Finally, the expression is conditioned by the modified Bessel function of the
second kind $\mathcal{K}$ whose order is set by $m-n$.  Regarding
$f_\epsilon$, \citet{novikov_distribution_1997} considered a cascade mechanism
and the ratio between the initial size $l_1$ and the resulting size $l$ of a
fragmenting droplet, where $a=\langle y \rangle$ and $\sigma=\langle
(y-a)^2\rangle$. It is worth noting that, even if $f_\epsilon$ relies on the
cascade concept initially derived by \citet{richardson_weather_1922} and used in
the seminal papers of \citet{kolmogorov_energy_1941,kolmogorov_local_1941}, the
infinitely divisible nature of this distribution ensures that it is at no point
close to a logarithmic normal distribution resulting from the Central Limit
Theorem.

A systematic fit campaign is carried out to test the three distributions.
Details of the procedure are given in appendix \ref{app.fit_cmpgn}. Figure
\ref{fig:dns.fit_sizepdf} gives the best fits produced by $f_\Gamma$, $f_{\Gamma
B}$ and $f_\epsilon$ in both fragmentation regimes at the time instant
$t/T_a=15$ while Table \ref{tab:dns.fit_sizepdf_r2} gives the corresponding
final parameters and the square of the Pearson coefficient $r^2$.
Qualitatively, the three theoretical distributions capture well the size PDF in
both regimes and describe with a good accuracy the right-hand tail on the
available range of sizes. No relevant comment can be drawn about the left-hand
tail as no physical droplet sizes are available in the DNS, see section
\ref{sec:dns.dropstat}, and this range was discarded in the fit procedure. The
slight differences between the distributions mainly lie in the description of
the main mode. In both regimes, $f_\epsilon$ performs slightly better in
capturing the main mode and the short left-hand tail. Additionally, the manual
fit of the PDF modes for the \swi{} regime indicates the good performance of
$f_\Gamma$ to describe the mode separately. Quantitatively, the values of $r^2$
bring a sharp light on the performance of each theoretical distribution.  For
both regimes, the law exposed by \citet{kooij_what_2018} shows $r^2$ values the
closest to 1 with a mean computed over the two better fits equal to 1.00025.
Then follows the $\Gamma$ law and the distribution derived by
\citet{novikov_distribution_1997} with mean $r^2$ values respectively equal to
0.9245 and 0.878. Thus $f_{\Gamma B}$ better describes the size PDF in the flow,
close to the nozzle, while $f_\Gamma$ describes each mode separately. Meanwhile,
$f_\epsilon$ shows a correct performance close to the nozzle, which achieves its
good performance for describing multimodal size PDFs far away from the nozzle in
the \swi{} regime \citep{vallon_multimodal_2021}.

\begin{figure} 
  \centering
  
  \begin{subfigure}[c]{0.49\textwidth} 
    \centering
    \includegraphics[width=\textwidth]{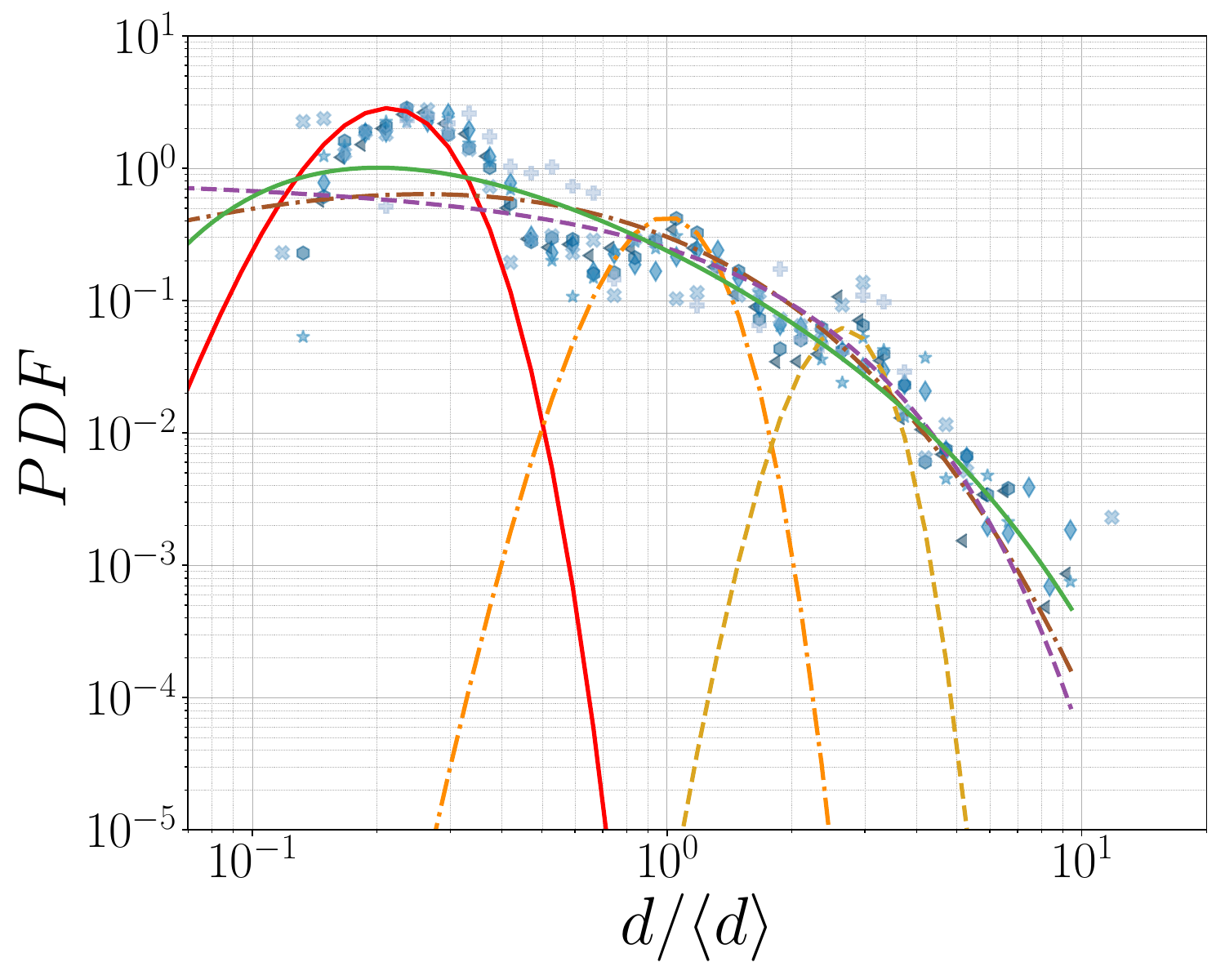} 
    \caption{}
    \label{fig:dns.droppdf_fit_pdf_size_tm15_swi} 
  \end{subfigure}%
  \begin{subfigure}[c]{0.49\textwidth} 
    \centering
    \includegraphics[width=\textwidth]{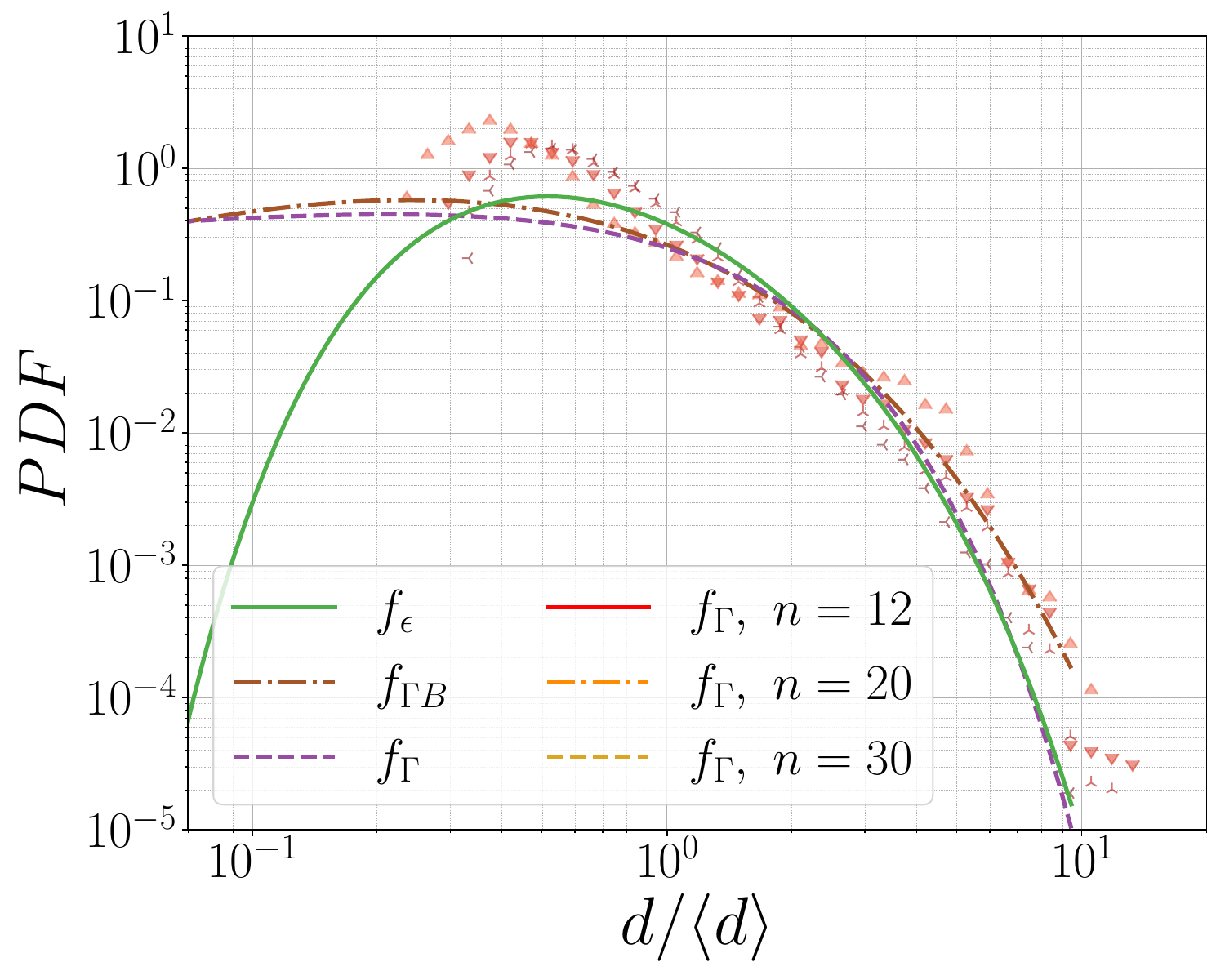} 
    \caption{}
    \label{fig:dns.droppdf_fit_pdf_size_tm15_ato} 
  \end{subfigure} 

  \caption{
    Color on-line. Fit of $\mathcal{P}_{d/\langle d\rangle}$ by $f_\Gamma$,
    $f_{\Gamma B}$ and $f_\epsilon$ in the \swi{} regime (a) and the atomisation
    regime (b) at $t/T_a=15$. The fit procedure is carried on the data shown
    here and the best fit is represented over $d/\langle d\rangle
    \in[10^{-2},20]$.  
  } 
  \label{fig:dns.fit_sizepdf} 
\end{figure}

\begin{table} \begin{center} \def~{\hphantom{0}}
    \begin{tabular}{c|ccc|cccc|ccccc}

        Regime& $f_\Gamma$& & & $f_{\Gamma B}$& & & & $f_\epsilon$& & & &    \\
        & $C$ & $n$ & $r^2$ & $C$ & $m$ & $n$ & $r^2$ & $C$ & $a$ & $\sigma$ &
        $r^2$  \\[3pt] SWI & 0.704 & 0.932 & 0.919 & 0.741 & 2.513 & 2.513 &
        0.998 & 0.751 & 0.921 & 1.111 & 0.829   \\ ATO& 0.594 & 1.269 & 0.930 &
      0.660 & 2.411 & 2.411 & 1.007 & 0.670 & 1.058 & 0.670 & 0.927  \\
    \end{tabular} 
    \caption{ 
      Final parameters and $r^2$, truncated at the third decimal, for the best
      fits given by $f_\Gamma$, $f_{\Gamma B}$ and $f_\epsilon$ at $t/T_a=15$.
      ATO: atomisation, SWI:\swi{}. $C$ is a prefactor applied to each function
      during the fit procedure.
    } 
    \label{tab:dns.fit_sizepdf_r2} 
  \end{center} 
\end{table}

\section{Dynamics of the jet and the droplets: a two speed fragmentation
\label{sec:dns.two_speed_frag}} 

This section brings explanations about the specific features of the PDF of the
droplet axial velocity, in connection with the vortex ring theory, and about the
joint distribution of the droplet size and velocity. It also analyses the
distribution of the droplets in the Reynolds - Ohnesorge phase space and compares
it to experiments. 

\subsection{The axial velocity PDF and the jet head vortex ring
\label{sec:dns.vortexring}}

The analysis of the distribution of the axial velocity of the droplets in
section \ref{sec:dns.pdf_size_velocity} highlighted the existence of droplets
showing negative velocities and velocities larger than $U_{inj}$, two infrequent 
features for a jet fragmentation. In order to investigate those two
characteristics, it is interesting to have a glance on the spatial
distribution of the droplets such that $u_x/U_{inj}<0$ or $u_x/U_{inj}>1$. To do
so, the cylindrical coordinates $(x/d_n,r/d_n,\theta)$ are preferred to the
Cartesian coordinates. 

\begin{figure}
  \centering
  \includegraphics[width=0.4\textwidth]{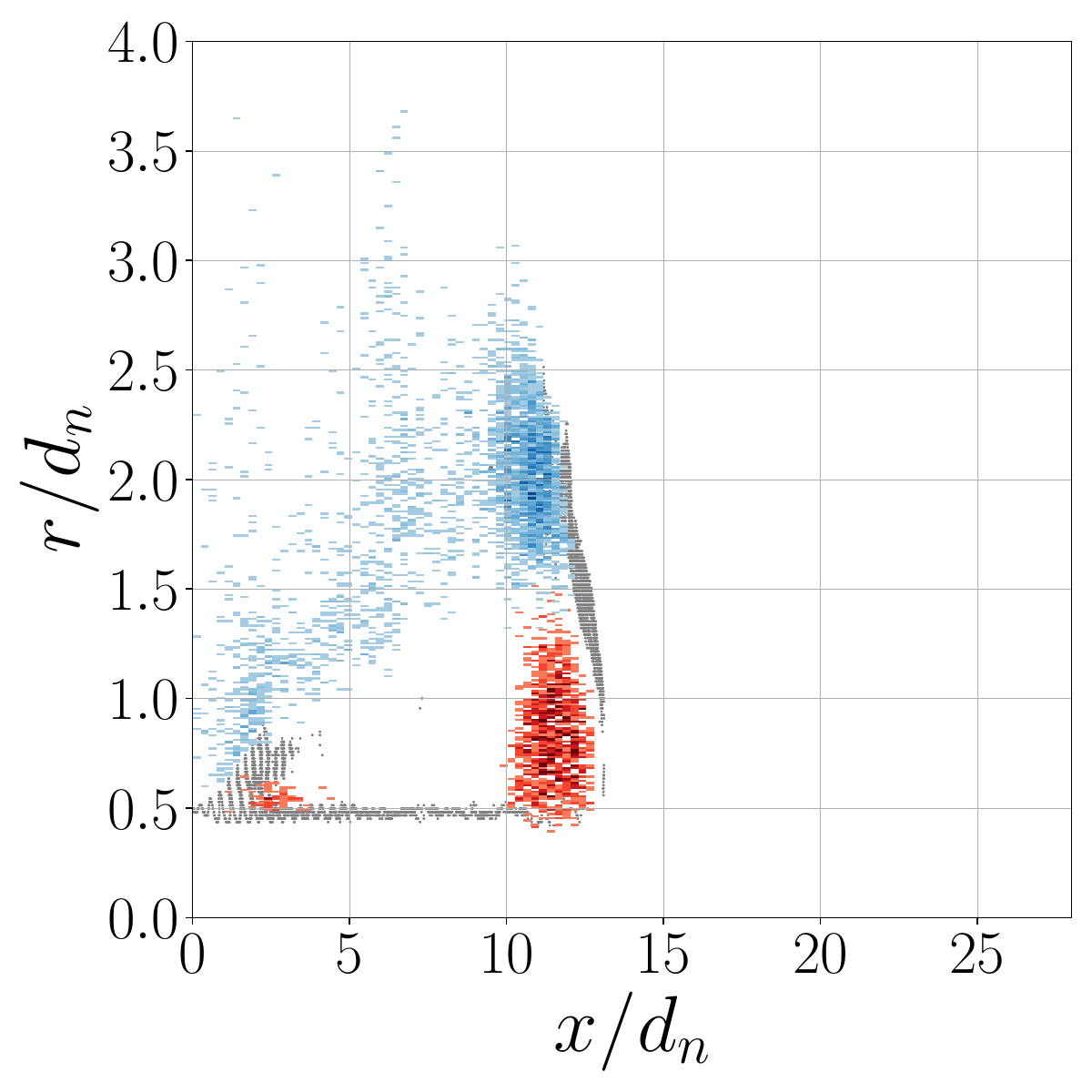}
  \includegraphics[width=0.49\textwidth]{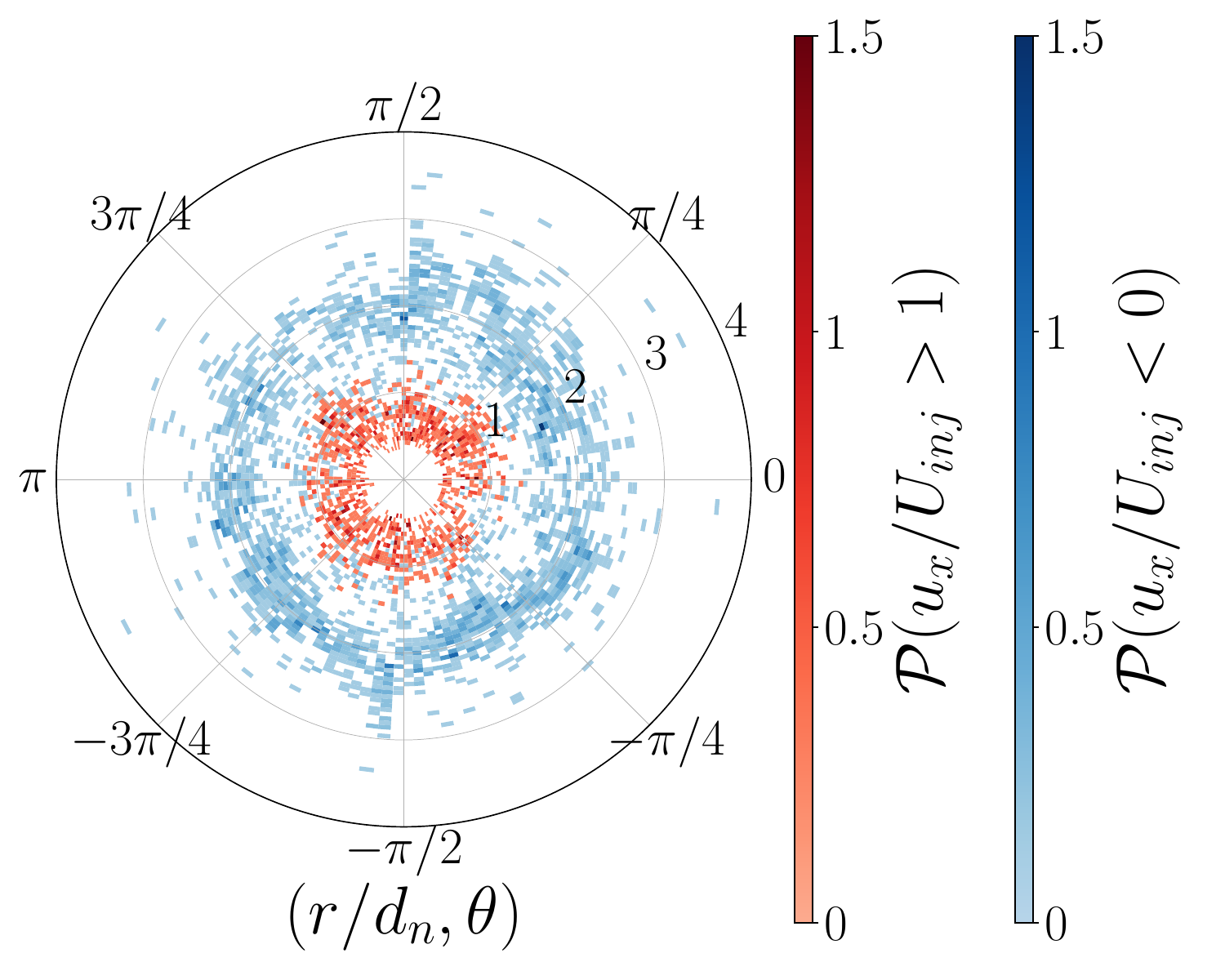}
  \caption{ 
    Spatial evolution of the probabilities $\mathcal{P}(u_x/U_{inj}<0)$ and
    $\mathcal{P}(u_x/U_{inj}>1)$ for $We_2=99.8$ (DNS 8) at $t/T_a=15$ in
    cylindrical coordinates $(x/d_n,r/d_n,\theta)$. For each 2D graph, the
    probabilities are integrated on the third direction. On the $(x/d_n,r/d_n)$,
    the gray bullets represent the mean jet interface and few droplets, see
    appendix \ref{app.mean_interface} for the details. 
  }
  \label{fig:dns.uxfeatures_spatialdistrib} \end{figure}

Figure \ref{fig:dns.uxfeatures_spatialdistrib} gives the spatial evolution in
cylindrical coordinates of the probabilities $\mathcal{P}(u_x/U_{inj}<0)$ and
$\mathcal{P}(u_x/U_{inj}>1)$. For each 2D graph, the probabilities are
integrated on the third direction, e.g. along the $\theta$ direction for the
$(x/d_n,r/d_n)$ graph. Note that the liquid core starts at $r/d_n=0.5$ and that
the jet extends up to $x/d_n\approx12.5$.  In the $(x/d_n,r/d_n)$ space, the
droplets appear to be located in four regions. The ones being faster than
$U_{inj}$ are preferentially located next to the nozzle $\left( 0<x/d_n<0.5,
r/d_n = 0.5 \right)$ and at the backside of the jet head up to half of the head
sheet extension $\left( 10<x/d_n<15, 0.5 < r/d_n < 1.5 \right)$. The former are
due to the jet forcing.  Indeed, the forcing described in section
\ref{sec:dns.phys_config} is sinusoidal with a mean equal to $U_{inj}$ and some
droplets issued from the corolla fragmentation can show velocities larger than
$U_{inj}$. The droplets showing negative velocities are preferentially located
at the backside of the jet head from the half of the head extension up to its
edge and located on a tail expanding over $x/d_n\in[0,10]$ and
$r/d_n\in[0.5,2.5]$.  The negative velocity or the velocity larger than
$U_{inj}$ of the droplets located at the downstream face of the jet head can be
connected to the recirculation occurring behind it.  Finally, the negative
velocities along the tail towards $(x/d_n=0,r/d_n=0.5)$ can correspond to some
droplets ejected from the recirculation region, with $r/d_n$ increasing because
of the increasing radius of the jet head in the time range $t/T_a\in[0,15]$. The
spatial distribution in the $(r/d_n,\theta)$ space shows homogeneity along the
$\theta$ direction and a clear distinction between the two droplet groups along
the $r$ direction. The velocities larger than $U_{inj}$ are concentrated in the
boundary layer region, $r/d_n\in[0.5,1]$, while the negative velocities spread
over it, $r/d_n\in[1.5,2.5]$.

\begin{figure} 
  \centering
  \includegraphics[width=0.49\textwidth]{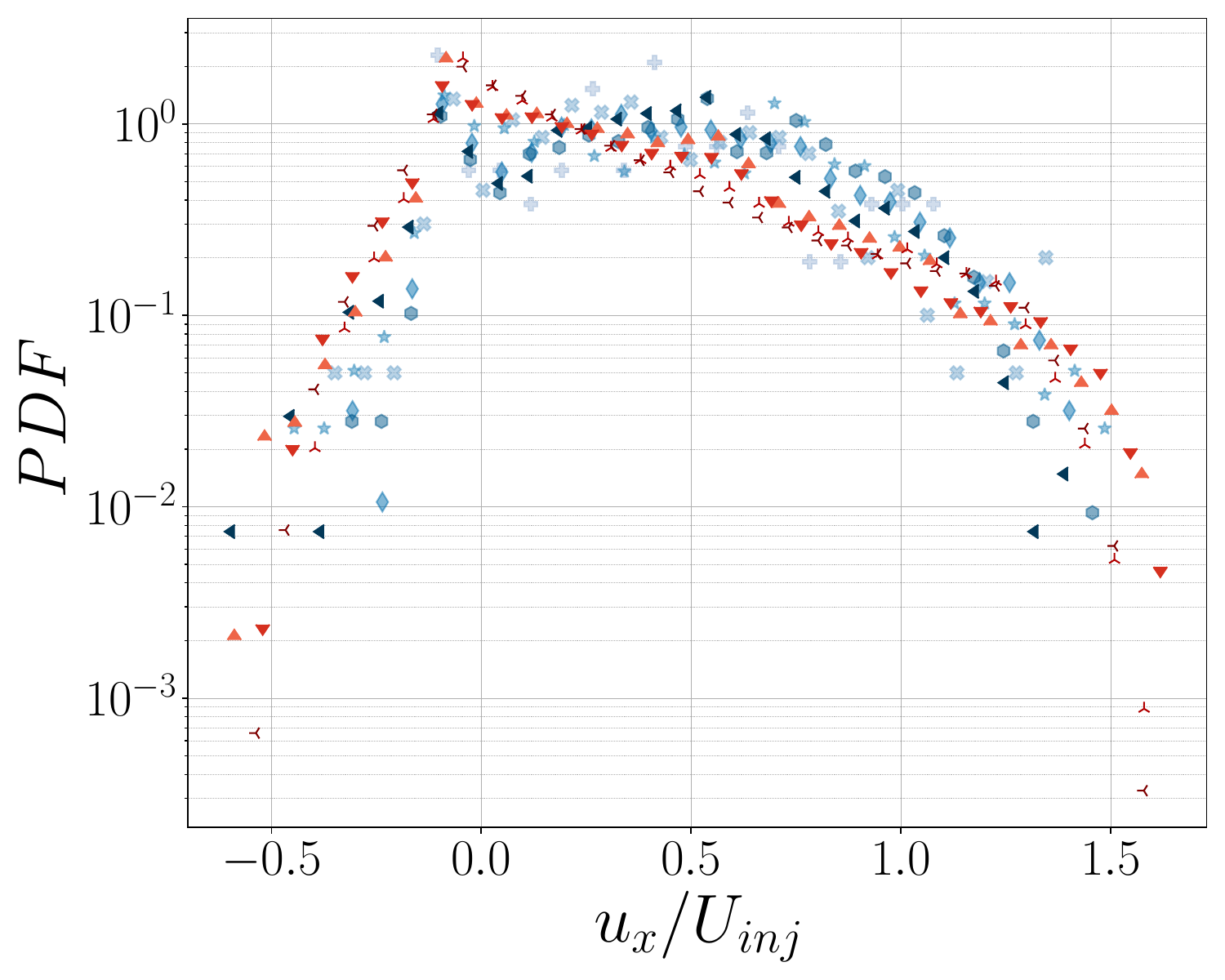} 
  \caption{ 
    Distribution of $u_x/U_{inj}$ at $t/T_a=15$.
    The blue and red colours denote the \swi{} and
    atomisation regimes. The colour code denoting the DNS is the same as in Figure
    \ref{fig:dns.front_evolution}.
  }
\label{fig:dns.pdf_ux_uinj}
\end{figure}

Now that the droplets with, at first sight, unexpected axial velocities,
$u_x/U_{inj}<0$ and $u_x/U_{inj}>1$, are located in the recirculation region
behind the jet head, assessing this recirculation would help to explain why such
velocities are reached. Looking at the distribution of $u_x/U_{inj}$ is a time
saver for this purpose, as it quantifies in a straightforward manner the range
of velocities relatively to $U_{inj}$ happening in this region.
$\mathcal{P}_{u_x/U_{inj}}$ is given in Figure \ref{fig:dns.pdf_ux_uinj} and the
ranges of unexpected velocities are  $u_x/U_{inj}\in[-0.5,0]$ and
$u_x/U_{inj}\in[1,1.5]$.

Assuming that the recirculation observed behind the jet head behaves as a vortex
ring behind a plate, it is possible to use the developments of
\citet{saffman_vortex_1992} describing the dynamics of such unsteady
objects to express the velocity at the vortex core $u_c$ in terms of the plate
velocity $U_d$, see appendix \ref{app.vortex_dyn}: 

\begin{equation} u_c = \bigg(\frac{c}{R}\bigg)^{-1}
\frac{2}{\pi^2\sqrt{2/3}}~U_d \label{eq:dns.vortexring_edgevelocity}
\end{equation}

\noindent where $c$ and $R$ are respectively the core radius and the vortex
radius. For a uniform core $c/R=0.19$, for a hollow core $c/R=0.14$, leading to
$u_c$ respectively equal to $1.31~U_{d}$ and $1.77~U_{d}$. Taking $c/R=0.165$, the
mean value between 0.19 and 0.14, $u_c= 1.504~U_{d}$. In our flow, the jet head
can be approximated as a disc behind which a vortex ring develops.  In section
\ref{sec:dns.flowcharac_dropstat}, we observed that the jet head has the same
velocity as $U_{inj}$, so $U_d=U_{inj}$.  The velocity at the core edge then
equals $\pm1.5~U_{inj}$ and corresponds to the range of unexpected droplet
velocities, $u_x/U_{inj}\in[-0.5,0]$ and $u_x/U_{inj}\in[1,1.5]$. Thus, the
negative velocities and velocities larger than $U_{inj}$ result from the vortex
ring dynamics taking place at the downstream side of the jet head.

\subsection{Joint distribution of the droplet size and axial velocity \label{sec:dns.jpdf_size_velocity}}

\begin{figure}
  \centering
  \begin{subfigure}[c]{0.49\textwidth}
            \centering
            \includegraphics[width=\textwidth]{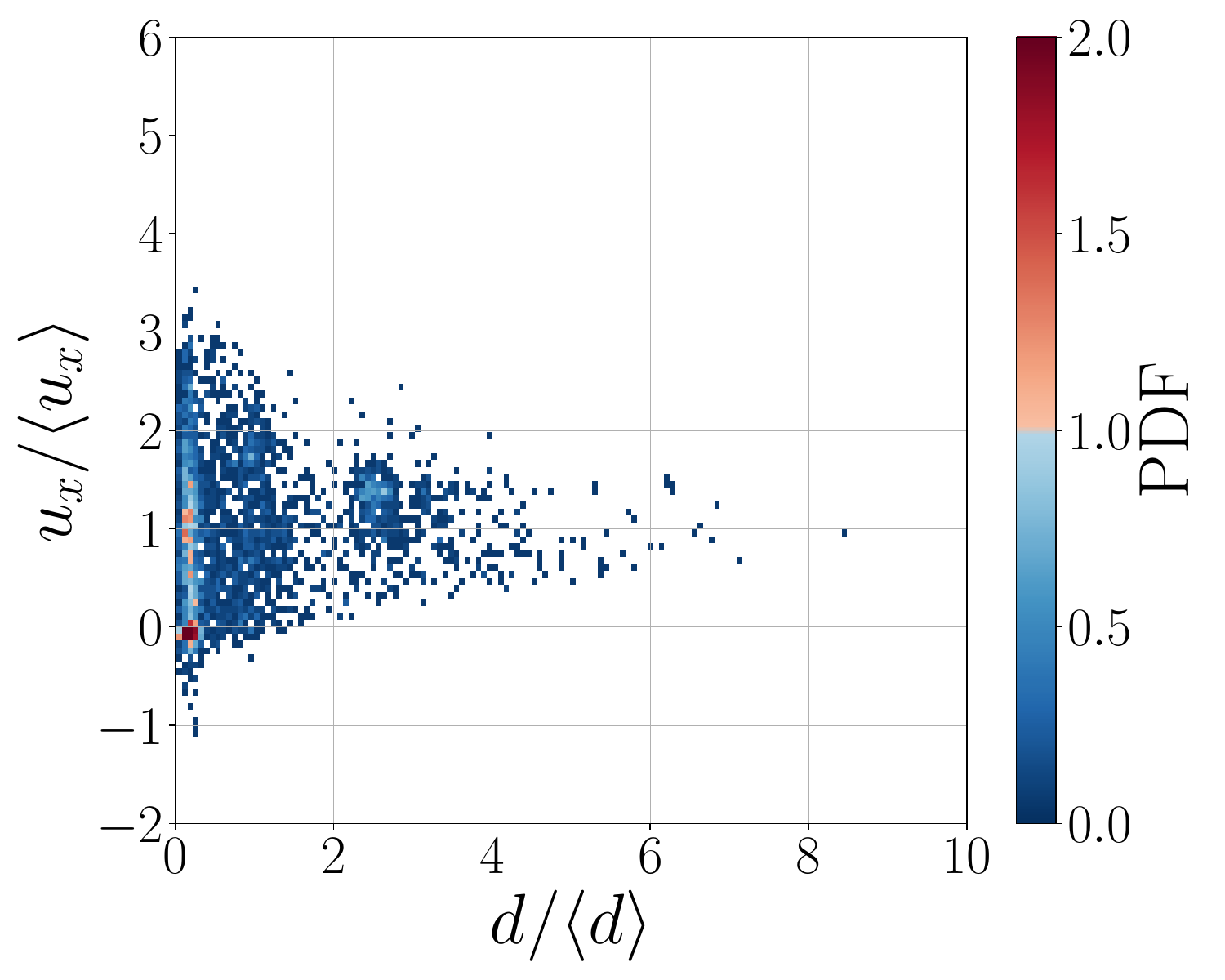}
                \caption{}
                \label{fig:dns.jpdf_size-ux_dns6_tm30}
  \end{subfigure}%
  \begin{subfigure}[c]{0.49\textwidth}
            \centering
            \includegraphics[width=\textwidth]{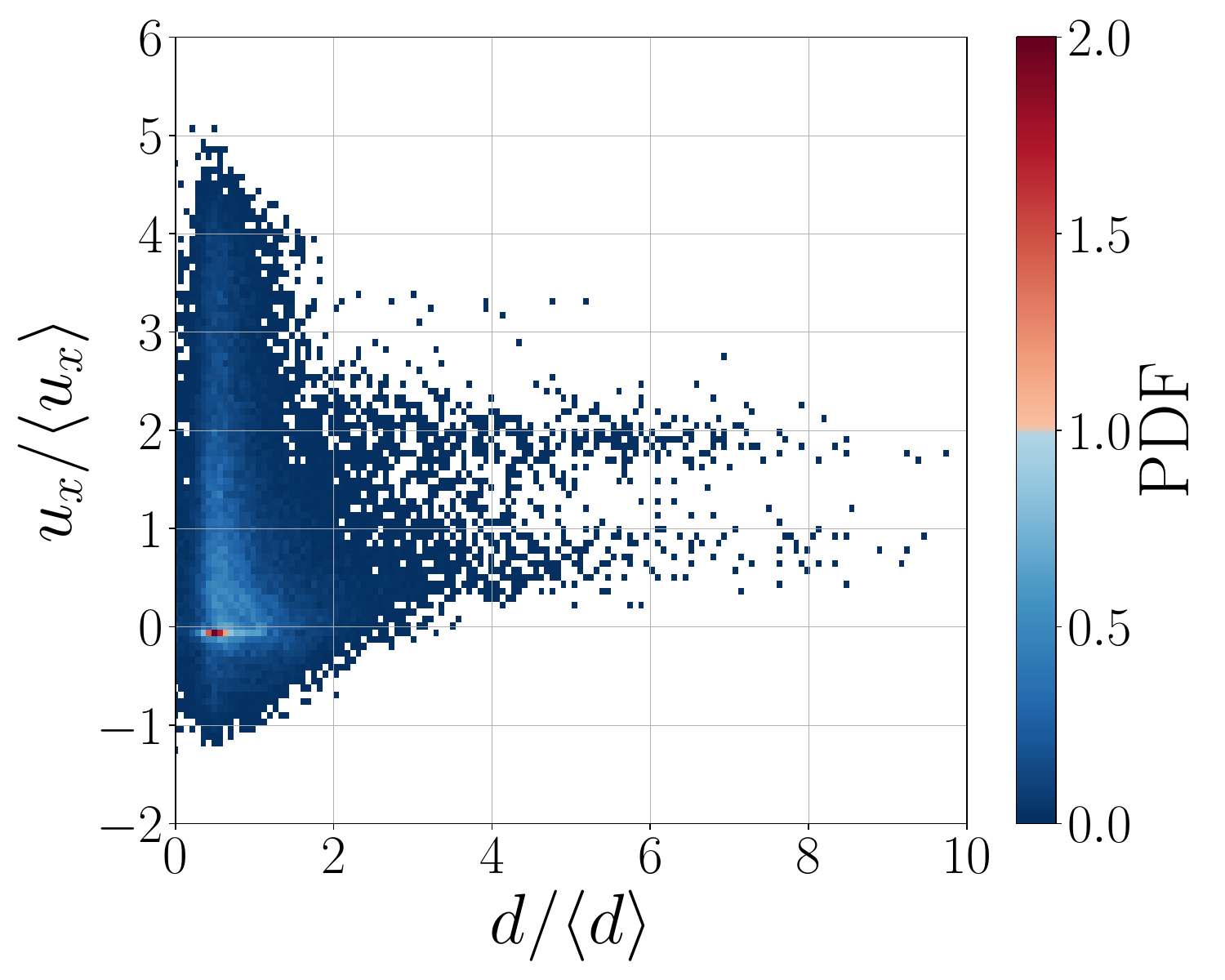}
                \caption{}
                \label{fig:dns.jpdf_size-ux_dns9_tm15}
  \end{subfigure}%
  \caption{
    Joint distributions of the size and the axial velocity of the droplets for
    (a) $We_2=40$ (DNS $6$) in \swi{} regime at $t/T_a=30$ and for
    (b) $We_2=130.3$ (DNS $9$) in atomisation regime at $t/T_a=15$.
  }
  \label{fig:dns.jpdf_size_ux}
\end{figure}

The fragmentation of a jet or droplets is governed by aerodynamic and surface
tension forces. Depending on the equilibrium between those, droplets
of a given size and velocity result. Those two quantities influence each other
comparably to a two-way coupling mechanism. Thus, looking at the joint
distribution of the size and the axial velocity could bring extra information to
the analysis of the marginal PDF $\mathcal{P}_{d/\langle d \rangle}$ and
$\mathcal{P}_{u_x/\langle u_x \rangle}$. Figure \ref{fig:dns.jpdf_size_ux} gives the
joint distribution of $d/\langle d \rangle$ and $u_x/\langle u_x \rangle$ for
$We_2\in\{40,130\}$ (DNS 6 and 9), respectively in the \swi{} regime and the atomisation
regime.

It is possible to recover the characteristics of the marginal PDF in the joint
distributions. For instance, in the joint distribution for $We_2=40$ (DNS 6), three
patches are noticeable along the size axis and correspond to the three modes of
$\mathcal{P}_{d/\langle d \rangle}$, $d/\langle d \rangle \in\{ 0.2, 1, 2.5 \}$.
In addition, the negative velocities and velocities larger than $U_{inj}$
described in section \ref{sec:dns.pdf_size_velocity} and explained in section
\ref{sec:dns.vortexring} are also noticeable. Observing so is expected as the
marginal PDFs are simply the integration of the joint PDF on the size or the
velocity axes.

In the \swi{} regime, the negative axial velocities and the velocities larger
than $U_{inj}$ are preferentially observed for the first two size modes, while
the third size mode is concentrated around $( d/\langle d \rangle = 2.5,
u_x/\langle u_x \rangle \approx 1.5)$ and the tail, from $d/\langle d\rangle
\approx 3$ and towards large sizes, seems to be centered on $u_x/\langle u_x
\rangle =1$.  Globally, the smaller droplets appear to have, at the same time, a
dispersion being large along the velocity axis and being short on the size axis.
Conversely, the larger droplets show a large dispersion along the size axis and
a short one along the velocity axis. This corresponds to the literature and the
common behaviors of tracers and ballistic objects which are classically
observed.  In comparison, even if tracers and ballistic objects are visible as
well, the aspect of the joint distribution in the atomisation regime is
different. Once again, the negative velocities and the velocities larger than
$U_{inj}$ are preferentially observed for the smaller droplets. However, the
distribution shows two tails along the size axis, one centered on $u_x/\langle
u_x \rangle \approx 0.75$ and the second one centered on $u_x/\langle u_x
\rangle\approx2$.  Thus, the smaller and larger droplets still respectively
behave like tracers and ballistic objets, but the ballistic objects show two
traveling velocities.  Conversely to the experimental observations
\citep{vallon_multimodal_2021}, the joint distributions do not show a clear
elbow shape. Also, drawing a third group of droplets showing a similar
dispersion along the size and velocity axes, as in the experimental analysis,
seems less manifest here. Such a group could be extrapolated from the joint
distribution and the second size mode, $d/\langle d \rangle=1$, in the \swi{}
regime and for $d/\langle d \rangle \in [1.5,2.5]$, while the velocity spans
over $u_x/\langle u_x \rangle \in[0,2]$ in both cases.

\begin{figure}
  \centering
  \includegraphics[width=0.4\textwidth]{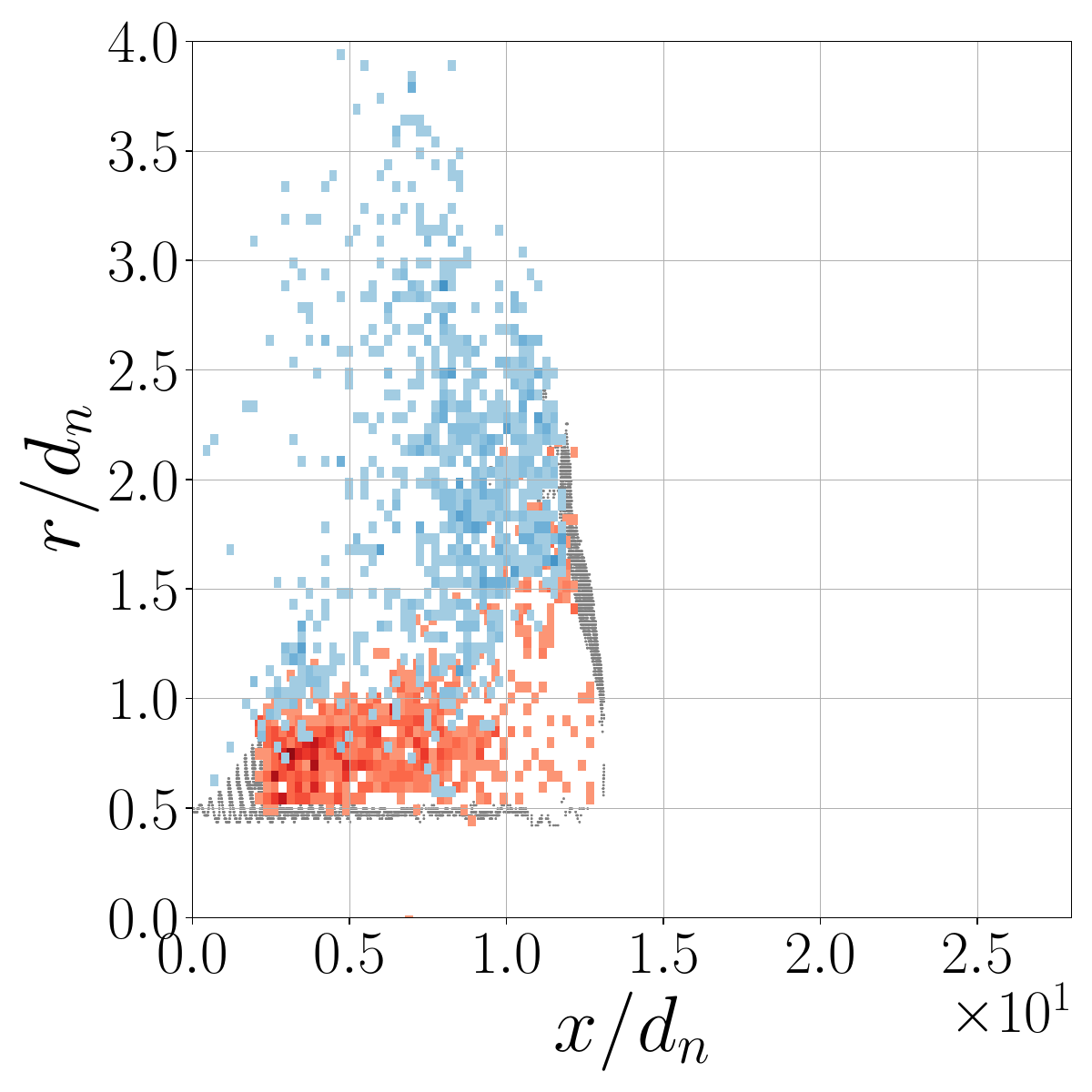}
  \includegraphics[width=0.49\textwidth]{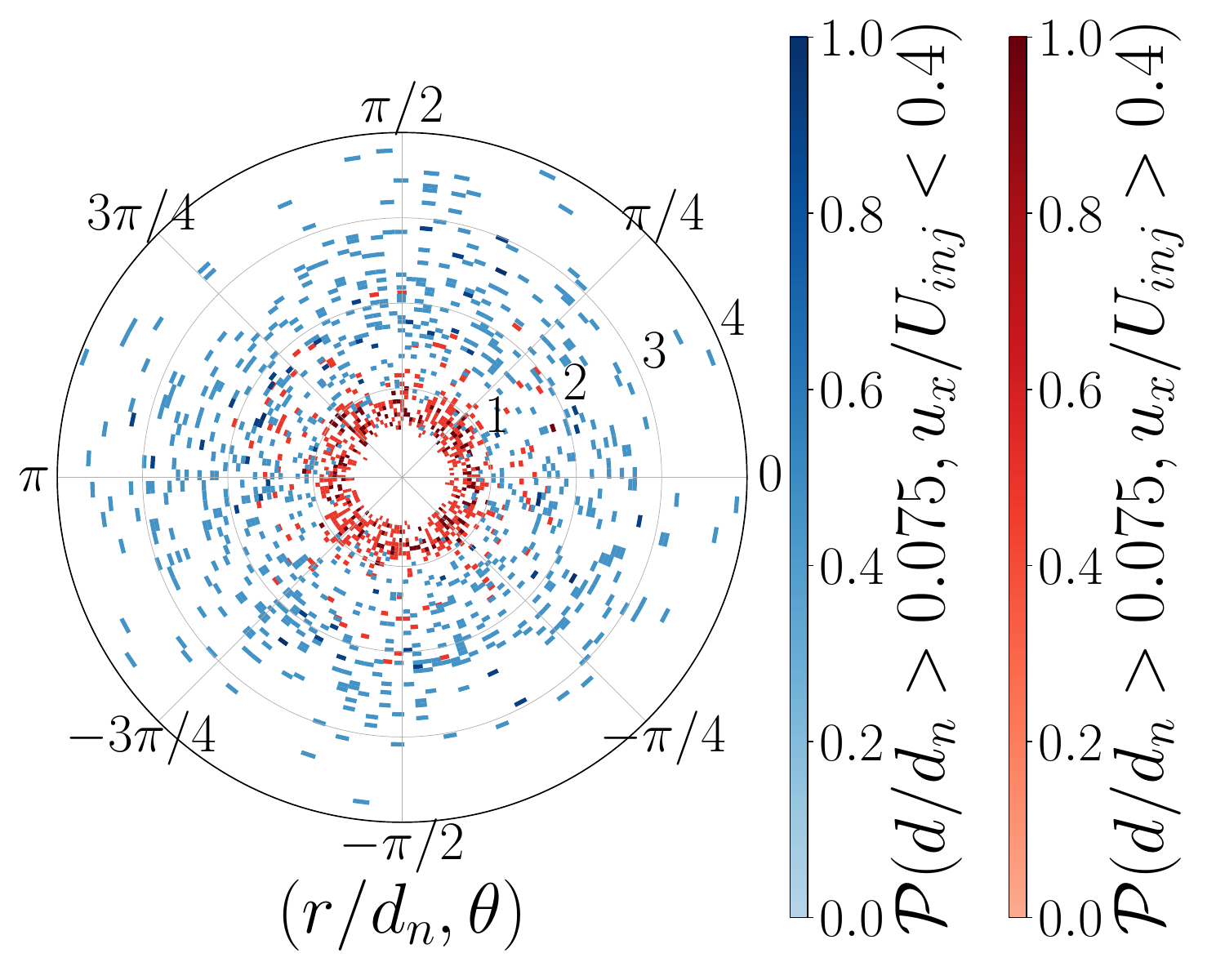}
  \caption{
    Spatial evolution of the probabilities $\mathcal{P}(d/d_n > 0.075,
    u_x/U_{inj}<0.4)$ (blue) and $\mathcal{P}(d/d_n > 0.075, u_x/U_{inj}>0.4)$
    (red) for $We_2=99.8$ (DNS 8) at $t/T_a=15$ in cylindrical coordinates. For
    each 2D graph, the probabilities are integrated on the third direction. On
    the $(x/d_n,r/d_n)$ graph, the gray boundary represents the mean jet interface and
    few droplets, see appendix \ref{app.mean_interface} for the details.
  }
  \label{fig:dns.jpdf_large-d_features}
\end{figure}

Similarly to section \ref{sec:dns.vortexring}, it is possible to check out the
spatial distribution of the droplets corresponding to the tails of the joint
distribution $\mathcal{P}_{(d/\langle d \rangle, u_x/\langle u_x \rangle)}$ for
$We_2=130$ (DNS $9$) in the atomisation regime. Those droplets are such that
$d/\langle d \rangle > 4$ and are distinguished by their axial velocity being
larger or smaller than $1.5\langle u_x \rangle$. Alike the PDF of the axial
velocity of the droplets, the joint distribution shows the same feature for the
larger droplets for all the DNS in the atomisation regime. Once again, it is
more practical to express the conditions on the size and the velocity
independently of the arithmetic average, but relatively to the injection
conditions. Thus, the condition on the size writes as $d/d_n > 0.075$ and
$0.4~U_{inj}$ is considered to be the threshold to distinguish the two tails.
Figure \ref{fig:dns.jpdf_large-d_features} gives the spatial evolution in
cylindrical coordinates of the probabilities $\mathcal{P}(d/d_n > 0.075,
u_x/U_{inj}<0.4)$ and $\mathcal{P}(d/d_n > 0.075, u_x/U_{inj}>0.4)$. As for
Figure \ref{fig:dns.uxfeatures_spatialdistrib}, the liquid core starts at
$r/d_n=0.5$ and the jet extends up to $x/d_n\approx12.5$. The droplets from each
tail appear to exist in specific regions of the space. The large, fast droplets,
$d/d_n > 0.075$ and $u_x/U_{inj}>0.4$, are preferentially located in the
boundary layer region, $r/d_n \in [0.5,1]$, from the nozzle to the jet head. The
large, slow droplets, for which $d/d_n > 0.075$ and $u_x/U_{inj} < 0.4$, are
located on the downstream side of the jet head and around the maximal head sheet
extension. The two groups show some overlapping in the recirculation region. It
is possible that some droplets are caught in the vortex circulation, even if
they preferentially behave as ballistic objects. In the $(r/d_n,\theta)$ space,
the distributions are homogeneous along the azimuthal axis, which respects the
flow symmetry, and the same distribution along the $r/d_n$-axis appears between
the two groups. Thus, the two tails of the joint distribution of the size and
the velocity come from the existence of two sources of fragmentation in the
flow: the head sheet edge and the corollas developing from the jet forcing.

\subsection{Governing parameters at the droplet scale \label{sec:dns.govprm_dropscale}}

\begin{figure}
  \centering
  \begin{subfigure}[c]{0.49\textwidth}
            \centering
            \includegraphics[width=\textwidth]{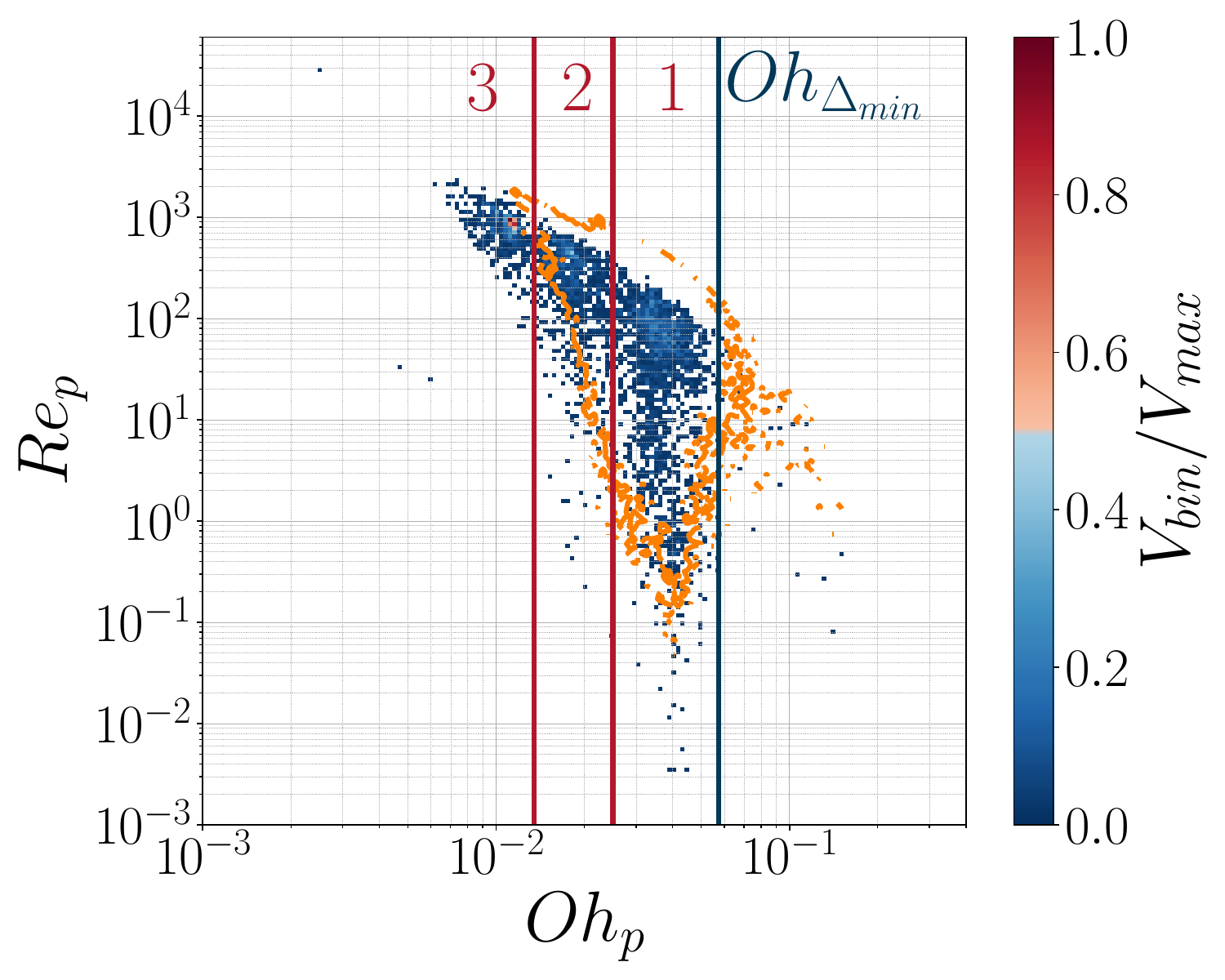}
                \caption{}
                \label{fig:dns.mapRO_map_rep_ohp_dns6_tm30_avgnone}
  \end{subfigure}%
  \begin{subfigure}[c]{0.49\textwidth}
            \centering
            \includegraphics[width=\textwidth]{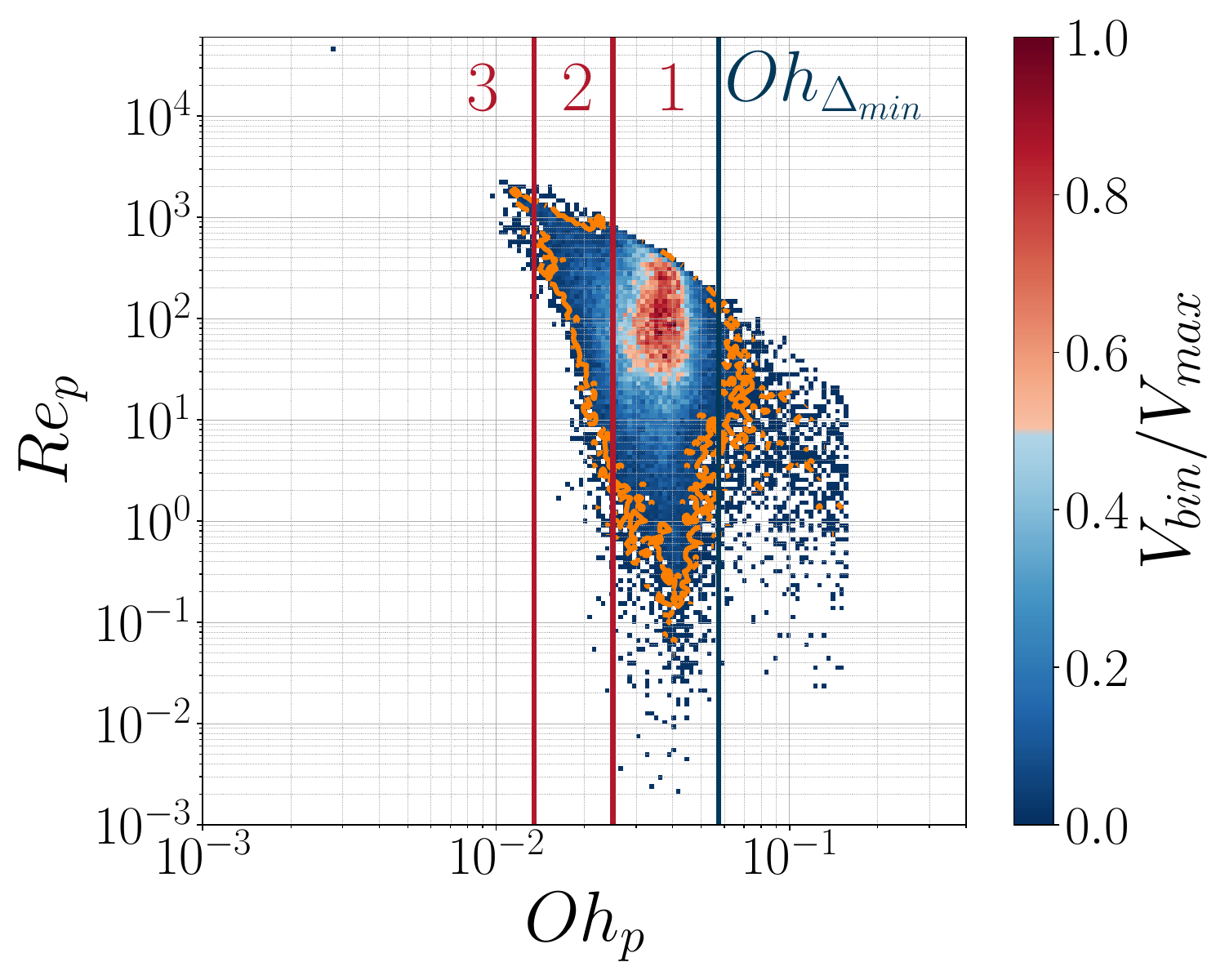}
                \caption{}
                \label{fig:dns.mapRO_map_rep_ohp_dns10_tm15_avgnone}
  \end{subfigure}%
  \caption[
    Joint volume histogram of $Re_p$ and $Oh_p$ for the droplet population for
    $We_2=40$ at $t/T_a=30$ and for $We_2=165$ at $t/T_a=15$.
  ]{
    Joint volume histogram of $Re_p$ and $Oh_p$ for the droplet population for
    $We_2=40$ (DNS 6) at $t/T_a=30$ (a) and for $We_2=165$ (DNS 10) at
    $t/T_a=15$ (b). The Ohnesorge number corresponding to the smallest grid
    cell, $Oh_{\Delta_{min}}$, is indicated by the vertical blue line. The
    vertical red lines indicate the limits between the 3 modes of the size PDF
    for $We_2=40$ (DNS 6), figure \ref{fig:dns.droppdf_size_tm25}. The orange line
    represents the isovalue $V_{bin}/V_{max}=0.03$ for $We_2=165$ (DNS 10).
  }
  \label{fig:dns.map_Rep_Ohp_num}
\end{figure}

The joint distribution of the size and the axial velocity of the droplets gives
some insights on the droplet dynamics in the flow. However, compared with the
experiments, the numerical distributions show a slightly more complex trend and
do not allow to characterise droplets with different behaviors on the basis of
the marginal PDF characteristics. Beyond this, it would be interesting to have a
glance on the flow perceived by a droplet as well as the droplet deformation
resulting from the droplet-flow interaction. Without detailing the flow around
each droplet down to the smallest scales, it is possible to characterise such a
flow by considering its governing parameters, the particulate Reynolds number
and the particulate Ohnesorge number, respectively expressed as:

\begin{equation}
  Re_p=\frac{|u_{p,x}-U_{g,x}| d}{\nu_l},~~~~Oh_p = \frac{\mu_l}{\sqrt{\rho_l\sigma d}}
  \label{eq:dns.rep_ohp}
\end{equation}

\noindent where $d$, $u_{p,x}$ and $|u_{p,x}-U_{g,x}|$ are the particle
diameter, the particle axial velocity and its relative velocity compared to
$U_{g,x}$, the $x$ component of the gas phase velocity averaged over the domain.
The particulate Reynolds number not only brings light on the balance between the
inertial and viscosity forces at the scale of a droplet but it also brings
information on the product of the droplet relative velocity and its diameter. By
concatenating the size and the velocity of a droplet, the latter quantity could
be seen as a potential of fragmentation. The higher the product $d\cdot |u_{p,x}
- U_{g,x}|$ is, the more likely the droplet will fragment in multiple elements.
It also enables to distinguish the droplet-flow interactions between droplets
having the same size but different relative velocities or, equivalently, having
the same relative velocity and different sizes. In addition, the particulate
Ohnesorge number characterises the ratio between the viscosity forces and the
product of the inertial and surface tension forces. This dimensionless number is
usually used to characterise droplet deformation in a given flow. The larger is
$Oh_p$, the less deformable is the droplet. Thus, even if they give global
information on the droplet-scale flow, the combinations of $Re_p$ and $Oh_p$
could help to characterise the droplet behaviors depending on their possible
deformation and potential of fragmentation. Figure \ref{fig:dns.map_Rep_Ohp_num}
gives the normalised joint volume histogram of $Re_p$ and $Oh_p$ of the droplet
population for $We_2\in\{40,165\}$ (DNS 6 and 10) respectively at $t/T_a=30$ and
$t/T_a=15$. Note that the quantity which is plotted is the total volume of the
droplets contained in a bin $(\Delta Re_p,\Delta Oh_p)$, denoted $V_{bin}$, and
normalised by the maximum value of $V_{bin}$, denoted $V_{max}$.

First of all, the $Oh_p$ values larger than $Oh_{\Delta_{min}}$ correspond to
the droplets smaller than the smallest cell size $\Delta_{min}$ and are not
physically relevant. Considering the pair $(Re_p,Oh_p)$ reshapes drastically the
droplet data. Regarding $We_2=165$ (DNS 10), the two peaks present for the large
sizes as well as the peaks around large velocities and negative velocities for
the small sizes in figure  \ref{fig:dns.jpdf_size_ux} do not appear anymore in
the joint volume histogram of the particulate dimensionless numbers. In
addition, while the trends of the size-velocity joint distributions are
significantly different between the two fragmentation regimes, the limits of the
joint volume histogram appear not only to be regular but also follow similar
trends between the two fragmentation regimes, as shown by the comparison of the
joint histogram for $We_2=40$ (DNS 6) and the edge contour for $We_2=165$ (DNS
10).  Regarding the histogram values, different modes appear in the joint
histogram of each DNS. For the DNS 6, in the swi{} regime, it is possible to
denote the three size modes, observed in section
\ref{sec:dns.pdf_size_velocity}, denoted from 1 to 3 and separated at
$Oh_p\in\{1.35\times 10^{-2}, 2.5\times 10^{-2}\}$ by the red vertical lines.
Each droplet group shows some dispersion along the $Re_p$ direction, dispersion
which increases when the droplet size decreases. The population thus shows three
subgroups whose dynamics seems to mainly be governed by their size. From those
three size subgroups, only the modes 1 and 2 remain in the joint volume
histogram of the DNS 10, in the atomisation regime. The mode of large sizes,
mode 3, does not exist in the atomisation regime because the corolla issued from
the forcing cannot develop nor create rim leading to the generation of such
droplet sizes. For the latter DNS, the mode 1, existing at large $Oh_p$ and
indicated by the red region in figure
\ref{fig:dns.mapRO_map_rep_ohp_dns10_tm15_avgnone}, gains in importance and is
the main size mode in the atomisation regime, existing for $Oh_p\in
[2.5,5]\times 10^{-2}$ and $Re_p\in [30,400]$.  Additionally, the dispersion of
the modes for moderate and small $Oh_p$ increases between the two regimes while
respecting the similar outer limits, as the droplet data spread over all the
space delimited by the edge contour for $We_2=165$ (DNS 10).  Finally, it is worth noting
the absence of droplets in the region of large particulate Reynolds and small
particulate Ohnesorge, $(Re_p,Oh_p)\in([10^3,10^4],[1,7]\times 10^{-3})$, i.e.
droplets whose size and axial velocity are of the order of $d_n$ and $U_{inj}$.

\begin{figure}
  \centering
  \begin{subfigure}[c]{0.49\textwidth}
            \centering
            \includegraphics[width=\textwidth]{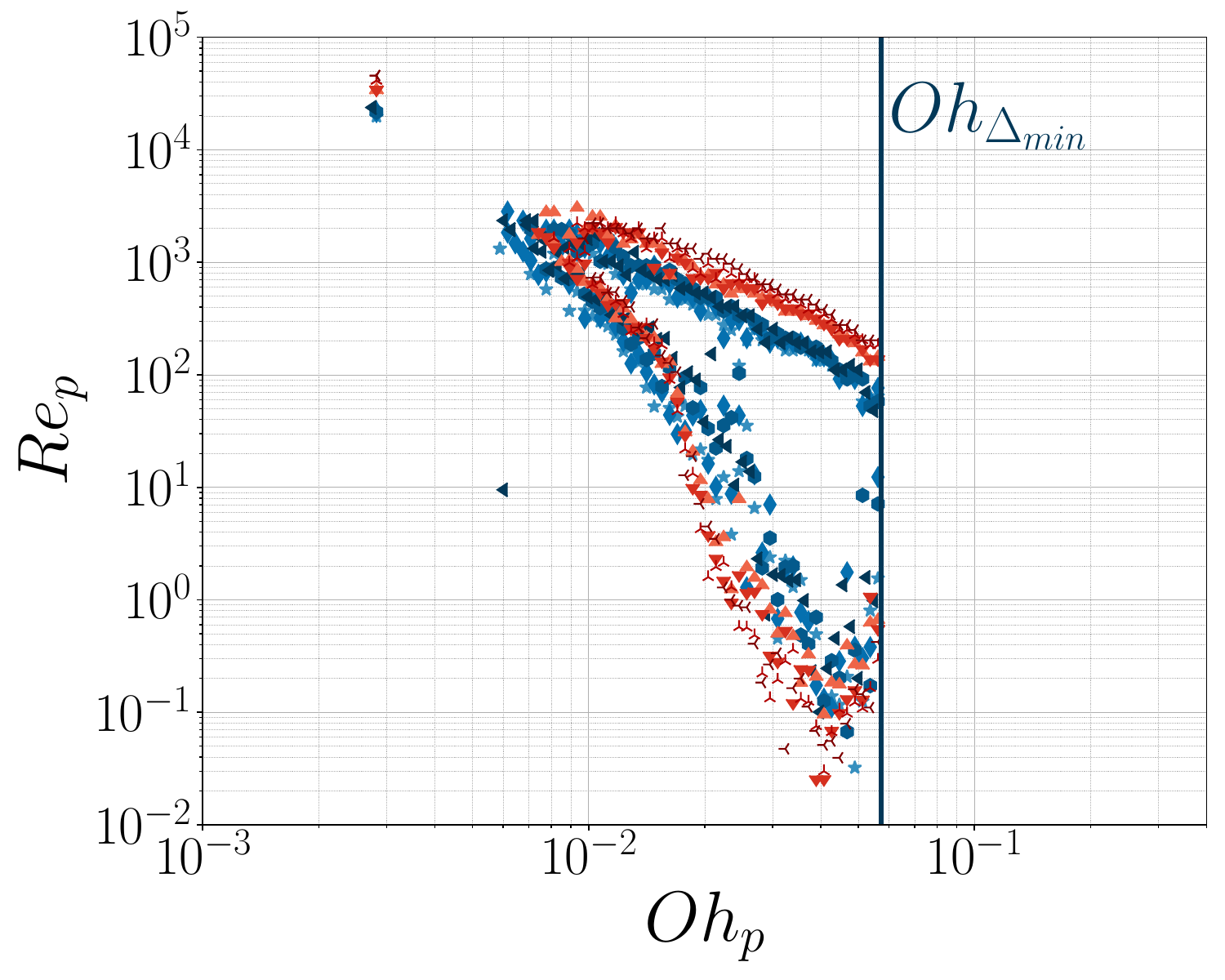}
                \caption{}
                \label{fig:dns.mapRO_contour_avgnone-none_tm15}
  \end{subfigure}%
  \begin{subfigure}[c]{0.49\textwidth}
            \centering
            \includegraphics[width=\textwidth]{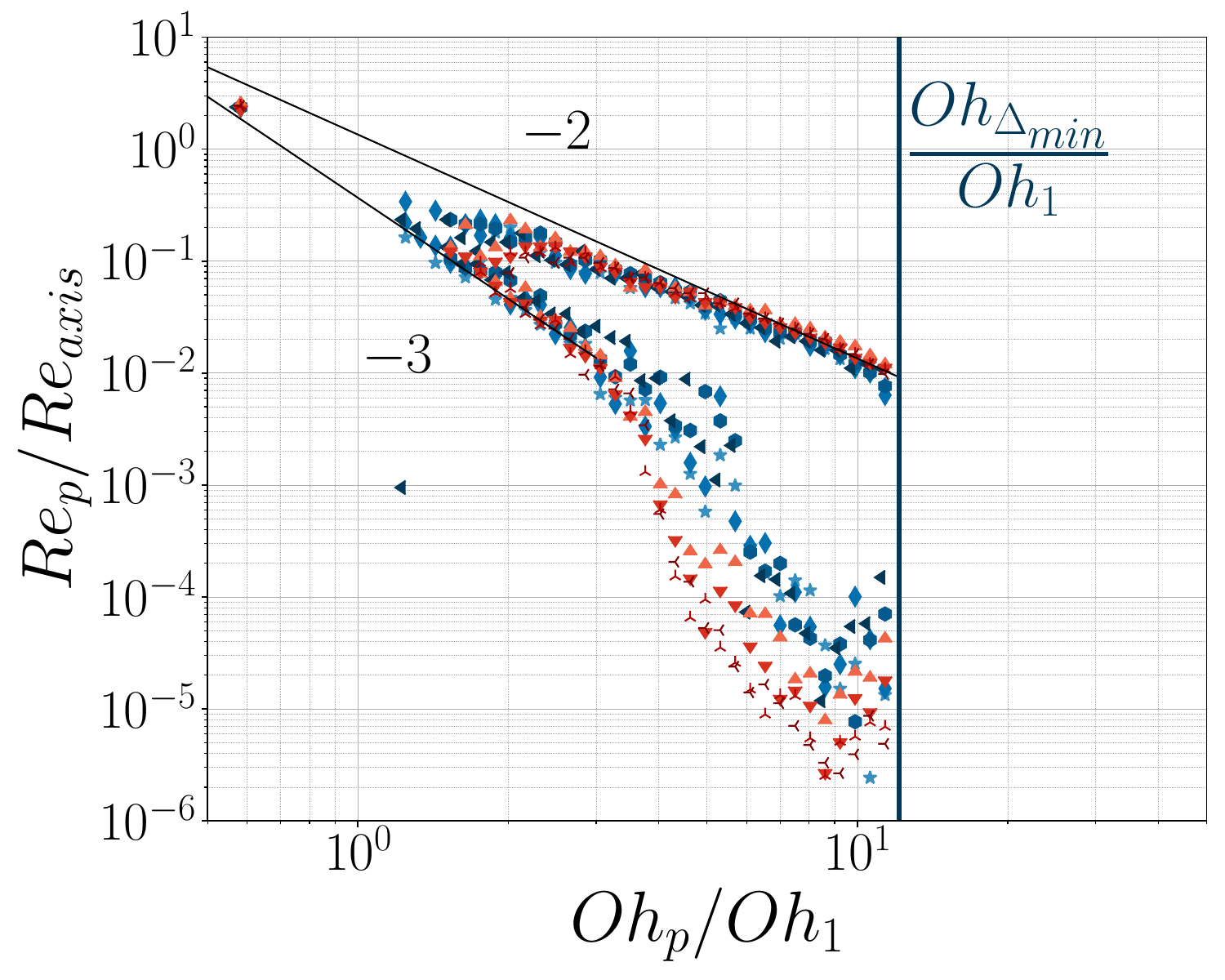}
                \caption{}
                \label{fig:dns.mapRO_contour_avgOhL-ReAxis_tm15}
  \end{subfigure}%
  \caption{
    Color on-line. Edges of (a) the joint volume histograms and of (b) the
    normalised joint volume histograms at $t/T_a=15$. The blue and red colours
    denote the \swi{} and atomisation regimes. The colour code denoting the DNS
    is the same as in Figure \ref{fig:dns.front_evolution}.
  }
  \label{fig:dns.contour_Rep_Ohp_num}
\end{figure}

The comparison of the edge contour for $We_2=165$ (DNS 10) with the joint volume
histogram for $We_2=40$ (DNS 6) given by figure \ref{fig:dns.map_Rep_Ohp_num}
suggests that the joint histograms follow similar borders regardless of the
fragmentation regime.  Figure \ref{fig:dns.contour_Rep_Ohp_num} dives in a more
detailed comparison of the joint histogram borders by superposing the edges for
all the DNS and proposes a normalisation of the two dimensionless numbers. The
edges are obtained by sampling each joint histogram along the $Oh_p$ direction
and keeping for each sample the maximum of the ordinates and the ordinate of the
percentile at 7 $\%$. This technique enables to discard the outlier points
existing at small $Re_p$.  From the edges of the non normalised joint
histograms, it appears clearly that the joint histograms evolve in the same
phase space for both fragmentation regimes and that the borders only show a
slight evolution with the gaseous Weber number $We_2$.  Note that the isolated
points in the top left corner are the $Re_p$ and $Oh_p$ values corresponding to the
liquid core.  Those points depart from $Re_1$ and $Oh_1$ because the liquid core
has a volume larger than that of a sphere of diameter $d_n$. As a reminder, the
injection dimensionless numbers are given in Table \ref{tab:dns.uinj_we_re_f_sr}.

In order to normalise $Re_p$ and $Oh_p$, one can choose the Reynolds number on
the jet axis $Re_{axis}$, computed with the nozzle diameter $d_n$ and
$u_{x,axis}$ the jet velocity on the $x$-axis, and the injection Ohnesorge
number $Oh_1$ computed with $d_n$. In our simulation, the velocity of the jet
along the $x$-axis does not show any diminution, thus $u_{x,axis}=U_{inj}$ and
$Re_{axis}= Re_1$. Making the distinction between the injection velocity and the
jet velocity on the axis might appear auxiliary here but is relevant for
comparing the numerical data with experiments. Once normalised, all the upper
and lower borders of the joint histograms collapse. Not only those borders
collapse but also show a power law dependency. For $Re_p / Re_{axis} \geq
10^{-2} $, the former scales such that $Re_p/Re_{axis} = 1.35 (Oh_p/Oh_1)^{-2}$
and the latter scales such that $Re_p/Re_{axis}=0.37 (Oh_p/Oh_1)^{-3}$. The
collapse however does not hold for the borders on the range $Oh_p/Oh_1\in[4,10]$
and $Re_p/Re_{axis} \leq 10^{-2}$, which could simply result from the difference
in the relative velocity and the extreme values along the $Re_p$-axis.
Additionally, not only the isolated points corresponding to the liquid core
collapse, but also they lie within the space delimited by the two power laws.
Thus, a spray developing after the pinching region, in the same configuration,
could show a phase space delimited by the same power laws and spreading from the
liquid core towards the smallest droplets.  Finally, as all the contours
collapse in figure \ref{fig:dns.mapRO_contour_avgOhL-ReAxis_tm15}, it is
possible to add that this cascade in the phase space
$(Re_p/Re_{axis},Oh_p/Oh_1)$ is independent of the gaseous weber number
$We_2$.

Overall, the comparison of the joint volume histogram in the \swi{} regime,
$We_2=40$ (DNS 6), and in the atomisation regime, $We_2=165$ (DNS 10), indicates
that the dominant modes of the droplet population evolve with $We_2$.
Particularly, the mode for small $Oh_p$, i.e. large droplet sizes, does not
exist in the atomisation regime.  Also, the atomisation regime presents a larger
dispersion in $Oh_p$ and $Re_p$.  This is expected as the increase in $We_2$
creates aerodynamic conditions in which large droplets are very unlikely to
survive, or even be generated, and the increase of the relative velocity between
the gas and the liquid induces an increase of the deviation of the size and the
axial velocity, see section \ref{sec:dns.pdf_size_velocity}.  This analysis
highlights the possibility to reshape the size and velocity data of the droplets
into a regular shape, even if the size-velocity joint distribution shows
irregular boundaries and infrequent features. Additionally, it indicates that
the joint histogram values of the particulate dimensionless numbers evolve with
$We_2$ while respecting outer borders which are largely independent of $We_2$.
Normalising the particulate dimensionless numbers by the injection dimensionless
numbers shows that the droplets exist over a steady phase space, delimited by
power  and exponential laws. Finally, such joint volume histogram opens the way
for qualifying the different flow regimes undergone by the droplets and the
consequent fragmentation mechanisms. 

\subsection{Droplet phase space: simulations and experiments \label{sec:dns.dropphasespace_num_exp}}

\begin{figure}
  \centering
  \includegraphics[width=0.7\textwidth]{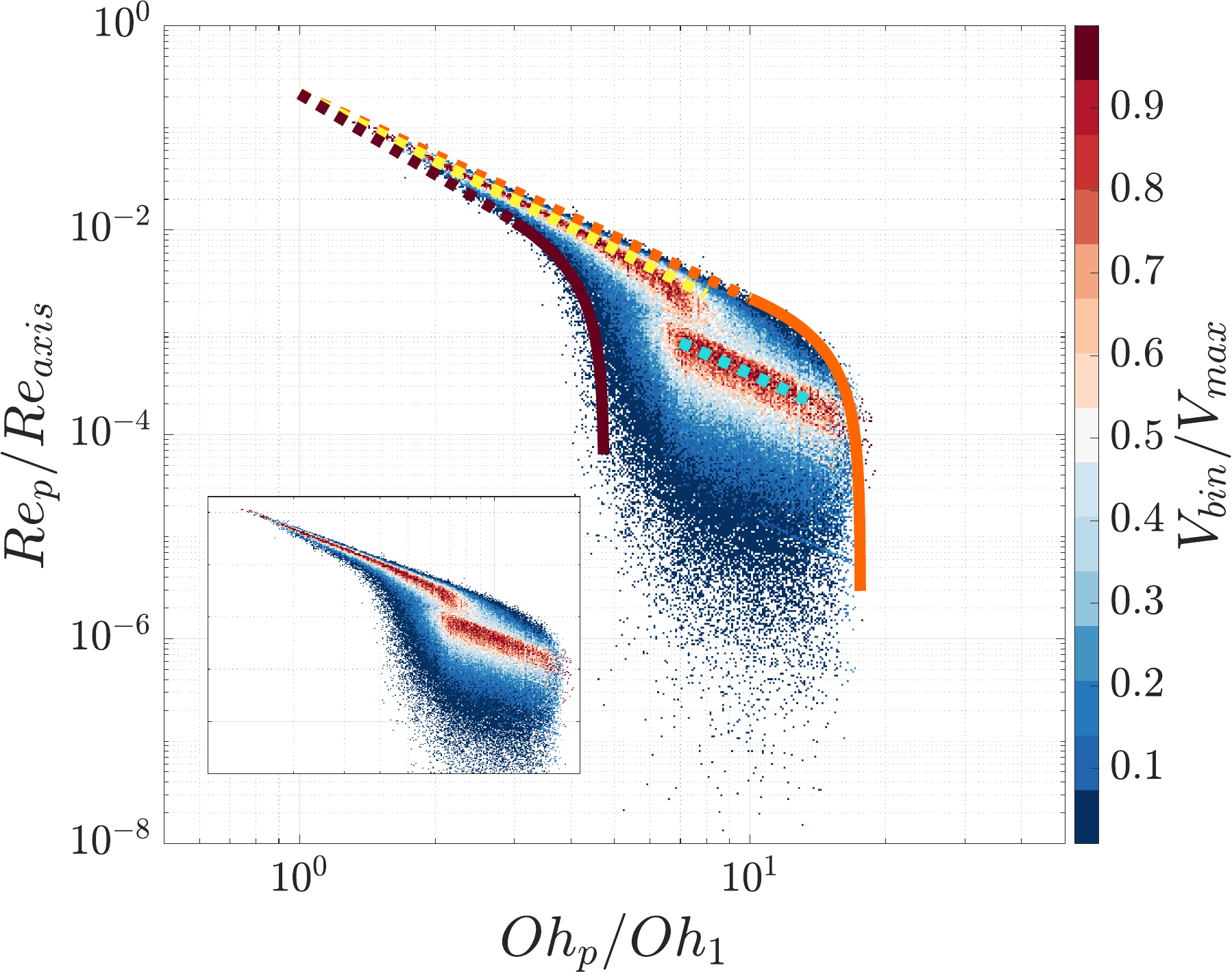}
  \caption[
    Experimental joint volume histogram of $Re_p/Re_{axis}$ and
    $Oh_p/Oh_1$ at $x/d_n=800$.
  ]{
    Experimental joint volume histogram of $Re_p/Re_{axis}$ and
    $Oh_p/Oh_1$ at $x/d_n=800$. The dash-dot lines represent the power
    law scalings while the solid lines represent the exponential scalings. The
    insert recalls the joint volume histogram without the
    modelled borders. 
  }
  \label{fig:dns.map_Rep_Ohp_exp}
\end{figure}

\citet{vallon_multimodal_2021} proposed, among others, a detailed analysis of
the experimental joint distribution of the size and axial velocity of the
droplets in the case of a water jet injected into quiescent air at $We_2=24$ and
lying in the \swi{} regime. The experimental apparatus used to perform
simultaneous measurements of the size and the velocity of the droplets is
detailed by \citet{felis_experimental_2020}. The originality of that
experimental campaign lies in the simultaneity of the DTV size-velocity
measurements and the distance where they were carried out: from 400 to 800
nozzle diameters along the jet axis. Complementarily, and following the insights
of section \ref{sec:dns.govprm_dropscale}, it is possible to look at the
experimental joint volume histogram of the particulate Reynolds and Ohnesorge
numbers. Figure \ref{fig:dns.map_Rep_Ohp_exp} gives the joint volume histogram
for $Re_p/Re_{axis}$ and $Oh_p/Oh_1$ derived from the experimental measurements
of \citet{felis_experimental_2020}. Note that the mean velocity of the liquid
phase on the jet axis $u_{x,axis}$ is no longer equal to the injection velocity,
$U_{inj} = 35~\textnormal{m/s}$, but has decreased by $20\%$ at $x/d_n=800$.

Once again, the borders of the joint volume histogram are well-defined and can
be easily modelled. The upper and lower borders split into two scalings. For
$Re_p/Re_{axis}\geqslant O(10^{-3})$, the borders follow a power law while they
follow an exponential scaling for smaller values of $Re_p/Re_{axis}$.  The upper
and lower borders are respectively denoted $\mathcal{B}_{up}$ and
$\mathcal{B}_{low}$ and their scaling is such that:

\begin{equation}
  \mathcal{B}_{up}:\left\{
    \begin{array}{ll}
      \frac{Re_p}{Re_{axis}}=0.215 \Big(\frac{Oh_p}{Oh_1}\Big)^{-2}, & \forall~
      Oh_p/Oh_1\in[1,10]\\[2pt]
      \frac{Re_p}{Re_{axis}}= \exp\bigg(-0.1 \Big(\frac{Oh_p}{Oh_1}+45.1\Big) \bigg) -
      1.90\times 10^{-3}, & \forall~
      Oh_p/Oh_1\in[10,20]\\[2pt]
  \end{array} \right.
  \label{eq:dns.map_rep_oh_exp_bordersUP}
\end{equation}

\begin{equation}
  \mathcal{B}_{low}:\left\{
    \begin{array}{ll}
      \frac{Re_p}{Re_{axis}}=0.215 \Big(\frac{Oh_p}{Oh_1}\Big)^{-2.61}, &
      \forall~
      Oh_p/Oh_1\in[1,3]\\[2pt]
      \frac{Re_p}{Re_{axis}}= \exp\bigg(-0.6 \Big(\frac{Oh_p}{Oh_1}+3.65\Big) \bigg) -
      6.5\times 10^{-3}, & \forall~
      Oh_p/Oh_1\in[3,5]\\[2pt]
  \end{array} \right.
  \label{eq:dns.map_rep_oh_borderLOW}
\end{equation}

\noindent Note that, for a given value of $Oh_p$, the upper border describes the
fastest droplets at a given size while, for a given value of $Re_p$, it
describes the smallest droplets at a given velocity. Thus, the upper border can
be seen as the border describing the smallest and fastest droplets in a given
region of the phase space, the reverse logic holds for the lower border.
Additionally, two main ``paths'' can be distinguished in the joint histogram.
The first one lies in the power law region and the second one in the exponential
region, respectively denoted $\mathcal{P}_1$ and $\mathcal{P}_2$, both of
them follow a power law scaling such that:

\begin{equation}
  \begin{array}{ll}
      \mathcal{P}_1:~ \frac{Re_p}{Re_{axis}} = 0.215 \Big(
  \frac{Oh_p}{Oh_1}\Big)^{-2.175}, & \forall~ Oh_p/Oh_1\in[1,7]\\[2pt]

  \mathcal{P}_2:~ \frac{Re_p}{Re_{axis}} = 0.039 \Big(
  \frac{Oh_p}{Oh_1}\Big)^{-2}, & \forall~ Oh_p/Oh_1\in[7,14]\\[2pt]
  \end{array}
  \label{eq:dns.map_rep_oh_exp_paths}
\end{equation}

Let us focus on the borders scaling as a power law. Starting from the expression of
$Oh_p$, it is possible to rewrite $Re_p$ as $Re_p=\sigma^{-1}\mu_l|u_{p,x}-U_{g,x}|
Oh_p^{-2}$. Knowing that in this region $Re_p=C~Oh_p^{-2-\alpha}$ with
$C\in\mathbb{R}$, we then have $|u_{p,x}-U_{g,x}|=\sigma C ~Oh_p^{-\alpha}$,
which is equivalent to $|u_{p,x}-U_{g,x}| \propto d^{\alpha/2}$. The droplets
then show a velocity relative to the gas phase which increases with the droplet
size. The coefficient $\alpha$ necessarily lies in $\mathbb{R}^+$. Indeed, a
negative $\alpha$ would mean that the relative velocity of a droplet decreases
when its size increases and consequently that larger objects would be more
sensitive to the gas phase flow, which goes against the observation of ballistic
objects in fragmentation flows. Consequently, the borders scaling as $Oh^{-2}$
seem to result from dynamical limits.  Regarding the lower border scaling as a
power law, we have $\alpha \in \{0.4,0.45,0.5,0.56,0.61\}$ for
$x/d_n\in\{400,500,600,700,800\}$. Inferring a rule on the evolution of the
upper bound of $\alpha$ from the experimental data seems reckless.

Regarding the exponential scaling of the borders, it is interesting to note the
existence of an offset along $Oh_p$ and $Re_p$. The droplets being on the upper
border preferentially have a smaller size and a larger relative velocity, while
those on the lower border have a larger size and a smaller relative velocity.
In order to have all the droplets lying in a stable configuration, which
corresponds to the region where $Re_p<O(10^{-3})$, the difference in the droplet
dynamics has to be accounted for. This is what the offsets along $Oh_p$ and
$Re_p$ in the exponential scaling enable to do.

\begin{figure}
  \centering
  \begin{subfigure}[c]{0.49\textwidth}
            \centering
            \includegraphics[width=\textwidth]{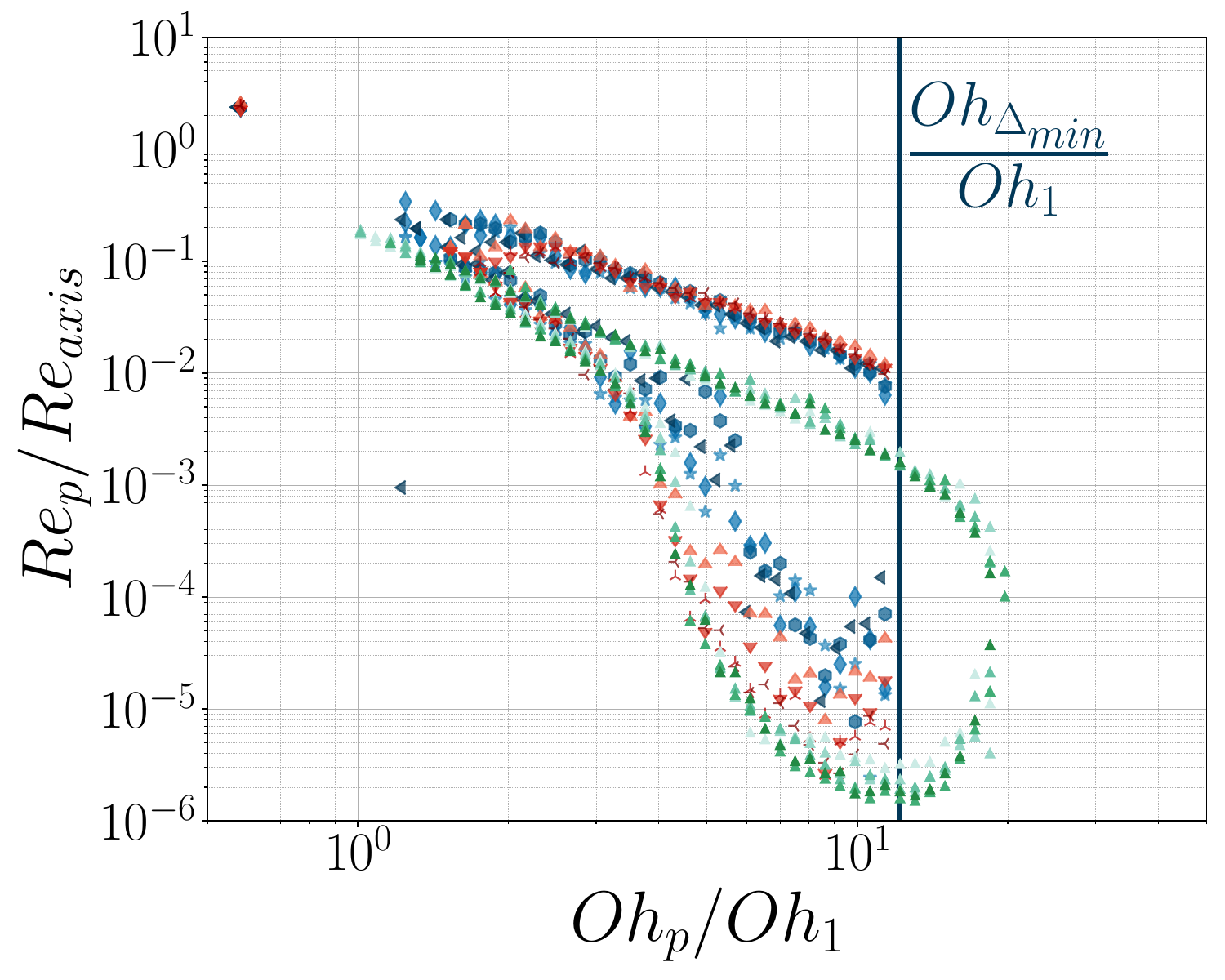}
                \caption{}
                \label{fig:dns.mapRO_contour_comparison_exp_num_tm15}
  \end{subfigure}%
  \begin{subfigure}[c]{0.49\textwidth}
            \centering
            \includegraphics[width=\textwidth]{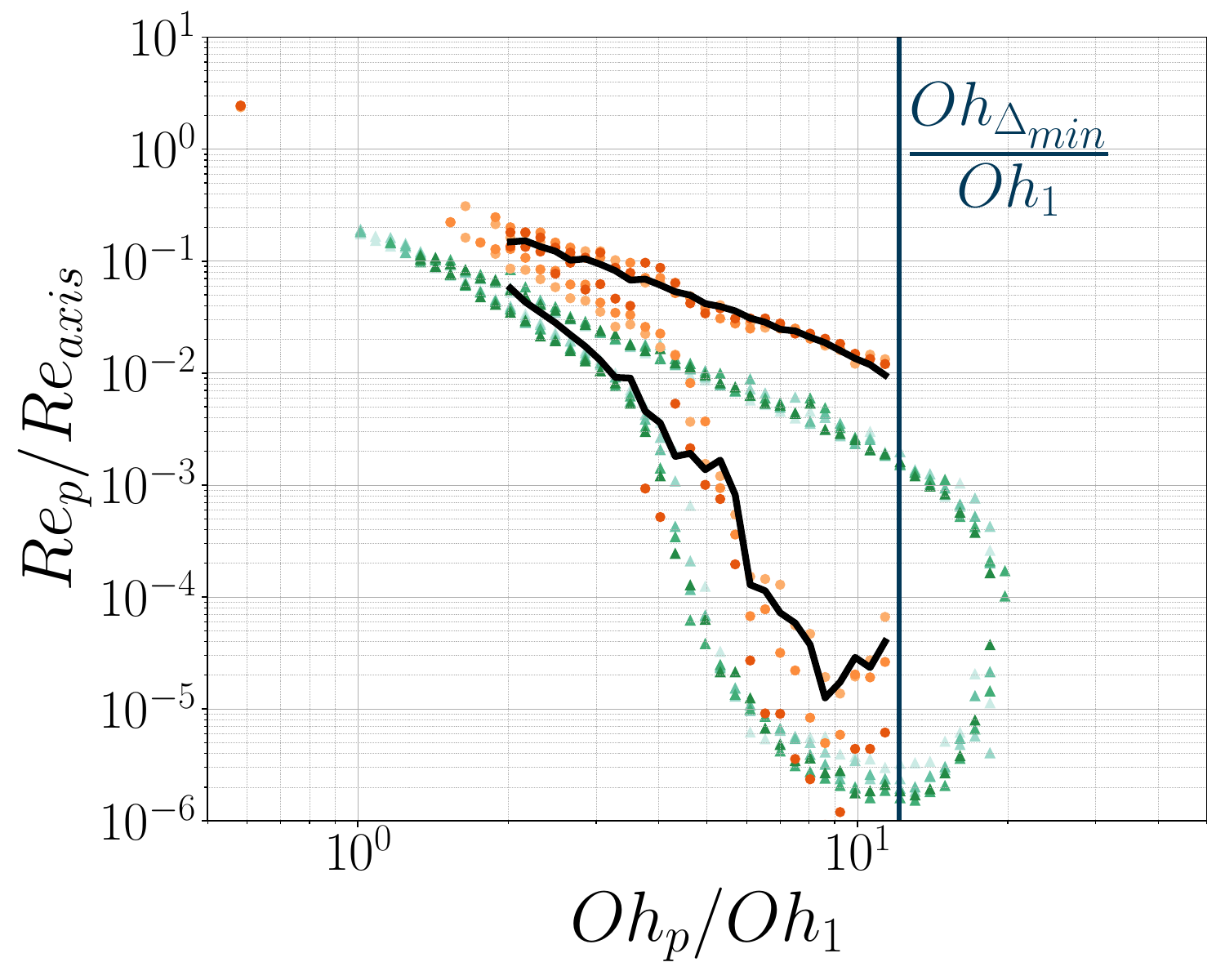}
                \caption{}
                \label{fig:dns.mapRO_contour_comparison_exp_num_visc_tm15}
  \end{subfigure}%
  \caption{
    Color on-line. Comparison of the borders of the joint volume histograms
    obtained from the DNS campaign and from the experimental data of
    \citet{felis_experimental_2020}.  The blue and red colours denote the \swi{}
    and atomisation regimes. The color code denoting the DNS is the same as in
    Figure \ref{fig:dns.front_evolution}. The green triangles represent the
    experimental data and the orange bullets represent DNS with $U_{inj}=2.216
    \textnormal{m/s}$ and different density ratios,
    $\rho_1/\rho_2=82.5,110,165$. In (b), the black solid line indicates the
    mean value of the numerical joint histogram edges for $We_2\in[26,165]$
    (blue and red).
  }
  \label{fig:dns.map_Rep_Ohp_edges_exp_num}
\end{figure}

Figure \ref{fig:dns.mapRO_contour_comparison_exp_num_tm15} compares the edge
contour of the experimental and numerical joint volume histograms. When
comparing the original experimental histograms and the numerical ones, it
appears that the phase spaces in which the droplets evolve show ranges of
existence being very similar between the experiments and the simulations. Even
if an offset along the $Re_p / Re_{axis}$-axis exists, they lie in the same
$Oh_p / Oh_1$ range.  By multiplying the edge contour of the
experimental joint volume histogram by 3, the edges of the numerical and the
experimental data collapse. Thus, we have $(Re_p/Re_{axis})_{num} = C\times
(Re_p/Re_{axis})_{exp}$, where $C\approx 3$, which leads to:

\begin{equation}
  \frac{|u_{p,x}-U_{g,x}|_{num}~d_{num}}{|u_{p,x}-U_{g,x}|_{exp}~d_{exp}} =
  C~
  \frac{
    d_{n,num}\times u_{x,axis,num}
  }{
    d_{n,exp}\times u_{x,axis,exp}
  }
\end{equation}

\noindent and results to
\begin{equation}
  |u_{p,x}-U_{g,x}|_{num}~d_{num}\approx1.8~|u_{p,x}-U_{g,x}|_{exp}~d_{exp}
  \label{eq:ratio_dns-exp}
\end{equation}

\noindent with $u_{x,axis,exp}=0.8\times U_{inj,exp}=28~\textnormal{m/s}$ and
$u_{x,axis,num}=U_{inj,num}=4.5~\textnormal{m/s}$. The numerical and
experimental contours lie in the same range of $Oh_p/Oh_1$. Thus, it can be assumed
that $d_{exp}/d_{n,exp} \approx d_{num}/d_{n,num}$ which implies:

\begin{equation}
  \frac{d_{num}}{d_{exp}} \approx 3.73,~ ~ ~
  ~ \frac{|u_{p,x} - U_{g,x}|_{num}}{|u_{p,x} - U_{g,x}|_{exp}} \approx 0.48.
  \label{eq:ratio_dns-exp_size-velocity}
\end{equation}

\noindent The experimental and numerical mean sizes are respectively $\mean{d}_{exp}=95
~\mu\textnormal{m}$, averaged over the 5 $x/d_n$ positions, and
$\mean{d}_{num}\approx300 ~\mu\textnormal{m}$, at $t/T_a=34$ in the \swi{} regime.
The ratio of the means equals 3.16, thus $\mean{d}_{num}/\mean{d}_{exp}\approx
d_{num}/d_{exp}$ and verifies the previous result. 

Explaining why the last three ratios appear, Eqs. \ref{eq:ratio_dns-exp} and
\ref{eq:ratio_dns-exp_size-velocity}, must be made carefully. On the one
hand, the way the measurements of the size and the velocity of the droplets is
carried out greatly differs between the experiments and the simulations.  In the
simulations, once a droplet is detected thanks to the tag function of Basilisk,
see section \ref{sec:dns.num_methods}, its volume and velocity are computed as
the volume average in 3D of the cell values contained in the droplet. In the
experiments, the measurements of the droplet size and velocity are carried out
with a 2D laser sheet thanks to DTV. In addition, the measurement of the mean
gas phase velocity also differs. While numerically it results from the velocity
average over all the cells in the gas phase, the experimental mean gas phase
velocity is estimated from LDV measurements at different radial positions and
then averaged along those positions. On the other hand, it is important to keep
in mind that the experimental and numerical data correspond to two drastically
different physical spaces. The former were measured for $x/d_n=800$ and the
latter for $x/d_n\approx20$. 

With these limits in mind, different hypotheses can be sketched. As the
measurements are carried in two drastically different regions of the fragmenting
jet, the three ratios could reveal some dynamics occurring at the overall jet
scale, for instance the overall slowdown of the droplet population when the
droplet spray moves towards larger $x/d_n$. If it was the case, it could be
expected that the edges of the experimental joint volume histograms would be
translated towards smaller $Re_p/Re_{axis}$, see Figure
\ref{fig:dns.mapRO_contour_comparison_exp_num_tm15}, as they span over a
distance of 400 $d_n$. But, no such translation is noticeable for the
experimental edges. 

The difference along the $Re_p$-axis between the experimental and numerical
joint histograms could also raise from the difference of the density ratios
considered in the simulations and experiments.  Indeed, since gravity is not
considered, the effects of density are accounted by the Reynolds and
Ohnesorge numbers.  Figure
\ref{fig:dns.mapRO_contour_comparison_exp_num_visc_tm15} gives the evolution of
the joint histogram edges obtained for 3 density ratios,
$\rho_l/\rho_g\in\{82.5,110,165\}$, such that $We_2=40$, see appendix
\ref{app.compl_dns}. These density ratios remain small compared to
experimental values, $O(1000)$ for water injected in air, but are already quite
computationally expensive. No translation of the joint histogram edges towards
smaller $Re_p/Re_{axis}$ values is noticeable, thus discarding this hypothesis
too.  The choice of the normalisation for the Reynolds number could also be an
explanation. Using the Reynolds number computed over $d_n$ and the averaged
velocity of the dispersed liquid phase, instead of $u_{axis}$, could help to
make the edges of the joint volume histograms collapse, both experimentally and
numerically. 

Notwithstanding those limitations and differences, it still seems legitimate to
conclude that the joint histogram edges are self similar. This conclusion only
holds for the edges and not for the joint histogram values, which evolve very
differently between the experiments and the simulations. 

Yet, the two jet flows differ in terms of fragmentation mechanisms. In the
experimental flow, the bag breakup fragmentation plays an important role while
it is totally absent in the numerical flows. In the experiments, the droplets
undergoing bag breakup originate from the liquid core pinch and are
characterised by a large size and axial velocity. As the liquid core is still
developing in the simulations, the absence of such droplets is expected. Even if
the joint histogram edges are self similar, they slightly differ for large
values of particulate Reynolds and Ohnesorge numbers where the experimental
joint histograms exhibit a well-defined tail. Besides, section
\ref{sec:dns.sizepdf_model} shows that the ligament-mediated fragmentation
describes well the droplet fragmentation in the numerical flows.  Thus, it is
tempting to conclude that the droplets are likely to undergo a bag breakup
fragmentation when $Oh_p/Oh_1<2$ and a ligament mediated fragmentation when
$Oh_p/Oh_1\geqslant2$.

\section{Conclusion \label{sec:ccl}}

In this work, the droplet population generated by the fragmentation of a round
jet in a quiescent gas medium was studied numerically for different gaseous
Weber numbers $We_2$ spanning the \swi{} regime and part of the atomisation
regime. At first, the statistical moments of the size, the axial velocity and
the radial velocity were depicted and their evolution with $We_2$ was detailed.
The study of the distribution of the droplet size shows the existence of three
modes in the \swi{} regime, while only one mode exists in the atomisation
regime. Complementary, the size distribution shows two exponential decays
connected by a transition region scaling as a power law.  In this near-field
(close to the nozzle) fragmentation, the size distribution is better modelled by
the law derived by \citet{kooij_what_2018} in the context of ligament-mediated
fragmentation than by the law derived by \citet{novikov_distribution_1997} in
the framework of turbulence intermittency.  This could, at first sight, raise
from the difference in the fragmentation mechanisms occurring in the region
close to the nozzle, studied here, and the region far away from the nozzle
studied in \citet{vallon_liquid_2021}.  On the side of the axial velocity
distribution, additionally to elucidating the scaling of the axial velocity
distribution tails, the origin of the droplet velocities being negative and
larger than the injection velocity $U_{inj}$ is explained thanks to the vortex
ring theory of \citet{saffman_vortex_1992}, vortex ring which sustains the
recirculation region on the downstream side of the jet head. The existence of a
double tail along the size direction for the size-velocity joint distribution is
also explained by spatially separating the droplets evolving in the boundary
layer and those ejected from the jet head. 

The analysis also scaled down to the flow perceived by the droplets with the
study of the droplet volume histogram over the phase space of the particulate
Reynolds and Ohnesorge numbers. Properly scaled by the injection Ohnesorge
number $Oh_1$ and the Reynolds number computed on the jet  axis $Re_{axis}$, the
boundaries of the joint volume histograms from the DNS collapse, thus indicating
the weak dependence of the joint histogram boundaries on the gaseous Weber
number. The collapse is also obtained between the numerical and the experimental
joint volume histograms, with a slight correction along the Reynolds axis for
the far-field experimental one. This highlights the existence of a phase space properly
bounded which contains the whole droplet population as well as the jet liquid
core. Advantages could be taken from this result for modeling the turbulent jet
fragmentation in terms of particulate dimensionless numbers or for improving the
model of mass transfer proposed by \citet{vallon_liquid_2021}.  Overall, the
good accuracy of the statistical properties of the droplet population with the
theoretical models, as well as with the experimental data, validate the accuracy
of the simulations within the numerical limitations.

Further work could be done regarding the droplet dynamics and geometry. Now that
the interface of the jet and the droplet population are described, it could be
possible to focus on the size distribution resulting from specific fragmentation
mechanisms, like the fragmentation of the rims in the \swi{}
regime. This could help to understand the origin of the 3 modes observed for the
size distribution in this regime. Also, performing a Lagrangian tracking of the
rims in the DNS lying in the \swi{} regime would enable to verify the
break-up of such toroidal ligaments and compare the resulting size distribution
with the $\Gamma$ distribution from the ligament-mediated fragmentation theory.
Finally, a statistical analysis of the ligament geometry in the atomisation
regime, specifically in the DNS with the highest $We_2$, could help to better
describe the ligament size and corrugation distributions over the jet
fragmentation.


\backsection[Acknowledments]{
  This work was granted access to the HPC resources of CINES under the
  allocation 2019-A0072B11103  made by GENCI.
}

\backsection[Funding]{
  This research received no specific grant from any funding agency, commercial
  or not-for-profit sectors.
}

\backsection[Declaration of interests]{
  The authors report no conflict of interest.
}

\backsection[Author ORCIDs]{
  \\ R. Vallon, https://orcid.org/0000-0003-0770-787X; 
  \\ M. Abid, https://orcid.org/0000-0002-0438-4182;
  \\ F. Anselmet, https://orcid.org/0000-0001-6443-7437
}

\backsection[Author contributions]{
  R.V. performed the direct numerical simulations, developed the data analysis
  scripts, carried most of the data analysis and wrote the manuscript. M.A.
  provided scientific and technical supervisions, including key insights
  regarding the fragmentation and vortex theories. F.A. provided scientific
  supervision. F.A. and M.A.  contributed equally to designing the work and
  proofreading the manuscript. All authors contributed equally to reaching
  conclusions.
}

\newpage
\appendix 

\section{Computation of the most unstable mode \label{app.unstable_mode}}

Following the work of \citet{yang_asymmetric_1992} on the growth of waves in
round jets, it is possible to characterise the most unstable axisymmetric mode.
The author studied the stability of an infinitesimal perturbation on the surface
of a round jet of radius $a$. The configuration is the same as described in
\ref{sec:dns.phys_config}. Additionally, the gas phase can be injected at a
velocity $U_2$ and the fluids are incompressible and inviscid. In this
section, the injection velocity previously denoted $U_{inj}$ is denoted $U_1$.

The velocity and pressure fields can be split into an averaged part and a
fluctuation part: $\textbf{u}_i = \textbf{U}_i + \textbf{u}_i'$ and $p_i = P_i +
p_i'$, where $i\in\{1,2\}$ respectively denotes the liquid and gaseous phases.
Injecting this decomposition into the governing equations, expressed in
cylindrical coordinates $(r,\theta,z)$, and applying the divergence operator
gives the pressure disturbance equation :

\begin{equation}   \nabla^2p_i' = 0,~~~~\nabla^2=\frac{1}{r}
\frac{\partial}{\partial r}r\frac{\partial}{\partial r} +
\frac{1}{r^2}\frac{\partial^2}{\partial \theta^2} + \frac{\partial^2}{\partial
z^2} 
\label{eq:dns.unstabmode.perturb_pressure}
\end{equation}

\noindent Assuming a 3D disturbance with a normalised wavelength number $ka$ and
$m$ in the streamwise and azimuthal directions, the perturbed quantities are
$p_i'= p_i'(r)e^{i(kz+m\theta)+\alpha_{tg} t}$ and $\textbf{u}_i'=
\textbf{u}_i'(r)e^{i(kz+m\theta)+\alpha_{tg} t}$, where $\alpha_{tg}$ is the temporal
growth rate and $m$ introduces the non axisymmetric variations of the
disturbance. Eq. \ref{eq:dns.unstabmode.perturb_pressure} then becomes:
\begin{equation}
  \bigg( \frac{1}{r}\frac{\partial}{\partial r} r \frac{\partial}{\partial r}
  -\frac{m^2}{r^2} - k^2\bigg) p_i(r) = 0
\end{equation}

\noindent Resolving this equation gives a solution for $p_i(r)$, Eq.
\ref{eq:dns.unstabmode.pressure_solution_pi}, depending on the first and
second type modified Bessel functions of order m, respectively denoted $I_m$ and
$K_m$.  This solution can be used with the mass conservation equation for the
linearised perturbation to derive a solution for $\textbf{u}_i'$, Eq.
\ref{eq:dns.unstabmode.velocity_solution_ui}.

\begin{equation}
  p_i(r) = C_{i,1}I_m(kr) + C_{i,2} K_m(kr)
  \label{eq:dns.unstabmode.pressure_solution_pi}
\end{equation}

\begin{equation}
  \textbf{u}_i' = - \frac{\nabla\left( p_i(r)e^{i(kz+m\theta)+\alpha_{tg} t}
  \right)}{\rho_i(\alpha_{tg}+ikU_i)}
  \label{eq:dns.unstabmode.velocity_solution_ui}
\end{equation}

\noindent where the four constants $C_{i,1}$ and $C_{i,2}$ have to be derived
regarding the boundary conditions. The pressure is finite in the liquid at $r=0$
and in the gas when $r\rightarrow +\infty$, thus $C_{1,2}=C_{2,1}=0$. Let 
$\eta_1$ and $\eta_2$ denote the perturbed displacements of the interface and $\Delta
p_\sigma$ the pressure jump due to the surface tension $\sigma$. The pressure
follows $\Delta p_\sigma=\sigma(1/R_1+1/R_2)$ with $R_1$ and $R_2$ the principal
radii of curvature.  The remaining two constants can be derived from the
pressure jump, $p_1-p_2=\Delta p_\sigma$, and the interface displacement,
$\eta_1=\eta_2$. The perturbed displacements satisfy:

\begin{equation}
  v_i = \frac{\partial \eta_i}{\partial t} + U_i \frac{\partial \eta_i}{\partial
  x}
\end{equation}

\noindent with $v_i$ the velocity component in the radial direction. By letting
$\eta=\eta_1=\eta_2=\eta_0e^{i(kz+m\theta)+\alpha_{tg} t}$,
\citet{yang_asymmetric_1992} showed that to the first order of $\eta$,
$1/R_1+1/R_2 = 1/d_n-1/d_n^2\left[ 1-m^2-(ka)^2 \right]\eta_0$. The continuity
equations then become:

\begin{equation}
    C_{11} \bigg(I_m(ka) - \frac{\sigma [1-m^2-(ka)^2]	}{a^2}
    \frac{I'_m(ka)}{\rho_1(\alpha_{tg}+i k U_1)^2} \bigg) - C_{22} K_m(ka) = 0
\end{equation}

\begin{equation}
  C_{11} \frac{I'_m(ka)}{\rho_1(\alpha_{tg}+ i k U_1)^2} - C_{22} \frac{K'_m(ka)}{\rho_2(\alpha_{tg}+ikU_2)^2} = 0
\end{equation}

\noindent The latter equation system admits a non-trivial solution when its
determinant is zero. This condition gives the following dispersion
equation:

\begin{equation}
(\rho_{1m} + \rho_{2m}) \alpha_{tg}^2 + 2 i k \alpha_{tg} ( \rho_{2m} U_2 + \rho_{1m} U_1)
- k^2 (\rho_{2m} U_2^2 + \rho_{1m} U_1^2) - \frac{k \sigma}{a^2} [1 - m^2 -
(ka)^2 ] = 0 
  \label{eq:dns.unstabmode.dispersion}
\end{equation}

\noindent with: 

\begin{equation} 
  \left\{ 
    \begin{array}{ll} 
      \rho_{1m}  &= \gamma_m\rho_1 \\[2pt]
      \rho_{2m}  &= \beta_m\rho_2 \\[2pt] 
      \gamma_{m} &= kI_m(ka)/I_m'(ka) \\[2pt] 
      \beta_{m}  &= -kK_m(ka)/K_m'(ka) \\[2pt] 
      I_m'(ka) &= \frac{d I_m(kr)}{dr}_{r=a} \\[2pt] 
      K_m'(ka) &= \frac{dK_m(kr)}{dr}|_{r=a}
    \end{array} 
  \right.
\end{equation}

\noindent The dispersion equation, Eq. \ref{eq:dns.unstabmode.dispersion}, is a quadratic
equation in $\alpha_{tg}$ and the expression of the nondimensional temporal growth rate
for the $m$-th transversal mode can be derived from it:

\begin{equation}
    (\alpha_r^*)_m^2 = \frac{\gamma_m \beta_m Q \cdot (ka)^2}{(\gamma_m+\beta_m
    Q)^2} + \frac{ka}{We} \frac{1-m^2-(ka)^2}{\gamma_m + \beta_m Q}
  \label{eq:dns.unstabmode.dimensionless_growth_rate}
\end{equation}

\noindent where $(\alpha_r^*)_m^2=(\alpha_r)^2_m/[(U_1-U_2)^2/d_n^2]$,
$We=[d_n(U_1-U_2)\rho_1]/\sigma$ and $Q=\rho_2/\rho_1$. 

\section{Numerical performances of the DNS\label{app.num_perf}}

The computational performances can be tracked by checking the total number of
cells used for each DNS, $\mathcal{C}_{tot}$, the mean numerical velocity,
$\overline{\mathcal{V}_{num}}$, the maximal physical time, $t_{max}/T_a$, the
maximum jet elongation, $L_{j,max}/d_n$ and the total number of detected
droplets, $N_{tot}$. Table \ref{tab:dns.num_performances_full} summarizes the related
numerical performances. All the DNS are split into 3 runs and were computed on
the Occigen HPC (CINES, France).

\begin{table} 
  \begin{center} 
    \def~{\hphantom{0}} 
    \begin{tabular}{cccccc}
      DNS & $\mathcal{C}_{tot}~(10^6)$ & $\overline{\mathcal{V}_{num}}~\left(
      10^6 \textnormal{cells}/\textnormal{s} \right)$ & $t_{max}/T_a$
      & $L_{j,max}/d_n$ & $N_{tot}$ \\[3pt]
      1 & ~~4.43 & 0.59 & 34~~ & 28~~ & ~~~70 \\
      2 & ~29.29 & 0.93 & 34~~ & 28~~ & ~~459 \\
      3 & ~53.83 & 2.23 & 34~~ & 28~~ & ~1949 \\
      4 & ~42.15 & 1.96 & 34~~ & 28~~ & ~2182 \\
      5 & ~51.41 & 1.73 & 34~~ & 28~~ & ~2448 \\
      6 & ~34.13 & 1.92 & 34~~ & 28~~ & ~3545 \\
      7 & ~77.45 & 2.46 & 24.2 & 21.5 & ~9725 \\
      8 & 105.6~ & 2.59 & 20.0 & 17.0 & 18,478 \\
      9 & 141.7~ & 2.59 & 17.5 & 14.4 & 32922 \\
      10& 154.4~ & 2.56 & 16.5 & 14.2 & 45046 \\
    \end{tabular} 
    \caption{
      Numerical performances.
    }
  \label{tab:dns.num_performances_full} 
  \end{center} 
\end{table}

\section{Computation of the mean interface of the jet \label{app.mean_interface}}

Computing the mean interface of the jet is not straightforward and requires some
processing. To extract it, it is first necessary to compute the joint
distribution of the interface points in the physical space $(x/d_n,r/d_n)$. Once
computed, the mean interface can be extracted from the distribution by filtering
out the most probable interface points. This extraction step however relies on
the choice of a threshold. This threshold can be empirically chosen such that it
enables to depict the mean interface while discarding most of the interface
related to the droplets. 

\begin{figure}
  \centering
  \begin{subfigure}[c]{0.49\textwidth}
            \centering
            \includegraphics[width=\textwidth]{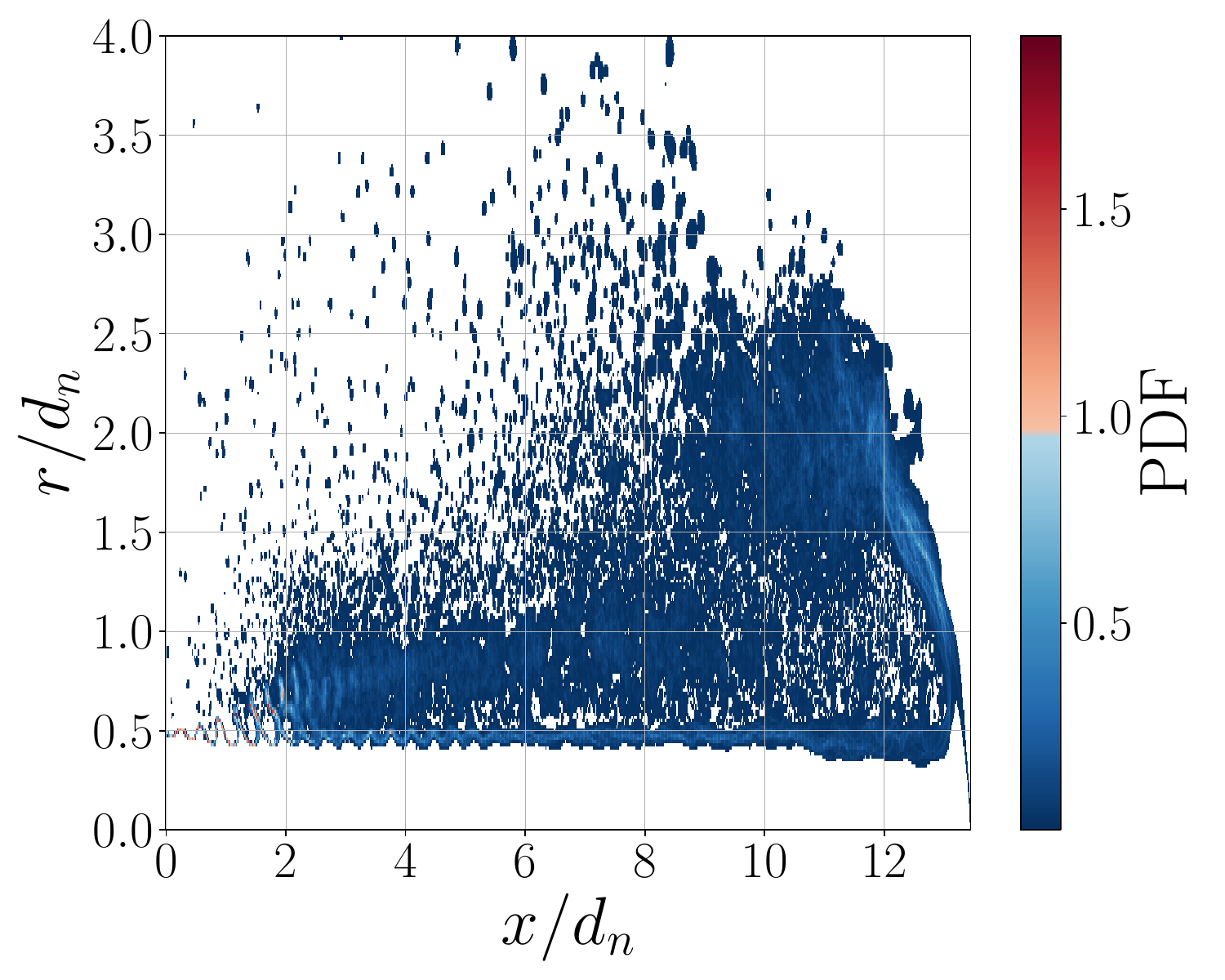}
                \caption{}
                \label{fig:dns.jpdf_interface}
  \end{subfigure}%
  \begin{subfigure}[c]{0.49\textwidth}
            \centering
            \includegraphics[width=\textwidth]{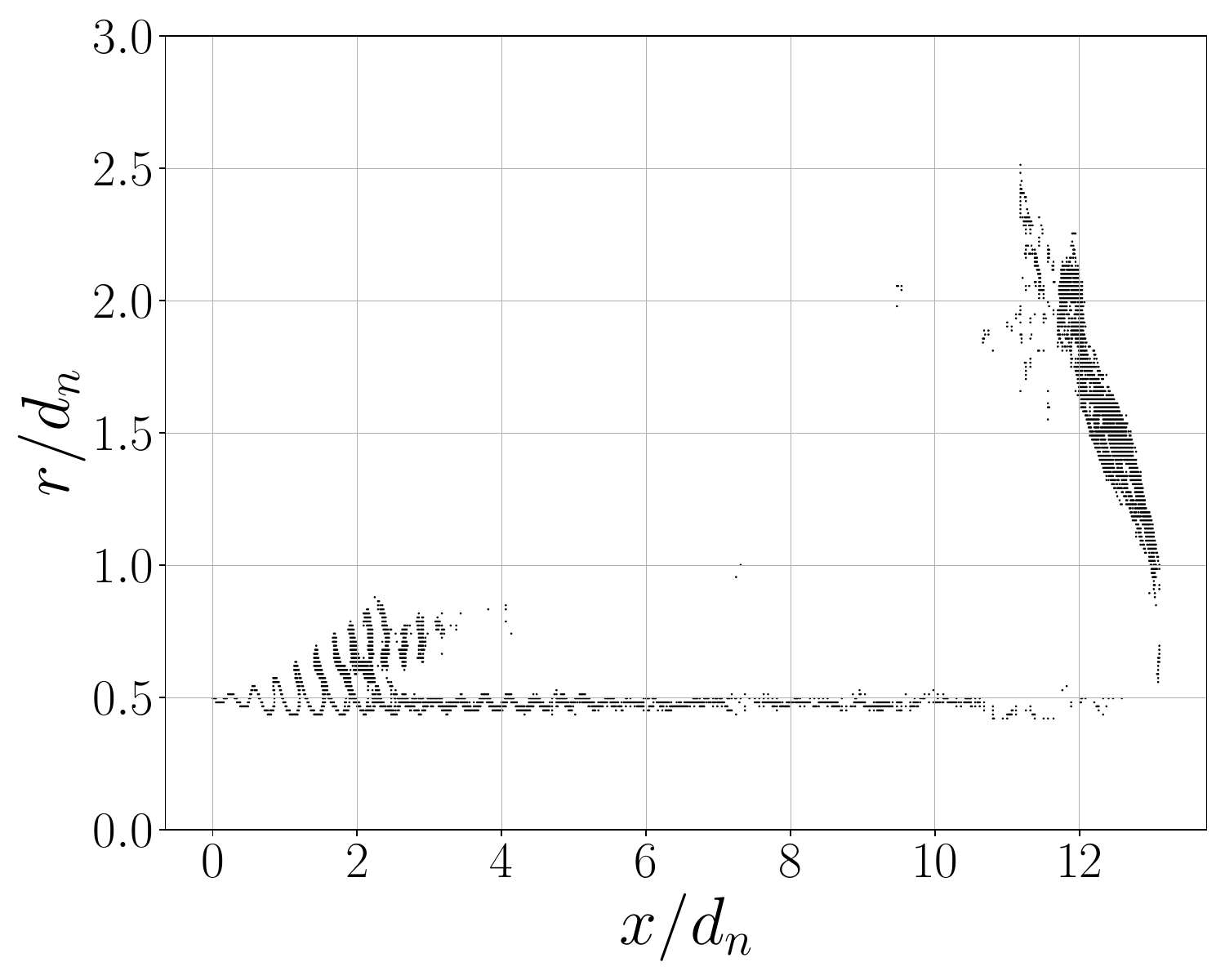}
                \caption{}
                \label{fig:dns.mean_interface}
  \end{subfigure}%
  \caption{
    Joint distribution of the interface points in the $(x/d_n,r/d_n)$ space for
    $We_2=99.8$ (DNS 8) at $t/T_a=15$ (a) and the interface filtered with a
    threshold of 0.2 (b).
  }
  \label{fig:jpdf_mean_interface}
\end{figure}

Figures  \ref{fig:dns.jpdf_interface} and \ref{fig:dns.mean_interface} respectively
give the joint distribution of the interface points across the $(x/d_n,r/d_n)$
space and the interface filtered from the joint distribution with a threshold of
0.2, i.e. the interface points with a probability larger than 0.2.  A way to
refine the mean interface of the jet would be to consider the interface of the
liquid core only, instead of considering all the interface points in the jet,
i.e. the liquid core and all the droplets. The droplets would be then naturally
discarded and the resulting interface would depict more precisely the mean
interface around the jet head. Even so, the method used here is satisfactory for
the following analysis.

\section{Temporal evolution of the mean values of the size, axial velocity and
radial velocity \label{app.temp_evol_mean}}

\begin{figure}
  \centering
  \begin{subfigure}[c]{0.48\textwidth}
            \centering
            \includegraphics[width=\textwidth]{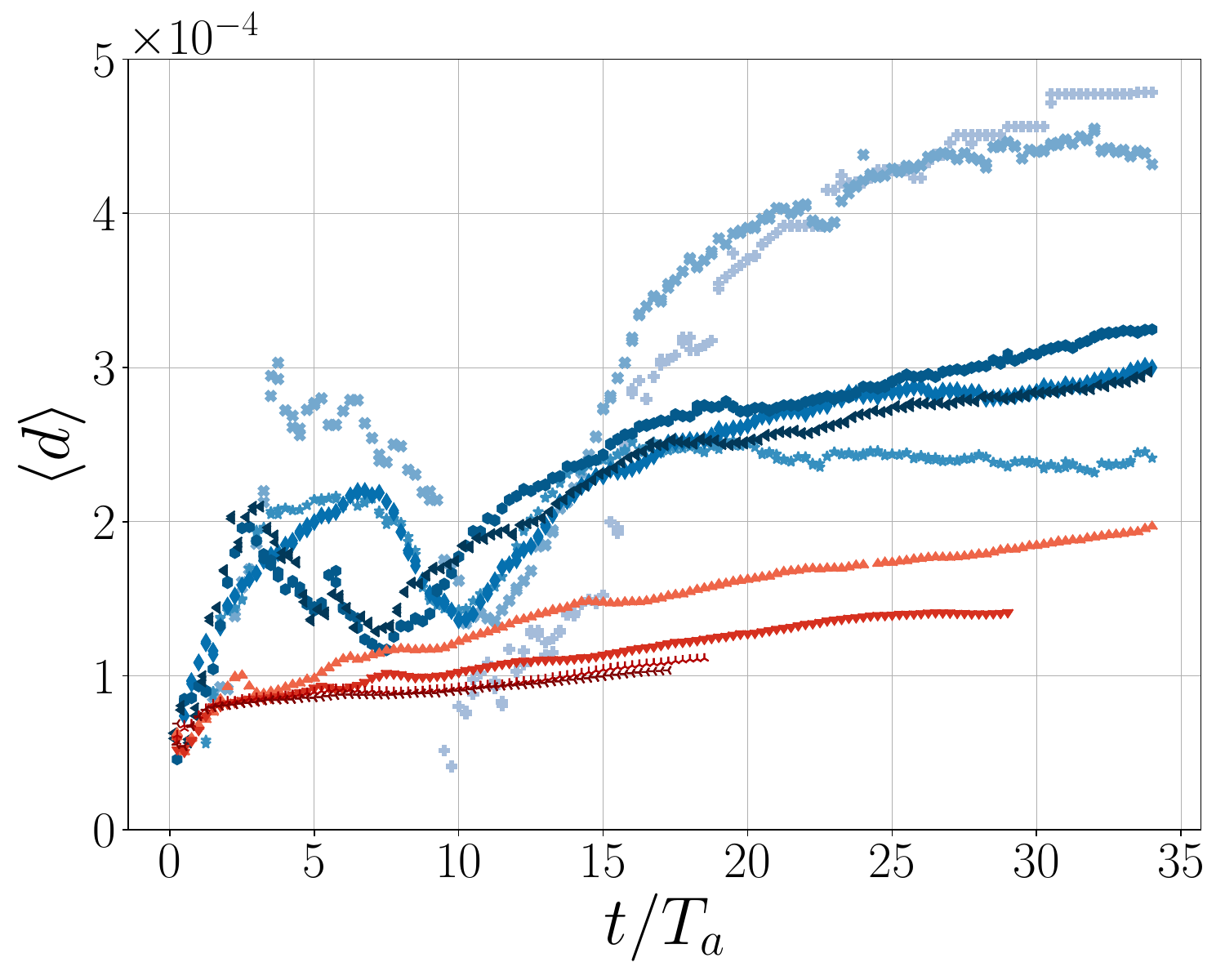}
                \caption{}
                \label{fig:dns.sizemean}
  \end{subfigure}%
  \begin{subfigure}[c]{0.48\textwidth}
            \centering
            \includegraphics[width=\textwidth]{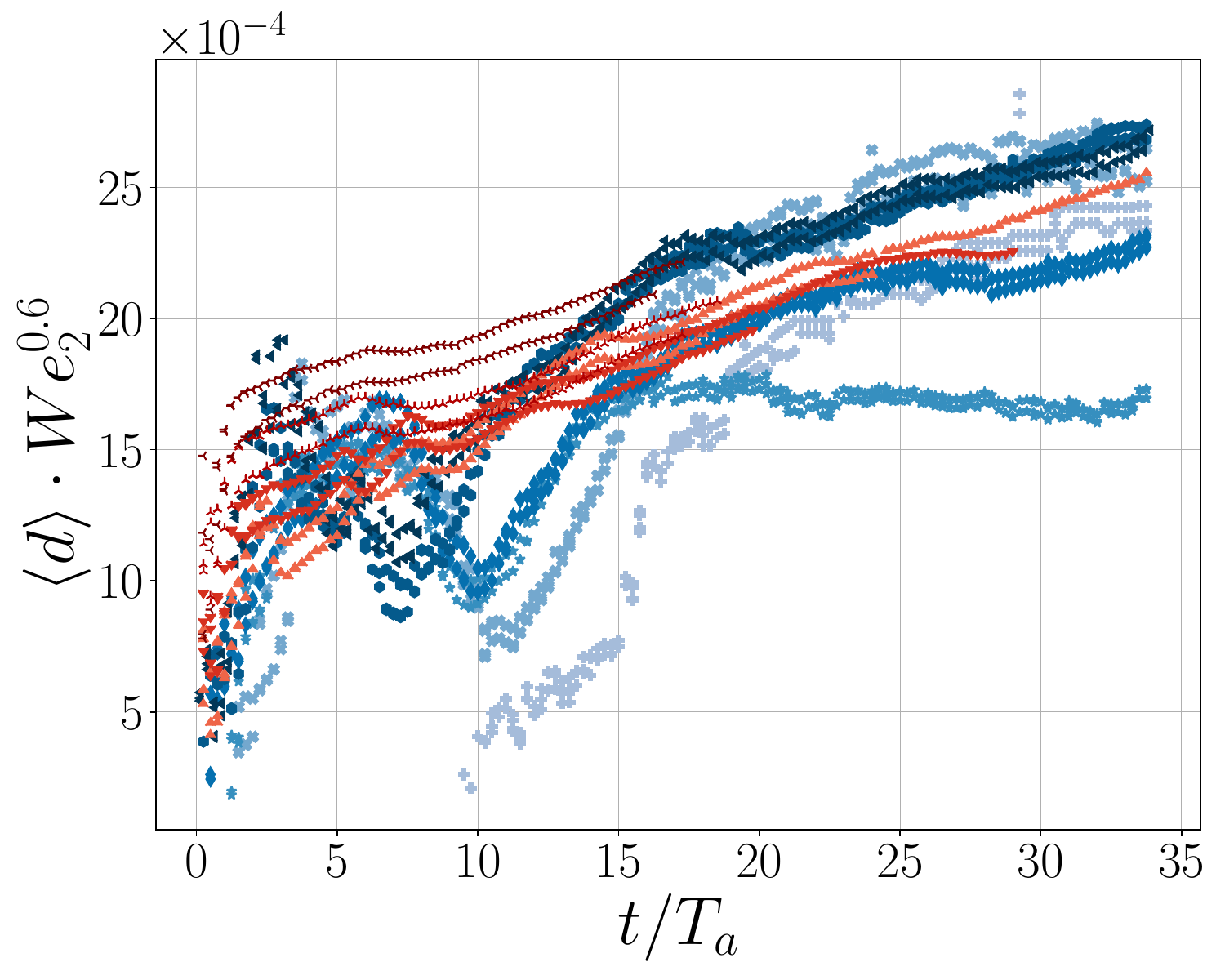}
                \caption{}
                \label{fig:dns.sizemean_scaled}
  \end{subfigure}%

  \begin{subfigure}[c]{0.48\textwidth}
            \centering
            \includegraphics[width=\textwidth]{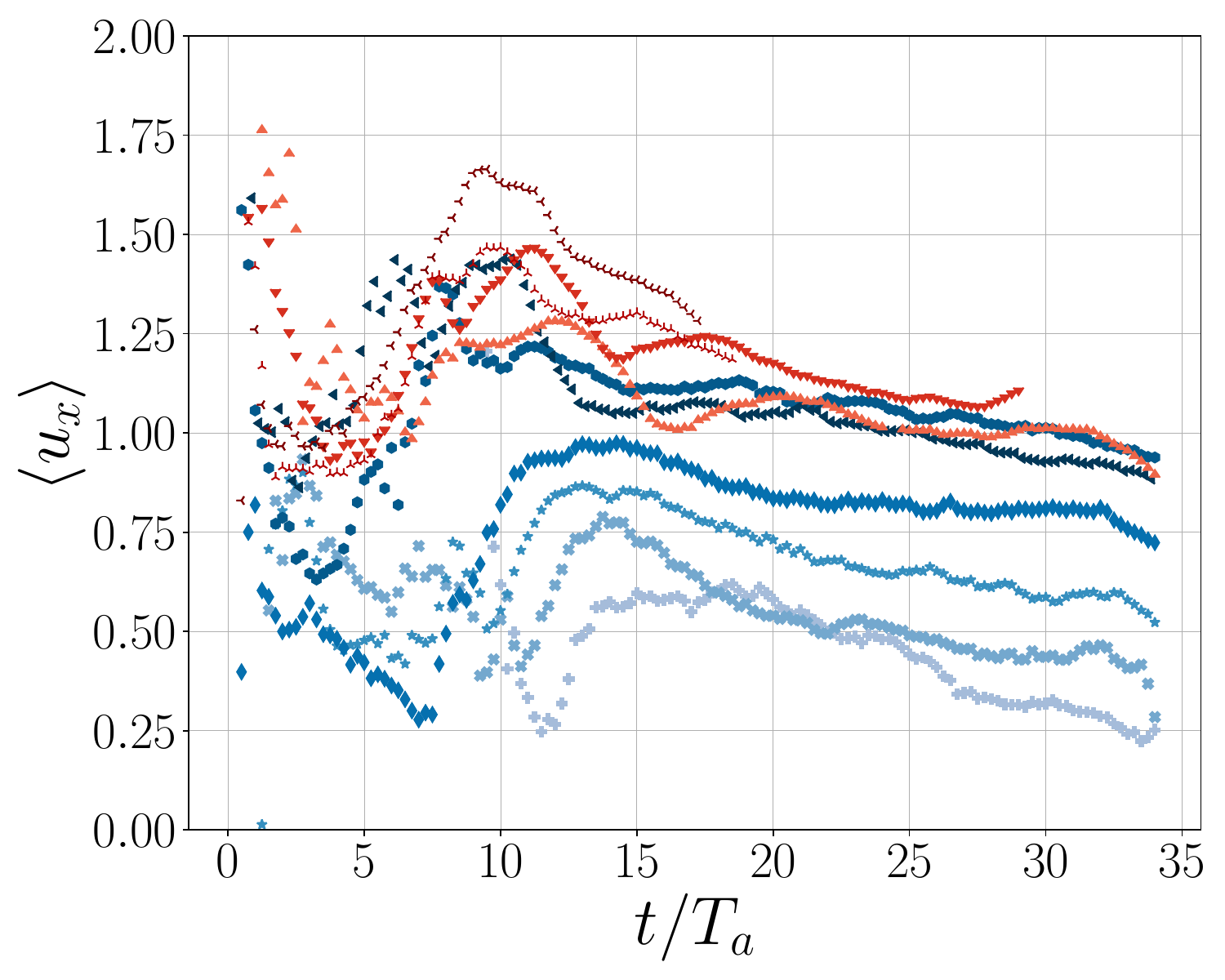}
                \caption{}
                \label{fig:dns.uxmean}
  \end{subfigure}%
  \begin{subfigure}[c]{0.48\textwidth}
            \centering
            \includegraphics[width=\textwidth]{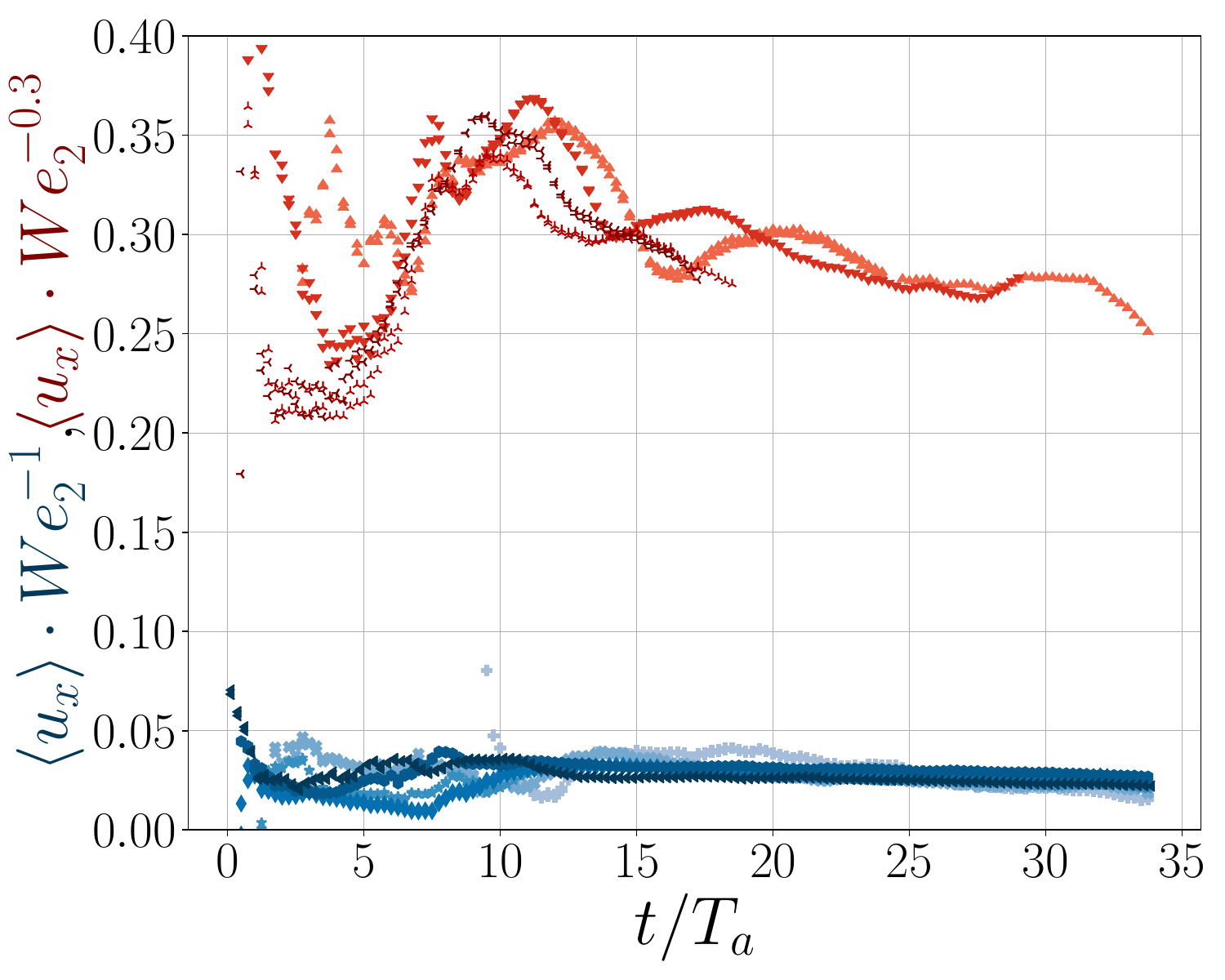}
                \caption{}
                \label{fig:dns.uxmean_scaled}
  \end{subfigure}%

  \begin{subfigure}[c]{0.48\textwidth}
            \centering
            \includegraphics[width=\textwidth]{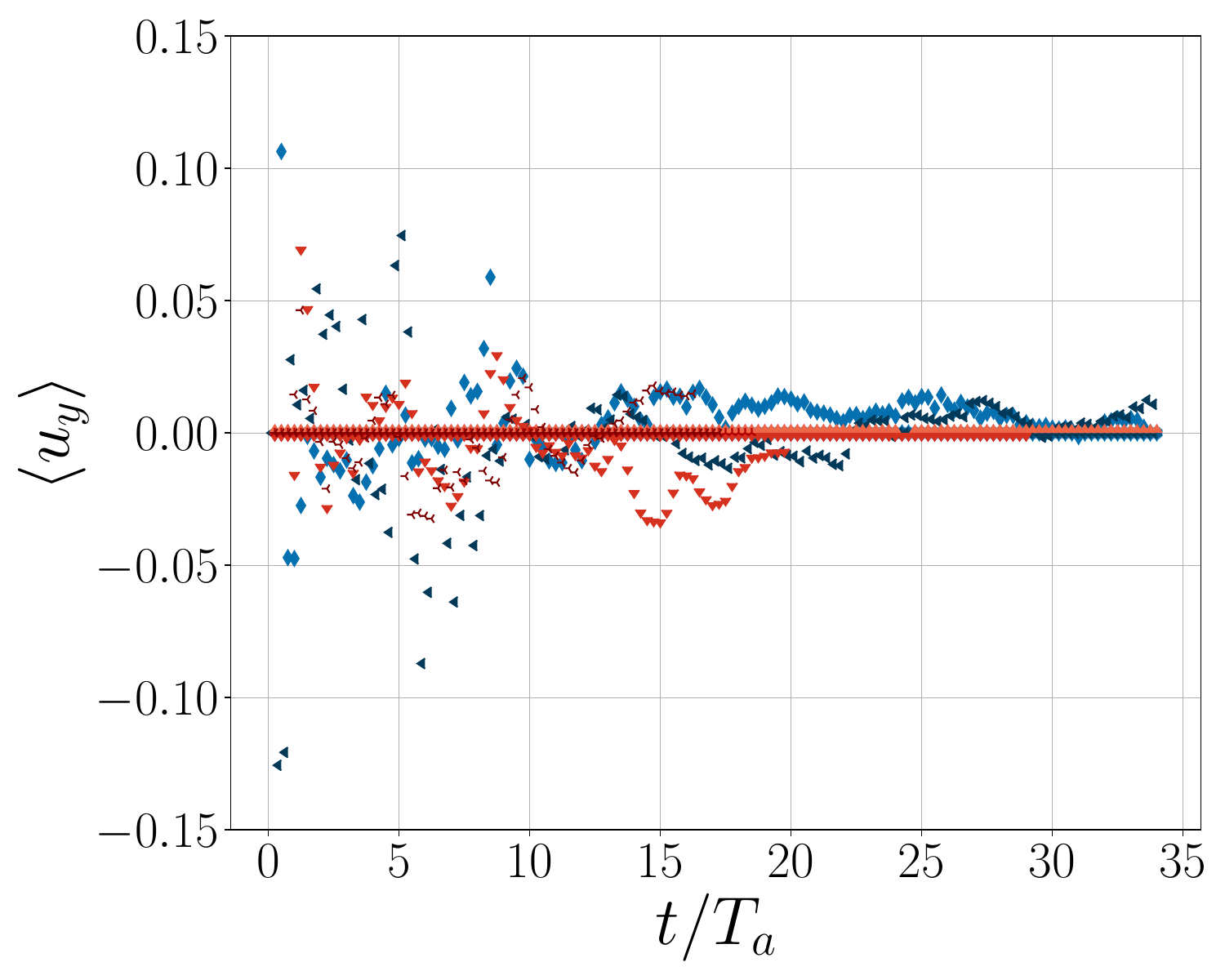}
                \caption{}
                \label{fig:dns.uymean}
  \end{subfigure}%

  \caption[
    Temporal evolution of the mean $\langle \cdot \rangle$, unscaled and scaled
    by $We_2$, of the droplet size $d$, the axial velocity $u_x$ and the
    transverse velocity $u_y$.
  ]{
    Temporal evolution of the mean $\langle \cdot \rangle$, unscaled (left) and
    scaled by $We_2$ (right) of the droplet size $d$ (a,b), the axial
    velocity $u_x$ (c,d) and the transverse velocity $u_y$ (e). The units of the
    variables are the SI base units.
  }
\label{fig:dns.drop_mean}
\end{figure}

Figure \ref{fig:dns.drop_mean} gives the temporal evolution of the mean values
of the size, the axial velocity and the radial velocity. Regarding the size and
the axial velocity, after reaching a peak value for $t/T_a\in[5,10]$, the mean
values increase relatively steadily within the time scope under consideration.
The time evolution of the mean of each DNS can be rescaled with $We_2$. On one
side, the mean size scaled by $We_2^{0.6}$ seems to evolve linearly with
$t/T_a$. On the other side, it is possible to collapse the time evolution of the
mean axial velocity for each regime by considering $\langle u_x
\rangle~We_2^{-1}$ for the \swi{} regime and $\langle u_x \rangle~
We_2^{-0.3}$ for the atomisation regime.  The evolution of $u_y$ is specific in
the sense that the flow is statistically axisymmetric and $\langle u_y \rangle$
should naturally be set to zero, which is verified here asymptotically.  Due to
the flow symmetry, the mean of $u_z$ behaves the same as the one of $u_y$. The
standard deviations, not shown here, reach a steady state faster than the mean
values for the size and the velocities.  

\section{Evolution of the velocity statistical moments with $We_2$ \label{app.stat_mom_vel}}

Regarding the distribution of $u_x$, all the four statistical moments increase
with $We_2$. The increase in the mean and standard deviation indicates that the
droplets are accelerated with $We_2$, which is obvious as $U_{inj}$ increases
meanwhile, and that the dispersion in terms of velocity is larger, which also
seems natural as the relative velocity between the injection and the gas phase
velocity increases too. The same observation holds to explain the evolution of
$u_x^{max}$ and $u_x^{min}$. Concurrently, the skewness is positive and
increases with $We_2$, thus the axial velocity distribution is right-tailed with
an increasing asymmetry. Compared to the skewness of the size distribution, the
skewness of the distribution of $u_x$ is much smaller and the distribution
should be moderately skewed.  Finally, the excess kurtosis not only increases but also
changes sign for $We_2\in[40,70]$ . The \swi{} regime is then characterised by a
negative excess kurtosis, which indicates tails being shorter and a peak being flatter
than the ones of the Normal distribution.  Conversely, the excess kurtosis in the
atomisation regime, for the values of $We_2$ under consideration, is positive,
indicating larger tails and a sharper peak compared to the Normal distribution.
Furthermore, the excess kurtosis is smaller than 3 and the distribution has tails
shorter than the ones of the Gaussian distribution. Thus, each fragmentation
regime shows a characteristic tail spanning for the distribution of $u_x$.

The interpretation of the evolution of the statistical moments for the
distribution of $u_y$ is straightforward. As discussed previously, the
statistical axisymmetry of the flow enforces a zero mean value as well as a
symmetric distribution of $u_y$ around its mean, i.e. a zero skewness. Those two
consequences of the flow symmetry are verified for each $We_2$ value and
highlighted by the evolution of $u_y^{min}$ and $u_y^{max}$. Similarly to the
distribution of $u_x$, the standard deviation increases with the gaseous Weber
number because of the increasing relative velocity between the liquid injection
and the gas phase and thus the shear.  The excess kurtosis, the one subtracted
by 3, remains stable and is positive. This indicates a steady behavior and tails
being larger than those of the Gaussian distribution.

\begin{figure}
  \centering

  \begin{subfigure}[c]{0.49\textwidth}
            \centering
            \includegraphics[width=\textwidth]{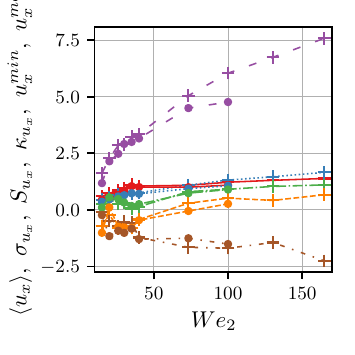}
                \caption{}
                \label{fig:dns.stat_ux_weg}
  \end{subfigure}%
  \begin{subfigure}[c]{0.49\textwidth}
            \centering
            \includegraphics[width=\textwidth]{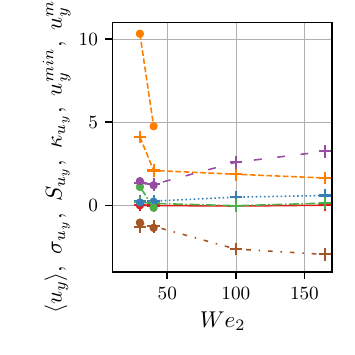}
                \caption{}
                \label{fig:dns.stat_uy_weg}
  \end{subfigure}%
  \caption{
    Evolution of $\langle \cdot \rangle$ (red), $\sigma$ (blue), $S$
    (green), $\kappa$ (orange), the minimum (purple) and maximum (brown) 
    against $We_2$ for the axial velocity $u_x$ (a) and the transversal
    velocity $u_y$ (b). The pluses ($+$) correspond to $t/T_a=15$ and the bullets
    ($\bullet$) to $t/T_a=25$. Note that $S$ and $\kappa$ are both dimensionless
    and that the dimensional variables are expressed with the SI base units.
  }
  \label{fig:dns.stat_vs_weg_vel}
\end{figure}

\section{Systematic fit campaign for testing the theoretical size distributions
\label{app.fit_cmpgn}}

The systematic fit campaign carried out to test the three distributions uses the
fitting algorithm of the \textsc{Ezyfit} toolbox developed by
\citet{moisy_ezyfit_2020} on \textsc{Matlab}. This algorithm is said to be able
to capture a given signal with a reference function when the parameters are set
with initial values of the same order as the final values.  Thus, the space of
initial values has to be explored sufficiently to ensure that the optimum set of
parameter values is captured for each theoretical distribution. To do so, the
fit campaign is performed in two phases. In the first phase, 23 combinations of
initial values are explored in linear and logarithmic modes, i.e. fitting the
signal or its logarithmic transform. In the second phase, the best fits in each
fitting mode and at the two time instant:s $t/T_a=\{15,25\}$ are selected and
tested a second time in order to improve the fit quality. Even if
$\mathcal{P}_{d/\langle d \rangle}$ shows several modes in the \swi{} regime, the
fit of the size distribution is carried out for the main mode only, i.e. with
only one theoretical distribution at a time. Finally, each theoretical PDF is
weighted by a coefficient $C$ which is let free in the fitting algorithm.
Generally speaking, a fit shows a good agreement with a given signal when the
Pearson coefficient $r$ is close to 1. One can also use $r^2$ as a more
discriminating criterion. 

\section{Vortex ring dynamics, deriving the velocity at the edge of the vortex
core \label{app.vortex_dyn}}

Assuming that the recirculation observed behind the jet head behaves as a vortex
ring behind a plate, it is possible to use the developments of \citet{saffman_vortex_1992}
which describe the dynamics of such unsteady objects. Let us consider a disc of
radius $a$ moving at a velocity $U_d$ in the direction normal to the disc
surface, denoted $x$ hereafter, a vortex ring
can develop on the downstream face and the velocity potential $\phi$ on the upstream
face follows:

\begin{equation}
  \phi = \mp\frac{2U_d}{\pi}\sqrt{a^2-r^2},~~ x=\pm0,~~ y^2+z^2=r^2<a^2  
  \label{eq:dns.vortexring_vel_potential}
\end{equation}

If the disc dissolves, the vortex ring remains with a strength $\kappa(r)=4 U_d
/ \pi \times r/\sqrt{a^2-r^2}$ and a vorticity $\omega=\kappa\theta\delta(x)$.
The amplitudes of the hydrodynamic impulse \footnote{The concept of
  hydrodynamic impulse has a long history in theoretical hydrodynamics having
  been described by \citet{lamb_hydrodynamics_1932}.  The advantage
  of the theory of hydrodynamic impulse is that it describes the physical origin
  of hydrodynamic forces and moments in terms of the vorticity generated at the
  body surface and its subsequent position in the fluid volume
\citep{holloway_hydrodynamic_2020}.} $I$ in the $x$ direction and the kinetic energy
$E$ are thus:

\begin{equation}
  I = \frac{1}{2} \int_{}^{}(\textnormal{\textbf{x}}\times \omega)_x~ dV =
  \frac{1}{2}\int_{0}^{a} 2\pi r^2 \kappa dr = 8U_d a^3 /3
  \label{eq:dns.vortexring_hydro_impulse}
\end{equation}

\begin{equation}
  E = \frac{1}{2} \int_{}^{} \phi \frac{\partial\phi}{\partial n} dS = 4 U_d^2
  a^3
  / 3
  \label{eq:dns.vortexring_energy_1}
\end{equation}

\noindent In addition, the circulation $\Gamma$ containing the disc while starting
and ending at the disc center is such that: 

\begin{equation}
  \Gamma = \int_{0}^{a}\kappa dr = [\phi]_{r=0}=4U_da/\pi
  \label{eq:dns.vortexring_circulation}
\end{equation}

\noindent Let us denote the vortex radius and the vortex core radius $R$ and $c$
and assume the conservation of the ring circulation and the hydrodynamic
impulse.  Knowing that the hydrodynamic impulse equals $\Gamma\pi R^2$
\citep{taylor_formation_1953}, the combination of Eqs.
\ref{eq:dns.vortexring_hydro_impulse} and
\ref{eq:dns.vortexring_circulation} results in $R=\sqrt{2/3}a$. Further
calculations give the expression of the vortex ring velocity $U_{vr}$ and of its
energy depending on $\Gamma$, $R$ and $c$:

\begin{equation}
  U_{vr}=\frac{\Gamma}{4\pi R}\Bigg[ \log\bigg( \frac{8R}{c}\bigg) - \frac{1}{2} +
    \int_{0}^{c}\bigg( \frac{\Gamma(s)}{\Gamma} \bigg)^2 \frac{ds}{s}+o\bigg(
\frac{c}{R} \bigg) \Bigg]
  \label{eq:dns.vortexring_velocity}
\end{equation}

\begin{equation}
  E=\frac{1}{2} \Gamma^2 R \Bigg[ \log\bigg( \frac{8R}{c}\bigg) - 2 +
    \int_{0}^{c}\bigg( \frac{\Gamma(s)}{\Gamma} \bigg)^2 \frac{ds}{s}+o\bigg(
\frac{c}{R} \bigg) \Bigg]
  \label{eq:dns.vortexring_energy_2}
\end{equation}

\noindent Combining the latter two equations with Eq. \ref{eq:dns.vortexring_energy_1}
enables to express the ratio of the vortex ring velocity $U_{vr}$ along $x$ and the disc
velocity $U_d$:

\begin{equation}
\frac{U_{vr}}{U_d} = \frac{1}{4} + \frac{1}{\pi^2} \bigg( \frac{3}{2}
\bigg)^{3/2} = 0.44
  \label{eq:dns.vortexring_vel_ratio}
\end{equation}

The question of the velocity at the edge of the vortex core remains and is of
most importance as it sets the droplet motion in the recirculation region. For a
uniform core, $c/R$ equals 0.19 while it equals $0.14$ in the case of a hollow
core. The velocity at the core edge, denoted $u_c$, can be expressed as a
function of the circulation $\Gamma$, $u_c=\Gamma / 2\pi c$. Using the
expression of $\Gamma$ given in Eq. \ref{eq:dns.vortexring_circulation} and $R=\sqrt{2/3}a$,
$u_c$ rewrites as: 

\begin{equation}
  u_c = \bigg(\frac{c}{R}\bigg)^{-1} \frac{2}{\pi^2\sqrt{2/3}}~U_d
\end{equation}

\section{Large density-ratio DNS \label{app.compl_dns}}

The comparison of the edges of the experimental and numerical joint volume
histogram over the space $(Oh_p/Oh_1, Re_p / Re_{axis})$, given in Figure
\ref{fig:dns.mapRO_contour_comparison_exp_num_tm15} indicates a discrepancy
along $Re_p / Re_{axis}$. In order to figure out if this discrepancy is due to
the difference in the density ratio, several DNS with larger density ratio
were ran. The gaseous Weber number is set to 40, as for the DNS 6. Except for
$\rho_2$, the parameters given by Table \ref{tab:dns.phys_prm} are kept the
same. Table \ref{tab:dns.uinj_we_re_f_sr_compl_dns} gives the injection velocity, $\rho_2$, the
corresponding $We_2$ and $Re_1$ along with the frequency $f$ of the most
unstable mode and the forcing Strouhal number $St$.

\begin{table} 
  \begin{center} 
    \def~{\hphantom{0}}
    \begin{tabular}{cccccccc}
    DNS & $U_{inj}$ & $\rho_2$ & $We_2$ & $Re_1$ & $f$ & $St$\\
    & $(\textnormal{m}/\textnormal{s})$ 
    & $(\textnormal{kg}/\textnormal{m}^{3})$ & & & $(\textnormal{kHz})$ & \\[4pt]
    11 & 2.714 & 1/82.5 & 40 & 12159 & 0.918 &1.52\\
    12 & 3.134 & 1/110  & 40 & 14040 & 0.915 & 1.31\\
    13 & 3.838 & 1/165  & 40 & 17195 & 0.911 & 1.06\\
  \end{tabular}
  \caption{
    Injection velocities, gaseous density and corresponding gas Weber and liquid
    Reynolds numbers along with the frequency $f$ of the most unstable mode and
    the corresponding forcing Strouhal number $St$.
  } 
\label{tab:dns.uinj_we_re_f_sr_compl_dns} 
\end{center} 
\end{table}

\clearpage
\bibliographystyle{jfm}
\bibliography{main}

\end{document}